\documentclass[longauth]{aa}

\usepackage{graphicx}
\usepackage{natbib}
\usepackage{scalerel}
\usepackage{comment}
\usepackage{multirow}
\usepackage{lastpage}

\usepackage[table]{xcolor}

\usepackage{txfonts}
\usepackage[pdfencoding=auto,psdextra]{hyperref}
\hypersetup{
    colorlinks=true,
    linkcolor=blue,
    filecolor=magenta,      
    urlcolor=blue,
    citecolor=blue
}
\makeatletter
\renewcommand*\aa@pageof{, page \thepage{} of \pageref*{LastPage}}
\makeatother

\usepackage[utf8]{inputenc}

\usepackage[switch, modulo]{lineno}
\linenumbers

\usepackage{euclid}

\begin{document}
%
%
\title{Euclid Quick Data Release (Q1)}
\subtitle{Exploring the complexity of quenching processes across time, environment, and mass through recently quenched galaxies}

\newcommand{\orcid}[1]{} 
\author{Euclid Collaboration: Z.~Mao\orcid{0000-0003-4016-845X}\thanks{\email{zhiying.mao@inaf.it}}\inst{\ref{aff1},\ref{aff2}}
\and L.~Pozzetti\orcid{0000-0001-7085-0412}\inst{\ref{aff1}}
\and A.~Enia\orcid{0000-0002-0200-2857}\inst{\ref{aff1}}
\and S.~Quai\orcid{0000-0002-0449-8163}\inst{\ref{aff3},\ref{aff1}}
\and W.~G.~Hartley\inst{\ref{aff4}}
\and M.~Bolzonella\orcid{0000-0003-3278-4607}\inst{\ref{aff1}}
\and F.~Gentile\orcid{0000-0002-8008-9871}\inst{\ref{aff5},\ref{aff1}}
\and G.~De~Lucia\orcid{0000-0002-6220-9104}\inst{\ref{aff6}}
\and M.~Moresco\orcid{0000-0002-7616-7136}\inst{\ref{aff3},\ref{aff1}}
\and N.~Borghi\orcid{0000-0002-2889-8997}\inst{\ref{aff3},\ref{aff1},\ref{aff7}}
\and C.~Saulder\orcid{0000-0002-0408-5633}\inst{\ref{aff8},\ref{aff9}}
\and C.~Nipoti\orcid{0000-0003-3121-6616}\inst{\ref{aff3}}
\and M.~Talia\orcid{0000-0003-4352-2063}\inst{\ref{aff3},\ref{aff1}}
\and P.~Corcho-Caballero\orcid{0000-0001-6327-7080}\inst{\ref{aff10}}
\and L.~Gabarra\orcid{0000-0002-8486-8856}\inst{\ref{aff11}}
\and Abdurro'uf\orcid{0000-0002-5258-8761}\inst{\ref{aff12}}
\and L.~Wang\orcid{0000-0002-6736-9158}\inst{\ref{aff13},\ref{aff10}}
\and L.~Bisigello\orcid{0000-0003-0492-4924}\inst{\ref{aff14}}
\and E.~Daddi\orcid{0000-0002-3331-9590}\inst{\ref{aff15}}
\and M.~Siudek\orcid{0000-0002-2949-2155}\inst{\ref{aff16},\ref{aff17},\ref{aff18}}
\and C.~D'Eugenio\orcid{0000-0001-7344-3126}\inst{\ref{aff19},\ref{aff5}}
\and G.~Zamorani\orcid{0000-0002-2318-301X}\inst{\ref{aff1}}
\and F.~Calura\orcid{0000-0002-6175-0871}\inst{\ref{aff1}}
\and H.~Fu\orcid{0009-0002-8051-1056}\inst{\ref{aff20},\ref{aff21}}
\and R.~Ragusa\inst{\ref{aff22}}
\and F.~Shankar\orcid{0000-0001-8973-5051}\inst{\ref{aff21}}
\and M.~Hirschmann\orcid{0000-0002-3301-3321}\inst{\ref{aff23}}
\and O.~Cucciati\orcid{0000-0002-9336-7551}\inst{\ref{aff1}}
\and T.~Kodama\orcid{0000-0002-2993-1576}\inst{\ref{aff2}}
\and S.~Belli\orcid{0000-0002-5615-6018}\inst{\ref{aff3}}
\and L.~Spinoglio\orcid{0000-0001-8840-1551}\inst{\ref{aff24}}
\and J.~R.~Weaver\orcid{0000-0003-1614-196X}\inst{\ref{aff25}}
\and B.~Altieri\orcid{0000-0003-3936-0284}\inst{\ref{aff26}}
\and S.~Andreon\orcid{0000-0002-2041-8784}\inst{\ref{aff27}}
\and N.~Auricchio\orcid{0000-0003-4444-8651}\inst{\ref{aff1}}
\and C.~Baccigalupi\orcid{0000-0002-8211-1630}\inst{\ref{aff28},\ref{aff6},\ref{aff29},\ref{aff30}}
\and M.~Baldi\orcid{0000-0003-4145-1943}\inst{\ref{aff31},\ref{aff1},\ref{aff7}}
\and A.~Balestra\orcid{0000-0002-6967-261X}\inst{\ref{aff14}}
\and S.~Bardelli\orcid{0000-0002-8900-0298}\inst{\ref{aff1}}
\and P.~Battaglia\orcid{0000-0002-7337-5909}\inst{\ref{aff1}}
\and A.~Biviano\orcid{0000-0002-0857-0732}\inst{\ref{aff6},\ref{aff28}}
\and E.~Branchini\orcid{0000-0002-0808-6908}\inst{\ref{aff32},\ref{aff33},\ref{aff27}}
\and M.~Brescia\orcid{0000-0001-9506-5680}\inst{\ref{aff34},\ref{aff22}}
\and S.~Camera\orcid{0000-0003-3399-3574}\inst{\ref{aff35},\ref{aff36},\ref{aff37}}
\and G.~Ca\~nas-Herrera\orcid{0000-0003-2796-2149}\inst{\ref{aff38}}
\and V.~Capobianco\orcid{0000-0002-3309-7692}\inst{\ref{aff37}}
\and C.~Carbone\orcid{0000-0003-0125-3563}\inst{\ref{aff39}}
\and J.~Carretero\orcid{0000-0002-3130-0204}\inst{\ref{aff40},\ref{aff41}}
\and S.~Casas\orcid{0000-0002-4751-5138}\inst{\ref{aff42},\ref{aff43}}
\and F.~J.~Castander\orcid{0000-0001-7316-4573}\inst{\ref{aff16},\ref{aff44}}
\and M.~Castellano\orcid{0000-0001-9875-8263}\inst{\ref{aff45}}
\and G.~Castignani\orcid{0000-0001-6831-0687}\inst{\ref{aff1}}
\and S.~Cavuoti\orcid{0000-0002-3787-4196}\inst{\ref{aff22},\ref{aff46}}
\and K.~C.~Chambers\orcid{0000-0001-6965-7789}\inst{\ref{aff47}}
\and A.~Cimatti\inst{\ref{aff48}}
\and C.~Colodro-Conde\inst{\ref{aff17}}
\and G.~Congedo\orcid{0000-0003-2508-0046}\inst{\ref{aff49}}
\and L.~Conversi\orcid{0000-0002-6710-8476}\inst{\ref{aff50},\ref{aff26}}
\and Y.~Copin\orcid{0000-0002-5317-7518}\inst{\ref{aff51}}
\and F.~Courbin\orcid{0000-0003-0758-6510}\inst{\ref{aff52},\ref{aff53},\ref{aff54}}
\and H.~M.~Courtois\orcid{0000-0003-0509-1776}\inst{\ref{aff55}}
\and M.~Cropper\orcid{0000-0003-4571-9468}\inst{\ref{aff56}}
\and A.~Da~Silva\orcid{0000-0002-6385-1609}\inst{\ref{aff57},\ref{aff58}}
\and H.~Degaudenzi\orcid{0000-0002-5887-6799}\inst{\ref{aff4}}
\and C.~Dolding\orcid{0009-0003-7199-6108}\inst{\ref{aff56}}
\and H.~Dole\orcid{0000-0002-9767-3839}\inst{\ref{aff59}}
\and F.~Dubath\orcid{0000-0002-6533-2810}\inst{\ref{aff4}}
\and X.~Dupac\inst{\ref{aff26}}
\and M.~Farina\orcid{0000-0002-3089-7846}\inst{\ref{aff24}}
\and R.~Farinelli\inst{\ref{aff1}}
\and S.~Ferriol\inst{\ref{aff51}}
\and P.~Fosalba\orcid{0000-0002-1510-5214}\inst{\ref{aff44},\ref{aff16}}
\and S.~Fotopoulou\orcid{0000-0002-9686-254X}\inst{\ref{aff60}}
\and M.~Frailis\orcid{0000-0002-7400-2135}\inst{\ref{aff6}}
\and E.~Franceschi\orcid{0000-0002-0585-6591}\inst{\ref{aff1}}
\and M.~Fumana\orcid{0000-0001-6787-5950}\inst{\ref{aff39}}
\and S.~Galeotta\orcid{0000-0002-3748-5115}\inst{\ref{aff6}}
\and K.~George\orcid{0000-0002-1734-8455}\inst{\ref{aff61}}
\and B.~Gillis\orcid{0000-0002-4478-1270}\inst{\ref{aff49}}
\and C.~Giocoli\orcid{0000-0002-9590-7961}\inst{\ref{aff1},\ref{aff7}}
\and J.~Gracia-Carpio\orcid{0000-0003-4689-3134}\inst{\ref{aff8}}
\and A.~Grazian\orcid{0000-0002-5688-0663}\inst{\ref{aff14}}
\and F.~Grupp\inst{\ref{aff8},\ref{aff9}}
\and S.~Gwyn\orcid{0000-0001-8221-8406}\inst{\ref{aff62}}
\and S.~V.~H.~Haugan\orcid{0000-0001-9648-7260}\inst{\ref{aff63}}
\and S.~Hemmati\orcid{0000-0003-2226-5395}\inst{\ref{aff64}}
\and J.~Hoar\inst{\ref{aff26}}
\and W.~Holmes\inst{\ref{aff65}}
\and A.~Hornstrup\orcid{0000-0002-3363-0936}\inst{\ref{aff66},\ref{aff67}}
\and M.~Huertas-Company\orcid{0000-0002-1416-8483}\inst{\ref{aff17},\ref{aff68},\ref{aff69}}
\and K.~Jahnke\orcid{0000-0003-3804-2137}\inst{\ref{aff70}}
\and M.~Jhabvala\inst{\ref{aff71}}
\and B.~Joachimi\orcid{0000-0001-7494-1303}\inst{\ref{aff72}}
\and S.~Kermiche\orcid{0000-0002-0302-5735}\inst{\ref{aff73}}
\and A.~Kiessling\orcid{0000-0002-2590-1273}\inst{\ref{aff65}}
\and B.~Kubik\orcid{0009-0006-5823-4880}\inst{\ref{aff51}}
\and M.~K\"ummel\orcid{0000-0003-2791-2117}\inst{\ref{aff9}}
\and M.~Kunz\orcid{0000-0002-3052-7394}\inst{\ref{aff74}}
\and H.~Kurki-Suonio\orcid{0000-0002-4618-3063}\inst{\ref{aff75},\ref{aff76}}
\and A.~M.~C.~Le~Brun\orcid{0000-0002-0936-4594}\inst{\ref{aff77}}
\and V.~Le~Brun\orcid{0000-0002-5027-1939}\inst{\ref{aff78}}
\and S.~Ligori\orcid{0000-0003-4172-4606}\inst{\ref{aff37}}
\and P.~B.~Lilje\orcid{0000-0003-4324-7794}\inst{\ref{aff63}}
\and V.~Lindholm\orcid{0000-0003-2317-5471}\inst{\ref{aff75},\ref{aff76}}
\and I.~Lloro\orcid{0000-0001-5966-1434}\inst{\ref{aff79}}
\and M.~Magliocchetti\orcid{0000-0001-9158-4838}\inst{\ref{aff24}}
\and G.~Mainetti\orcid{0000-0003-2384-2377}\inst{\ref{aff80}}
\and D.~Maino\inst{\ref{aff81},\ref{aff39},\ref{aff82}}
\and E.~Maiorano\orcid{0000-0003-2593-4355}\inst{\ref{aff1}}
\and O.~Mansutti\orcid{0000-0001-5758-4658}\inst{\ref{aff6}}
\and S.~Marcin\inst{\ref{aff83}}
\and O.~Marggraf\orcid{0000-0001-7242-3852}\inst{\ref{aff84}}
\and M.~Martinelli\orcid{0000-0002-6943-7732}\inst{\ref{aff45},\ref{aff85}}
\and N.~Martinet\orcid{0000-0003-2786-7790}\inst{\ref{aff78}}
\and F.~Marulli\orcid{0000-0002-8850-0303}\inst{\ref{aff3},\ref{aff1},\ref{aff7}}
\and R.~J.~Massey\orcid{0000-0002-6085-3780}\inst{\ref{aff86}}
\and N.~Mauri\orcid{0000-0001-8196-1548}\inst{\ref{aff48},\ref{aff7}}
\and S.~Maurogordato\inst{\ref{aff87}}
\and E.~Medinaceli\orcid{0000-0002-4040-7783}\inst{\ref{aff1}}
\and S.~Mei\orcid{0000-0002-2849-559X}\inst{\ref{aff88},\ref{aff89}}
\and M.~Meneghetti\orcid{0000-0003-1225-7084}\inst{\ref{aff1},\ref{aff7}}
\and E.~Merlin\orcid{0000-0001-6870-8900}\inst{\ref{aff45}}
\and G.~Meylan\inst{\ref{aff90}}
\and P.~Monaco\orcid{0000-0003-2083-7564}\inst{\ref{aff91},\ref{aff6},\ref{aff29},\ref{aff28}}
\and A.~Mora\orcid{0000-0002-1922-8529}\inst{\ref{aff92}}
\and C.~Moretti\orcid{0000-0003-3314-8936}\inst{\ref{aff6},\ref{aff28},\ref{aff29}}
\and L.~Moscardini\orcid{0000-0002-3473-6716}\inst{\ref{aff3},\ref{aff1},\ref{aff7}}
\and R.~Nakajima\orcid{0009-0009-1213-7040}\inst{\ref{aff84}}
\and C.~Neissner\orcid{0000-0001-8524-4968}\inst{\ref{aff93},\ref{aff41}}
\and S.-M.~Niemi\orcid{0009-0005-0247-0086}\inst{\ref{aff94}}
\and J.~W.~Nightingale\orcid{0000-0002-8987-7401}\inst{\ref{aff95}}
\and C.~Padilla\orcid{0000-0001-7951-0166}\inst{\ref{aff93}}
\and S.~Paltani\orcid{0000-0002-8108-9179}\inst{\ref{aff4}}
\and F.~Pasian\orcid{0000-0002-4869-3227}\inst{\ref{aff6}}
\and K.~Pedersen\inst{\ref{aff96}}
\and W.~J.~Percival\orcid{0000-0002-0644-5727}\inst{\ref{aff97},\ref{aff98},\ref{aff99}}
\and V.~Pettorino\orcid{0000-0002-4203-9320}\inst{\ref{aff94}}
\and A.~Pezzotta\orcid{0000-0003-0726-2268}\inst{\ref{aff27}}
\and S.~Pires\orcid{0000-0002-0249-2104}\inst{\ref{aff15}}
\and G.~Polenta\orcid{0000-0003-4067-9196}\inst{\ref{aff100}}
\and M.~Poncet\inst{\ref{aff101}}
\and L.~A.~Popa\inst{\ref{aff102}}
\and F.~Raison\orcid{0000-0002-7819-6918}\inst{\ref{aff8}}
\and A.~Renzi\orcid{0000-0001-9856-1970}\inst{\ref{aff103},\ref{aff104}}
\and J.~Rhodes\orcid{0000-0002-4485-8549}\inst{\ref{aff65}}
\and G.~Riccio\inst{\ref{aff22}}
\and I.~Risso\orcid{0000-0003-2525-7761}\inst{\ref{aff27},\ref{aff33}}
\and E.~Romelli\orcid{0000-0003-3069-9222}\inst{\ref{aff6}}
\and M.~Roncarelli\orcid{0000-0001-9587-7822}\inst{\ref{aff1}}
\and H.~J.~A.~Rottgering\orcid{0000-0001-8887-2257}\inst{\ref{aff38}}
\and B.~Rusholme\orcid{0000-0001-7648-4142}\inst{\ref{aff64}}
\and R.~Saglia\orcid{0000-0003-0378-7032}\inst{\ref{aff9},\ref{aff8}}
\and Z.~Sakr\orcid{0000-0002-4823-3757}\inst{\ref{aff105},\ref{aff106},\ref{aff107}}
\and D.~Sapone\orcid{0000-0001-7089-4503}\inst{\ref{aff108}}
\and B.~Sartoris\orcid{0000-0003-1337-5269}\inst{\ref{aff9},\ref{aff6}}
\and P.~Schneider\orcid{0000-0001-8561-2679}\inst{\ref{aff84}}
\and T.~Schrabback\orcid{0000-0002-6987-7834}\inst{\ref{aff109}}
\and M.~Scodeggio\inst{\ref{aff39}}
\and A.~Secroun\orcid{0000-0003-0505-3710}\inst{\ref{aff73}}
\and G.~Seidel\orcid{0000-0003-2907-353X}\inst{\ref{aff70}}
\and E.~Sihvola\orcid{0000-0003-1804-7715}\inst{\ref{aff110}}
\and P.~Simon\inst{\ref{aff84}}
\and C.~Sirignano\orcid{0000-0002-0995-7146}\inst{\ref{aff103},\ref{aff104}}
\and G.~Sirri\orcid{0000-0003-2626-2853}\inst{\ref{aff7}}
\and L.~Stanco\orcid{0000-0002-9706-5104}\inst{\ref{aff104}}
\and P.~Tallada-Cresp\'{i}\orcid{0000-0002-1336-8328}\inst{\ref{aff40},\ref{aff41}}
\and A.~N.~Taylor\inst{\ref{aff49}}
\and H.~I.~Teplitz\orcid{0000-0002-7064-5424}\inst{\ref{aff111}}
\and I.~Tereno\orcid{0000-0002-4537-6218}\inst{\ref{aff57},\ref{aff112}}
\and N.~Tessore\orcid{0000-0002-9696-7931}\inst{\ref{aff56}}
\and S.~Toft\orcid{0000-0003-3631-7176}\inst{\ref{aff113},\ref{aff114}}
\and R.~Toledo-Moreo\orcid{0000-0002-2997-4859}\inst{\ref{aff115}}
\and F.~Torradeflot\orcid{0000-0003-1160-1517}\inst{\ref{aff41},\ref{aff40}}
\and I.~Tutusaus\orcid{0000-0002-3199-0399}\inst{\ref{aff16},\ref{aff44},\ref{aff106}}
\and J.~Valiviita\orcid{0000-0001-6225-3693}\inst{\ref{aff75},\ref{aff76}}
\and T.~Vassallo\orcid{0000-0001-6512-6358}\inst{\ref{aff6},\ref{aff61}}
\and G.~Verdoes~Kleijn\orcid{0000-0001-5803-2580}\inst{\ref{aff10}}
\and A.~Veropalumbo\orcid{0000-0003-2387-1194}\inst{\ref{aff27},\ref{aff33},\ref{aff32}}
\and Y.~Wang\orcid{0000-0002-4749-2984}\inst{\ref{aff64}}
\and J.~Weller\orcid{0000-0002-8282-2010}\inst{\ref{aff9},\ref{aff8}}
\and A.~Zacchei\orcid{0000-0003-0396-1192}\inst{\ref{aff6},\ref{aff28}}
\and F.~M.~Zerbi\orcid{0000-0002-9996-973X}\inst{\ref{aff27}}
\and I.~A.~Zinchenko\orcid{0000-0002-2944-2449}\inst{\ref{aff116}}
\and E.~Zucca\orcid{0000-0002-5845-8132}\inst{\ref{aff1}}
\and V.~Allevato\orcid{0000-0001-7232-5152}\inst{\ref{aff22}}
\and M.~Ballardini\orcid{0000-0003-4481-3559}\inst{\ref{aff117},\ref{aff118},\ref{aff1}}
\and E.~Bozzo\orcid{0000-0002-8201-1525}\inst{\ref{aff4}}
\and C.~Burigana\orcid{0000-0002-3005-5796}\inst{\ref{aff119},\ref{aff120}}
\and A.~Cappi\inst{\ref{aff87},\ref{aff1}}
\and T.~Castro\orcid{0000-0002-6292-3228}\inst{\ref{aff6},\ref{aff29},\ref{aff28},\ref{aff121}}
\and J.~A.~Escartin~Vigo\inst{\ref{aff8}}
\and R.~Maoli\orcid{0000-0002-6065-3025}\inst{\ref{aff122},\ref{aff45}}
\and J.~Mart\'{i}n-Fleitas\orcid{0000-0002-8594-569X}\inst{\ref{aff123}}
\and R.~B.~Metcalf\orcid{0000-0003-3167-2574}\inst{\ref{aff3},\ref{aff1}}
\and A.~A.~Nucita\inst{\ref{aff124},\ref{aff125},\ref{aff126}}
\and M.~P\"ontinen\orcid{0000-0001-5442-2530}\inst{\ref{aff75}}
\and V.~Scottez\orcid{0009-0008-3864-940X}\inst{\ref{aff127},\ref{aff128}}
\and M.~Sereno\orcid{0000-0003-0302-0325}\inst{\ref{aff1},\ref{aff7}}
\and M.~Tenti\orcid{0000-0002-4254-5901}\inst{\ref{aff7}}
\and M.~Tucci\inst{\ref{aff4}}
\and M.~Viel\orcid{0000-0002-2642-5707}\inst{\ref{aff28},\ref{aff6},\ref{aff30},\ref{aff29},\ref{aff121}}
\and M.~Wiesmann\orcid{0009-0000-8199-5860}\inst{\ref{aff63}}
\and Y.~Akrami\orcid{0000-0002-2407-7956}\inst{\ref{aff105},\ref{aff129}}
\and I.~T.~Andika\orcid{0000-0001-6102-9526}\inst{\ref{aff61}}
\and S.~Anselmi\orcid{0000-0002-3579-9583}\inst{\ref{aff104},\ref{aff103},\ref{aff130}}
\and M.~Archidiacono\orcid{0000-0003-4952-9012}\inst{\ref{aff81},\ref{aff82}}
\and F.~Atrio-Barandela\orcid{0000-0002-2130-2513}\inst{\ref{aff131}}
\and L.~Bazzanini\orcid{0000-0003-0727-0137}\inst{\ref{aff117},\ref{aff1}}
\and D.~Bertacca\orcid{0000-0002-2490-7139}\inst{\ref{aff103},\ref{aff14},\ref{aff104}}
\and M.~Bethermin\orcid{0000-0002-3915-2015}\inst{\ref{aff132}}
\and F.~Beutler\orcid{0000-0003-0467-5438}\inst{\ref{aff49}}
\and L.~Blot\orcid{0000-0002-9622-7167}\inst{\ref{aff133},\ref{aff77}}
\and H.~B\"ohringer\orcid{0000-0001-8241-4204}\inst{\ref{aff8},\ref{aff61},\ref{aff134}}
\and M.~Bonici\orcid{0000-0002-8430-126X}\inst{\ref{aff97},\ref{aff39}}
\and S.~Borgani\orcid{0000-0001-6151-6439}\inst{\ref{aff91},\ref{aff28},\ref{aff6},\ref{aff29},\ref{aff121}}
\and M.~L.~Brown\orcid{0000-0002-0370-8077}\inst{\ref{aff135}}
\and S.~Bruton\orcid{0000-0002-6503-5218}\inst{\ref{aff136}}
\and A.~Calabro\orcid{0000-0003-2536-1614}\inst{\ref{aff45}}
\and B.~Camacho~Quevedo\orcid{0000-0002-8789-4232}\inst{\ref{aff28},\ref{aff30},\ref{aff6}}
\and F.~Caro\inst{\ref{aff45}}
\and C.~S.~Carvalho\inst{\ref{aff112}}
\and F.~Cogato\orcid{0000-0003-4632-6113}\inst{\ref{aff3},\ref{aff1}}
\and T.~E.~Collett\orcid{0000-0001-5564-3140}\inst{\ref{aff137}}
\and T.~Contini\orcid{0000-0003-0275-938X}\inst{\ref{aff106}}
\and A.~R.~Cooray\orcid{0000-0002-3892-0190}\inst{\ref{aff138}}
\and S.~Davini\orcid{0000-0003-3269-1718}\inst{\ref{aff33}}
\and G.~Desprez\orcid{0000-0001-8325-1742}\inst{\ref{aff10}}
\and A.~D\'iaz-S\'anchez\orcid{0000-0003-0748-4768}\inst{\ref{aff139}}
\and S.~Di~Domizio\orcid{0000-0003-2863-5895}\inst{\ref{aff32},\ref{aff33}}
\and J.~M.~Diego\orcid{0000-0001-9065-3926}\inst{\ref{aff140}}
\and P.-A.~Duc\orcid{0000-0003-3343-6284}\inst{\ref{aff132}}
\and V.~Duret\orcid{0009-0009-0383-4960}\inst{\ref{aff73}}
\and M.~Y.~Elkhashab\orcid{0000-0001-9306-2603}\inst{\ref{aff91},\ref{aff6},\ref{aff29},\ref{aff28}}
\and Y.~Fang\orcid{0000-0002-0334-6950}\inst{\ref{aff9}}
\and A.~Finoguenov\orcid{0000-0002-4606-5403}\inst{\ref{aff75}}
\and F.~Fontanot\orcid{0000-0003-4744-0188}\inst{\ref{aff6},\ref{aff28}}
\and A.~Franco\orcid{0000-0002-4761-366X}\inst{\ref{aff125},\ref{aff124},\ref{aff126}}
\and K.~Ganga\orcid{0000-0001-8159-8208}\inst{\ref{aff88}}
\and T.~Gasparetto\orcid{0000-0002-7913-4866}\inst{\ref{aff45}}
\and R.~Gavazzi\orcid{0000-0002-5540-6935}\inst{\ref{aff78},\ref{aff141}}
\and E.~Gaztanaga\orcid{0000-0001-9632-0815}\inst{\ref{aff16},\ref{aff44},\ref{aff137}}
\and F.~Giacomini\orcid{0000-0002-3129-2814}\inst{\ref{aff7}}
\and F.~Gianotti\orcid{0000-0003-4666-119X}\inst{\ref{aff1}}
\and G.~Gozaliasl\orcid{0000-0002-0236-919X}\inst{\ref{aff142},\ref{aff75}}
\and M.~Guidi\orcid{0000-0001-9408-1101}\inst{\ref{aff31},\ref{aff1}}
\and C.~M.~Gutierrez\orcid{0000-0001-7854-783X}\inst{\ref{aff17},\ref{aff18}}
\and A.~Hall\orcid{0000-0002-3139-8651}\inst{\ref{aff49}}
\and C.~Hern\'andez-Monteagudo\orcid{0000-0001-5471-9166}\inst{\ref{aff18},\ref{aff17}}
\and H.~Hildebrandt\orcid{0000-0002-9814-3338}\inst{\ref{aff143}}
\and J.~Hjorth\orcid{0000-0002-4571-2306}\inst{\ref{aff96}}
\and L.~K.~Hunt\orcid{0000-0001-9162-2371}\inst{\ref{aff144}}
\and J.~J.~E.~Kajava\orcid{0000-0002-3010-8333}\inst{\ref{aff145},\ref{aff146},\ref{aff147}}
\and Y.~Kang\orcid{0009-0000-8588-7250}\inst{\ref{aff4}}
\and V.~Kansal\orcid{0000-0002-4008-6078}\inst{\ref{aff148},\ref{aff149}}
\and D.~Karagiannis\orcid{0000-0002-4927-0816}\inst{\ref{aff117},\ref{aff150}}
\and K.~Kiiveri\inst{\ref{aff110}}
\and J.~Kim\orcid{0000-0003-2776-2761}\inst{\ref{aff11}}
\and C.~C.~Kirkpatrick\inst{\ref{aff110}}
\and S.~Kruk\orcid{0000-0001-8010-8879}\inst{\ref{aff26}}
\and F.~Lepori\orcid{0009-0000-5061-7138}\inst{\ref{aff151}}
\and G.~Leroy\orcid{0009-0004-2523-4425}\inst{\ref{aff152},\ref{aff86}}
\and G.~F.~Lesci\orcid{0000-0002-4607-2830}\inst{\ref{aff3},\ref{aff1}}
\and J.~Lesgourgues\orcid{0000-0001-7627-353X}\inst{\ref{aff42}}
\and T.~I.~Liaudat\orcid{0000-0002-9104-314X}\inst{\ref{aff153}}
\and S.~J.~Liu\orcid{0000-0001-7680-2139}\inst{\ref{aff24}}
\and X.~Lopez~Lopez\orcid{0009-0008-5194-5908}\inst{\ref{aff1}}
\and A.~Loureiro\orcid{0000-0002-4371-0876}\inst{\ref{aff154},\ref{aff155}}
\and G.~Maggio\orcid{0000-0003-4020-4836}\inst{\ref{aff6}}
\and E.~A.~Magnier\orcid{0000-0002-7965-2815}\inst{\ref{aff47}}
\and C.~Mancini\orcid{0000-0002-4297-0561}\inst{\ref{aff39}}
\and F.~Mannucci\orcid{0000-0002-4803-2381}\inst{\ref{aff144}}
\and C.~J.~A.~P.~Martins\orcid{0000-0002-4886-9261}\inst{\ref{aff156},\ref{aff157}}
\and L.~Maurin\orcid{0000-0002-8406-0857}\inst{\ref{aff59}}
\and M.~Miluzio\inst{\ref{aff26},\ref{aff158}}
\and G.~Morgante\inst{\ref{aff1}}
\and K.~Naidoo\orcid{0000-0002-9182-1802}\inst{\ref{aff137},\ref{aff70}}
\and A.~Navarro-Alsina\orcid{0000-0002-3173-2592}\inst{\ref{aff84}}
\and S.~Nesseris\orcid{0000-0002-0567-0324}\inst{\ref{aff105}}
\and D.~Paoletti\orcid{0000-0003-4761-6147}\inst{\ref{aff1},\ref{aff120}}
\and F.~Passalacqua\orcid{0000-0002-8606-4093}\inst{\ref{aff103},\ref{aff104}}
\and K.~Paterson\orcid{0000-0001-8340-3486}\inst{\ref{aff70}}
\and L.~Patrizii\inst{\ref{aff7}}
\and A.~Pisani\orcid{0000-0002-6146-4437}\inst{\ref{aff73}}
\and D.~Potter\orcid{0000-0002-0757-5195}\inst{\ref{aff159}}
\and G.~W.~Pratt\inst{\ref{aff15}}
\and M.~Radovich\orcid{0000-0002-3585-866X}\inst{\ref{aff14}}
\and G.~Rodighiero\orcid{0000-0002-9415-2296}\inst{\ref{aff103},\ref{aff14}}
\and W.~Roster\orcid{0000-0002-9149-6528}\inst{\ref{aff8}}
\and S.~Sacquegna\orcid{0000-0002-8433-6630}\inst{\ref{aff160}}
\and M.~Sahl\'en\orcid{0000-0003-0973-4804}\inst{\ref{aff161}}
\and D.~B.~Sanders\orcid{0000-0002-1233-9998}\inst{\ref{aff47}}
\and E.~Sarpa\orcid{0000-0002-1256-655X}\inst{\ref{aff30},\ref{aff121},\ref{aff6}}
\and C.~Scarlata\orcid{0000-0002-9136-8876}\inst{\ref{aff162}}
\and A.~Schneider\orcid{0000-0001-7055-8104}\inst{\ref{aff159}}
\and M.~Schultheis\inst{\ref{aff87}}
\and D.~Sciotti\orcid{0009-0008-4519-2620}\inst{\ref{aff45},\ref{aff85}}
\and E.~Sellentin\inst{\ref{aff163},\ref{aff38}}
\and L.~C.~Smith\orcid{0000-0002-3259-2771}\inst{\ref{aff164}}
\and J.~G.~Sorce\orcid{0000-0002-2307-2432}\inst{\ref{aff165},\ref{aff59}}
\and K.~Tanidis\orcid{0000-0001-9843-5130}\inst{\ref{aff166}}
\and F.~Tarsitano\orcid{0000-0002-5919-0238}\inst{\ref{aff167},\ref{aff4}}
\and G.~Testera\inst{\ref{aff33}}
\and R.~Teyssier\orcid{0000-0001-7689-0933}\inst{\ref{aff168}}
\and A.~Troja\orcid{0000-0003-0239-4595}\inst{\ref{aff103},\ref{aff104}}
\and A.~Venhola\orcid{0000-0001-6071-4564}\inst{\ref{aff169}}
\and D.~Vergani\orcid{0000-0003-0898-2216}\inst{\ref{aff1}}
\and G.~Verza\orcid{0000-0002-1886-8348}\thanks{Deceased}\inst{\ref{aff170},\ref{aff171}}
\and S.~Vinciguerra\orcid{0009-0005-4018-3184}\inst{\ref{aff78}}
\and N.~A.~Walton\orcid{0000-0003-3983-8778}\inst{\ref{aff164}}
\and A.~H.~Wright\orcid{0000-0001-7363-7932}\inst{\ref{aff143}}}
										   
\institute{INAF-Osservatorio di Astrofisica e Scienza dello Spazio di Bologna, Via Piero Gobetti 93/3, 40129 Bologna, Italy\label{aff1}
\and
Astronomical Institute, Tohoku University, Sendai 980-8578, Japan\label{aff2}
\and
Dipartimento di Fisica e Astronomia "Augusto Righi" - Alma Mater Studiorum Universit\`a di Bologna, via Piero Gobetti 93/2, 40129 Bologna, Italy\label{aff3}
\and
Department of Astronomy, University of Geneva, ch. d'Ecogia 16, 1290 Versoix, Switzerland\label{aff4}
\and
CEA Saclay, DFR/IRFU, Service d'Astrophysique, Bat. 709, 91191 Gif-sur-Yvette, France\label{aff5}
\and
INAF-Osservatorio Astronomico di Trieste, Via G. B. Tiepolo 11, 34143 Trieste, Italy\label{aff6}
\and
INFN-Sezione di Bologna, Viale Berti Pichat 6/2, 40127 Bologna, Italy\label{aff7}
\and
Max Planck Institute for Extraterrestrial Physics, Giessenbachstr. 1, 85748 Garching, Germany\label{aff8}
\and
Universit\"ats-Sternwarte M\"unchen, Fakult\"at f\"ur Physik, Ludwig-Maximilians-Universit\"at M\"unchen, Scheinerstr.~1, 81679 M\"unchen, Germany\label{aff9}
\and
Kapteyn Astronomical Institute, University of Groningen, PO Box 800, 9700 AV Groningen, The Netherlands\label{aff10}
\and
Department of Physics, University of Oxford, Keble Road, Oxford OX1 3RH, UK\label{aff11}
\and
Department of Astronomy, Indiana University, 727 East Third Street, Bloomington, IN 47405, USA\label{aff12}
\and
SRON Netherlands Institute for Space Research, Landleven 12, 9747 AD, Groningen, The Netherlands\label{aff13}
\and
INAF-Osservatorio Astronomico di Padova, Via dell'Osservatorio 5, 35122 Padova, Italy\label{aff14}
\and
Universit\'e Paris-Saclay, Universit\'e Paris Cit\'e, CEA, CNRS, AIM, 91191, Gif-sur-Yvette, France\label{aff15}
\and
Institute of Space Sciences (ICE, CSIC), Campus UAB, Carrer de Can Magrans, s/n, 08193 Barcelona, Spain\label{aff16}
\and
Instituto de Astrof\'{\i}sica de Canarias, E-38205 La Laguna, Tenerife, Spain\label{aff17}
\and
Universidad de La Laguna, Dpto. Astrof\'\i sica, E-38206 La Laguna, Tenerife, Spain\label{aff18}
\and
Institute de Physique du Globe de Paris, 1 Rue Jussieu, 75005, Paris, France\label{aff19}
\and
Center for Astronomy and Astrophysics and Department of Physics, Fudan University, Shanghai 200438, People's Republic of China\label{aff20}
\and
School of Physics \& Astronomy, University of Southampton, Highfield Campus, Southampton SO17 1BJ, UK\label{aff21}
\and
INAF-Osservatorio Astronomico di Capodimonte, Via Moiariello 16, 80131 Napoli, Italy\label{aff22}
\and
Institute of Physics, Laboratory for Galaxy Evolution, Ecole Polytechnique F\'ed\'erale de Lausanne, Observatoire de Sauverny, CH-1290 Versoix, Switzerland\label{aff23}
\and
INAF-Istituto di Astrofisica e Planetologia Spaziali, via del Fosso del Cavaliere, 100, 00100 Roma, Italy\label{aff24}
\and
MIT Kavli Institute for Astrophysics and Space Research, Massachusetts Institute of Technology, Cambridge, MA 02139, USA\label{aff25}
\and
ESAC/ESA, Camino Bajo del Castillo, s/n., Urb. Villafranca del Castillo, 28692 Villanueva de la Ca\~nada, Madrid, Spain\label{aff26}
\and
INAF-Osservatorio Astronomico di Brera, Via Brera 28, 20122 Milano, Italy\label{aff27}
\and
IFPU, Institute for Fundamental Physics of the Universe, via Beirut 2, 34151 Trieste, Italy\label{aff28}
\and
INFN, Sezione di Trieste, Via Valerio 2, 34127 Trieste TS, Italy\label{aff29}
\and
SISSA, International School for Advanced Studies, Via Bonomea 265, 34136 Trieste TS, Italy\label{aff30}
\and
Dipartimento di Fisica e Astronomia, Universit\`a di Bologna, Via Gobetti 93/2, 40129 Bologna, Italy\label{aff31}
\and
Dipartimento di Fisica, Universit\`a di Genova, Via Dodecaneso 33, 16146, Genova, Italy\label{aff32}
\and
INFN-Sezione di Genova, Via Dodecaneso 33, 16146, Genova, Italy\label{aff33}
\and
Department of Physics "E. Pancini", University Federico II, Via Cinthia 6, 80126, Napoli, Italy\label{aff34}
\and
Dipartimento di Fisica, Universit\`a degli Studi di Torino, Via P. Giuria 1, 10125 Torino, Italy\label{aff35}
\and
INFN-Sezione di Torino, Via P. Giuria 1, 10125 Torino, Italy\label{aff36}
\and
INAF-Osservatorio Astrofisico di Torino, Via Osservatorio 20, 10025 Pino Torinese (TO), Italy\label{aff37}
\and
Leiden Observatory, Leiden University, Einsteinweg 55, 2333 CC Leiden, The Netherlands\label{aff38}
\and
INAF-IASF Milano, Via Alfonso Corti 12, 20133 Milano, Italy\label{aff39}
\and
Centro de Investigaciones Energ\'eticas, Medioambientales y Tecnol\'ogicas (CIEMAT), Avenida Complutense 40, 28040 Madrid, Spain\label{aff40}
\and
Port d'Informaci\'{o} Cient\'{i}fica, Campus UAB, C. Albareda s/n, 08193 Bellaterra (Barcelona), Spain\label{aff41}
\and
Institute for Theoretical Particle Physics and Cosmology (TTK), RWTH Aachen University, 52056 Aachen, Germany\label{aff42}
\and
Deutsches Zentrum f\"ur Luft- und Raumfahrt e. V. (DLR), Linder H\"ohe, 51147 K\"oln, Germany\label{aff43}
\and
Institut d'Estudis Espacials de Catalunya (IEEC),  Edifici RDIT, Campus UPC, 08860 Castelldefels, Barcelona, Spain\label{aff44}
\and
INAF-Osservatorio Astronomico di Roma, Via Frascati 33, 00078 Monteporzio Catone, Italy\label{aff45}
\and
INFN -- Sezione di Napoli, Via Cinthia 6, 80126, Napoli, Italy\label{aff46}
\and
Institute for Astronomy, University of Hawaii, 2680 Woodlawn Drive, Honolulu, HI 96822, USA\label{aff47}
\and
Dipartimento di Fisica e Astronomia "Augusto Righi" - Alma Mater Studiorum Universit\`a di Bologna, Viale Berti Pichat 6/2, 40127 Bologna, Italy\label{aff48}
\and
Institute for Astronomy, University of Edinburgh, Royal Observatory, Blackford Hill, Edinburgh EH9 3HJ, UK\label{aff49}
\and
European Space Agency/ESRIN, Largo Galileo Galilei 1, 00044 Frascati, Roma, Italy\label{aff50}
\and
Universit\'e Claude Bernard Lyon 1, CNRS/IN2P3, IP2I Lyon, UMR 5822, Villeurbanne, F-69100, France\label{aff51}
\and
Institut de Ci\`{e}ncies del Cosmos (ICCUB), Universitat de Barcelona (IEEC-UB), Mart\'{i} i Franqu\`{e}s 1, 08028 Barcelona, Spain\label{aff52}
\and
Instituci\'o Catalana de Recerca i Estudis Avan\c{c}ats (ICREA), Passeig de Llu\'{\i}s Companys 23, 08010 Barcelona, Spain\label{aff53}
\and
Institut de Ciencies de l'Espai (IEEC-CSIC), Campus UAB, Carrer de Can Magrans, s/n Cerdanyola del Vall\'es, 08193 Barcelona, Spain\label{aff54}
\and
UCB Lyon 1, CNRS/IN2P3, IUF, IP2I Lyon, 4 rue Enrico Fermi, 69622 Villeurbanne, France\label{aff55}
\and
Mullard Space Science Laboratory, University College London, Holmbury St Mary, Dorking, Surrey RH5 6NT, UK\label{aff56}
\and
Departamento de F\'isica, Faculdade de Ci\^encias, Universidade de Lisboa, Edif\'icio C8, Campo Grande, PT1749-016 Lisboa, Portugal\label{aff57}
\and
Instituto de Astrof\'isica e Ci\^encias do Espa\c{c}o, Faculdade de Ci\^encias, Universidade de Lisboa, Campo Grande, 1749-016 Lisboa, Portugal\label{aff58}
\and
Universit\'e Paris-Saclay, CNRS, Institut d'astrophysique spatiale, 91405, Orsay, France\label{aff59}
\and
School of Physics, HH Wills Physics Laboratory, University of Bristol, Tyndall Avenue, Bristol, BS8 1TL, UK\label{aff60}
\and
University Observatory, LMU Faculty of Physics, Scheinerstr.~1, 81679 Munich, Germany\label{aff61}
\and
National Research Council, Herzberg Astronomy and Astrophysics Research Centre, 5071 W. Saanich Rd. Victoria, BC, V9E 2E7, Canada\label{aff62}
\and
Institute of Theoretical Astrophysics, University of Oslo, P.O. Box 1029 Blindern, 0315 Oslo, Norway\label{aff63}
\and
Caltech/IPAC, 1200 E. California Blvd., Pasadena, CA 91125, USA\label{aff64}
\and
Jet Propulsion Laboratory, California Institute of Technology, 4800 Oak Grove Drive, Pasadena, CA, 91109, USA\label{aff65}
\and
Technical University of Denmark, Elektrovej 327, 2800 Kgs. Lyngby, Denmark\label{aff66}
\and
Cosmic Dawn Center (DAWN), Denmark\label{aff67}
\and
Universit\'e PSL, Observatoire de Paris, Sorbonne Universit\'e, CNRS, LERMA, 75014, Paris, France\label{aff68}
\and
Universit\'e Paris-Cit\'e, 5 Rue Thomas Mann, 75013, Paris, France\label{aff69}
\and
Max-Planck-Institut f\"ur Astronomie, K\"onigstuhl 17, 69117 Heidelberg, Germany\label{aff70}
\and
NASA Goddard Space Flight Center, Greenbelt, MD 20771, USA\label{aff71}
\and
Department of Physics and Astronomy, University College London, Gower Street, London WC1E 6BT, UK\label{aff72}
\and
Aix-Marseille Universit\'e, CNRS/IN2P3, CPPM, Marseille, France\label{aff73}
\and
Universit\'e de Gen\`eve, D\'epartement de Physique Th\'eorique and Centre for Astroparticle Physics, 24 quai Ernest-Ansermet, CH-1211 Gen\`eve 4, Switzerland\label{aff74}
\and
Department of Physics, P.O. Box 64, University of Helsinki, 00014 Helsinki, Finland\label{aff75}
\and
Helsinki Institute of Physics, Gustaf H{\"a}llstr{\"o}min katu 2, University of Helsinki, 00014 Helsinki, Finland\label{aff76}
\and
Laboratoire d'etude de l'Univers et des phenomenes eXtremes, Observatoire de Paris, Universit\'e PSL, Sorbonne Universit\'e, CNRS, 92190 Meudon, France\label{aff77}
\and
Aix-Marseille Universit\'e, CNRS, CNES, LAM, Marseille, France\label{aff78}
\and
SKAO, Jodrell Bank, Lower Withington, Macclesfield SK11 9FT, UK\label{aff79}
\and
Centre de Calcul de l'IN2P3/CNRS, 21 avenue Pierre de Coubertin 69627 Villeurbanne Cedex, France\label{aff80}
\and
Dipartimento di Fisica "Aldo Pontremoli", Universit\`a degli Studi di Milano, Via Celoria 16, 20133 Milano, Italy\label{aff81}
\and
INFN-Sezione di Milano, Via Celoria 16, 20133 Milano, Italy\label{aff82}
\and
University of Applied Sciences and Arts of Northwestern Switzerland, School of Computer Science, 5210 Windisch, Switzerland\label{aff83}
\and
Universit\"at Bonn, Argelander-Institut f\"ur Astronomie, Auf dem H\"ugel 71, 53121 Bonn, Germany\label{aff84}
\and
INFN-Sezione di Roma, Piazzale Aldo Moro, 2 - c/o Dipartimento di Fisica, Edificio G. Marconi, 00185 Roma, Italy\label{aff85}
\and
Department of Physics, Institute for Computational Cosmology, Durham University, South Road, Durham, DH1 3LE, UK\label{aff86}
\and
Universit\'e C\^{o}te d'Azur, Observatoire de la C\^{o}te d'Azur, CNRS, Laboratoire Lagrange, Bd de l'Observatoire, CS 34229, 06304 Nice cedex 4, France\label{aff87}
\and
Universit\'e Paris Cit\'e, CNRS, Astroparticule et Cosmologie, 75013 Paris, France\label{aff88}
\and
CNRS-UCB International Research Laboratory, Centre Pierre Bin\'etruy, IRL2007, CPB-IN2P3, Berkeley, USA\label{aff89}
\and
Institute of Physics, Laboratory of Astrophysics, Ecole Polytechnique F\'ed\'erale de Lausanne (EPFL), Observatoire de Sauverny, 1290 Versoix, Switzerland\label{aff90}
\and
Dipartimento di Fisica - Sezione di Astronomia, Universit\`a di Trieste, Via Tiepolo 11, 34131 Trieste, Italy\label{aff91}
\and
Telespazio UK S.L. for European Space Agency (ESA), Camino bajo del Castillo, s/n, Urbanizacion Villafranca del Castillo, Villanueva de la Ca\~nada, 28692 Madrid, Spain\label{aff92}
\and
Institut de F\'{i}sica d'Altes Energies (IFAE), The Barcelona Institute of Science and Technology, Campus UAB, 08193 Bellaterra (Barcelona), Spain\label{aff93}
\and
European Space Agency/ESTEC, Keplerlaan 1, 2201 AZ Noordwijk, The Netherlands\label{aff94}
\and
School of Mathematics, Statistics and Physics, Newcastle University, Herschel Building, Newcastle-upon-Tyne, NE1 7RU, UK\label{aff95}
\and
DARK, Niels Bohr Institute, University of Copenhagen, Jagtvej 155, 2200 Copenhagen, Denmark\label{aff96}
\and
Waterloo Centre for Astrophysics, University of Waterloo, Waterloo, Ontario N2L 3G1, Canada\label{aff97}
\and
Department of Physics and Astronomy, University of Waterloo, Waterloo, Ontario N2L 3G1, Canada\label{aff98}
\and
Perimeter Institute for Theoretical Physics, Waterloo, Ontario N2L 2Y5, Canada\label{aff99}
\and
Space Science Data Center, Italian Space Agency, via del Politecnico snc, 00133 Roma, Italy\label{aff100}
\and
Centre National d'Etudes Spatiales -- Centre spatial de Toulouse, 18 avenue Edouard Belin, 31401 Toulouse Cedex 9, France\label{aff101}
\and
Institute of Space Science, Str. Atomistilor, nr. 409 M\u{a}gurele, Ilfov, 077125, Romania\label{aff102}
\and
Dipartimento di Fisica e Astronomia "G. Galilei", Universit\`a di Padova, Via Marzolo 8, 35131 Padova, Italy\label{aff103}
\and
INFN-Padova, Via Marzolo 8, 35131 Padova, Italy\label{aff104}
\and
Instituto de F\'isica Te\'orica UAM-CSIC, Campus de Cantoblanco, 28049 Madrid, Spain\label{aff105}
\and
Institut de Recherche en Astrophysique et Plan\'etologie (IRAP), Universit\'e de Toulouse, CNRS, UPS, CNES, 14 Av. Edouard Belin, 31400 Toulouse, France\label{aff106}
\and
Universit\'e St Joseph; Faculty of Sciences, Beirut, Lebanon\label{aff107}
\and
Departamento de F\'isica, FCFM, Universidad de Chile, Blanco Encalada 2008, Santiago, Chile\label{aff108}
\and
Universit\"at Innsbruck, Institut f\"ur Astro- und Teilchenphysik, Technikerstr. 25/8, 6020 Innsbruck, Austria\label{aff109}
\and
Department of Physics and Helsinki Institute of Physics, Gustaf H\"allstr\"omin katu 2, University of Helsinki, 00014 Helsinki, Finland\label{aff110}
\and
Infrared Processing and Analysis Center, California Institute of Technology, Pasadena, CA 91125, USA\label{aff111}
\and
Instituto de Astrof\'isica e Ci\^encias do Espa\c{c}o, Faculdade de Ci\^encias, Universidade de Lisboa, Tapada da Ajuda, 1349-018 Lisboa, Portugal\label{aff112}
\and
Cosmic Dawn Center (DAWN)\label{aff113}
\and
Niels Bohr Institute, University of Copenhagen, Jagtvej 128, 2200 Copenhagen, Denmark\label{aff114}
\and
Universidad Polit\'ecnica de Cartagena, Departamento de Electr\'onica y Tecnolog\'ia de Computadoras,  Plaza del Hospital 1, 30202 Cartagena, Spain\label{aff115}
\and
Astronomisches Rechen-Institut, Zentrum f\"ur Astronomie der Universit\"at Heidelberg, M\"onchhofstr. 12-14, 69120 Heidelberg, Germany\label{aff116}
\and
Dipartimento di Fisica e Scienze della Terra, Universit\`a degli Studi di Ferrara, Via Giuseppe Saragat 1, 44122 Ferrara, Italy\label{aff117}
\and
Istituto Nazionale di Fisica Nucleare, Sezione di Ferrara, Via Giuseppe Saragat 1, 44122 Ferrara, Italy\label{aff118}
\and
INAF, Istituto di Radioastronomia, Via Piero Gobetti 101, 40129 Bologna, Italy\label{aff119}
\and
INFN-Bologna, Via Irnerio 46, 40126 Bologna, Italy\label{aff120}
\and
ICSC - Centro Nazionale di Ricerca in High Performance Computing, Big Data e Quantum Computing, Via Magnanelli 2, Bologna, Italy\label{aff121}
\and
Dipartimento di Fisica, Sapienza Universit\`a di Roma, Piazzale Aldo Moro 2, 00185 Roma, Italy\label{aff122}
\and
Aurora Technology for European Space Agency (ESA), Camino bajo del Castillo, s/n, Urbanizacion Villafranca del Castillo, Villanueva de la Ca\~nada, 28692 Madrid, Spain\label{aff123}
\and
Department of Mathematics and Physics E. De Giorgi, University of Salento, Via per Arnesano, CP-I93, 73100, Lecce, Italy\label{aff124}
\and
INFN, Sezione di Lecce, Via per Arnesano, CP-193, 73100, Lecce, Italy\label{aff125}
\and
INAF-Sezione di Lecce, c/o Dipartimento Matematica e Fisica, Via per Arnesano, 73100, Lecce, Italy\label{aff126}
\and
Institut d'Astrophysique de Paris, 98bis Boulevard Arago, 75014, Paris, France\label{aff127}
\and
ICL, Junia, Universit\'e Catholique de Lille, LITL, 59000 Lille, France\label{aff128}
\and
CERCA/ISO, Department of Physics, Case Western Reserve University, 10900 Euclid Avenue, Cleveland, OH 44106, USA\label{aff129}
\and
Laboratoire Univers et Th\'eorie, Observatoire de Paris, Universit\'e PSL, Universit\'e Paris Cit\'e, CNRS, 92190 Meudon, France\label{aff130}
\and
Departamento de F{\'\i}sica Fundamental. Universidad de Salamanca. Plaza de la Merced s/n. 37008 Salamanca, Spain\label{aff131}
\and
Universit\'e de Strasbourg, CNRS, Observatoire astronomique de Strasbourg, UMR 7550, 67000 Strasbourg, France\label{aff132}
\and
Center for Data-Driven Discovery, Kavli IPMU (WPI), UTIAS, The University of Tokyo, Kashiwa, Chiba 277-8583, Japan\label{aff133}
\and
Max-Planck-Institut f\"ur Physik, Boltzmannstr. 8, 85748 Garching, Germany\label{aff134}
\and
Jodrell Bank Centre for Astrophysics, Department of Physics and Astronomy, University of Manchester, Oxford Road, Manchester M13 9PL, UK\label{aff135}
\and
California Institute of Technology, 1200 E California Blvd, Pasadena, CA 91125, USA\label{aff136}
\and
Institute of Cosmology and Gravitation, University of Portsmouth, Portsmouth PO1 3FX, UK\label{aff137}
\and
Department of Physics \& Astronomy, University of California Irvine, Irvine CA 92697, USA\label{aff138}
\and
Departamento F\'isica Aplicada, Universidad Polit\'ecnica de Cartagena, Campus Muralla del Mar, 30202 Cartagena, Murcia, Spain\label{aff139}
\and
Instituto de F\'isica de Cantabria, Edificio Juan Jord\'a, Avenida de los Castros, 39005 Santander, Spain\label{aff140}
\and
Institut d'Astrophysique de Paris, UMR 7095, CNRS, and Sorbonne Universit\'e, 98 bis boulevard Arago, 75014 Paris, France\label{aff141}
\and
Department of Computer Science, Aalto University, PO Box 15400, Espoo, FI-00 076, Finland\label{aff142}
\and
Ruhr University Bochum, Faculty of Physics and Astronomy, Astronomical Institute (AIRUB), German Centre for Cosmological Lensing (GCCL), 44780 Bochum, Germany\label{aff143}
\and
INAF-Osservatorio Astrofisico di Arcetri, Largo E. Fermi 5, 50125, Firenze, Italy\label{aff144}
\and
Department of Physics and Astronomy, Vesilinnantie 5, University of Turku, 20014 Turku, Finland\label{aff145}
\and
Finnish Centre for Astronomy with ESO (FINCA), Quantum, Vesilinnantie 5, University of Turku, 20014 Turku, Finland\label{aff146}
\and
Serco for European Space Agency (ESA), Camino bajo del Castillo, s/n, Urbanizacion Villafranca del Castillo, Villanueva de la Ca\~nada, 28692 Madrid, Spain\label{aff147}
\and
ARC Centre of Excellence for Dark Matter Particle Physics, Melbourne, Australia\label{aff148}
\and
Centre for Astrophysics \& Supercomputing, Swinburne University of Technology,  Hawthorn, Victoria 3122, Australia\label{aff149}
\and
Department of Physics and Astronomy, University of the Western Cape, Bellville, Cape Town, 7535, South Africa\label{aff150}
\and
Departement of Theoretical Physics, University of Geneva, Switzerland\label{aff151}
\and
Department of Physics, Centre for Extragalactic Astronomy, Durham University, South Road, Durham, DH1 3LE, UK\label{aff152}
\and
IRFU, CEA, Universit\'e Paris-Saclay 91191 Gif-sur-Yvette Cedex, France\label{aff153}
\and
Oskar Klein Centre for Cosmoparticle Physics, Department of Physics, Stockholm University, Stockholm, SE-106 91, Sweden\label{aff154}
\and
Astrophysics Group, Blackett Laboratory, Imperial College London, London SW7 2AZ, UK\label{aff155}
\and
Centro de Astrof\'{\i}sica da Universidade do Porto, Rua das Estrelas, 4150-762 Porto, Portugal\label{aff156}
\and
Instituto de Astrof\'isica e Ci\^encias do Espa\c{c}o, Universidade do Porto, CAUP, Rua das Estrelas, PT4150-762 Porto, Portugal\label{aff157}
\and
HE Space for European Space Agency (ESA), Camino bajo del Castillo, s/n, Urbanizacion Villafranca del Castillo, Villanueva de la Ca\~nada, 28692 Madrid, Spain\label{aff158}
\and
Department of Astrophysics, University of Zurich, Winterthurerstrasse 190, 8057 Zurich, Switzerland\label{aff159}
\and
INAF - Osservatorio Astronomico d'Abruzzo, Via Maggini, 64100, Teramo, Italy\label{aff160}
\and
Theoretical astrophysics, Department of Physics and Astronomy, Uppsala University, Box 516, 751 37 Uppsala, Sweden\label{aff161}
\and
Minnesota Institute for Astrophysics, University of Minnesota, 116 Church St SE, Minneapolis, MN 55455, USA\label{aff162}
\and
Mathematical Institute, University of Leiden, Einsteinweg 55, 2333 CA Leiden, The Netherlands\label{aff163}
\and
Institute of Astronomy, University of Cambridge, Madingley Road, Cambridge CB3 0HA, UK\label{aff164}
\and
Univ. Lille, CNRS, Centrale Lille, UMR 9189 CRIStAL, 59000 Lille, France\label{aff165}
\and
Center for Astrophysics and Cosmology, University of Nova Gorica, Nova Gorica, Slovenia\label{aff166}
\and
Institute for Particle Physics and Astrophysics, Dept. of Physics, ETH Zurich, Wolfgang-Pauli-Strasse 27, 8093 Zurich, Switzerland\label{aff167}
\and
Department of Astrophysical Sciences, Peyton Hall, Princeton University, Princeton, NJ 08544, USA\label{aff168}
\and
Space physics and astronomy research unit, University of Oulu, Pentti Kaiteran katu 1, FI-90014 Oulu, Finland\label{aff169}
\and
International Centre for Theoretical Physics (ICTP), Strada Costiera 11, 34151 Trieste, Italy\label{aff170}
\and
Center for Computational Astrophysics, Flatiron Institute, 162 5th Avenue, 10010, New York, NY, USA\label{aff171}}

\date{\today}

\abstract{Quenching is an essential process in galaxy evolution, and recently quenched galaxies (RQGs) are a key population for examining the underlying mechanisms. We present a photometric and spectroscopic analysis of RQGs using the rich data in the first Euclid Quick Data Release (Q1). Based on the rest-frame $UVJ$ diagram, we classify $5.6$ million star-forming (SFGs), $0.4$ million quiescent (QGs), and $0.4$ million recently quenched galaxies at $0.2<z<3.0$, with RQGs further separated into fast- and slow-quenching types (fRQGs and sRQGs) following model evolutionary tracks. We investigate the relative importance of the quenching process through the ratio between populations. The relative fractions of RQGs and their quenching efficiencies both decrease with cosmic time. At early epochs, fast quenching is more efficient in massive galaxies, whereas it becomes efficient only at low masses at later epochs. In contrast, slow quenching becomes more efficient at $z < 1.5$, progressing from high to intermediate masses with cosmic time. Most of the RQGs at $z<0.5$ reside in low-density environments, while they also exist in high-density environments at $z>0.5$, indicating that quenching in high-density environments occurs earlier than in low-density fields. RQGs show intermediate properties between SFGs and QGs in stellar mass, star-formation rate, specific star-formation rate, S\'ersic index, and radius. The star-formation history timescale $\tau$ of RQGs shows clear bimodality, peaking at $\tau< 0.05\,\mathrm{Gyr}$ for fRQGs, and at $\tau\sim 1\,\mathrm{Gyr}$ for sRQGs. In addition, our analysis reveals that most of the RQGs exhibit regular morphologies, excluding merger as a primary mechanism. We find that fRQGs still retain a disc-like morphology, suggesting they have not yet undergone a morphological transformation, unlike sRQGs, which have a more concentrated, early-type-like morphology. From our analysis, we infer that a complexity of different mechanisms drives quenching. Stacked spectra of a  subsample of $\sim45$ thousand photometrically classified galaxies with external spectroscopic redshift confirm that QGs are dominated by old, passive stellar populations, while sRQG spectrum shows weak \ha\ emission, indicating residual star-formation activity or AGN contribution. The recently-quenched nature of  RQGs is confirmed by strong Balmer absorption lines in stack of DESI spectra. The stacked \Euclid QG spectrum agrees well with previous templates and with stellar population models older than $5\,\mathrm{Gyr}$ at $z < 1$, despite the less prominent CN features. These results highlight the power of the \Euclid data in studying galaxy quenching, which will be further enhanced in future data releases. 
}
%
%
%
%
   \titlerunning{Quenching study through recently quenched galaxies}
   \authorrunning{Euclid Collaboration: Mao et al.}
   
   \maketitle
%
%
%
%
   
\section{\label{sc:Intro}Introduction}
In the galaxy population, there is a clear separation between, on the one hand, red, early-type, old, quiescent galaxies (QGs) and, on the other hand, blue, late-type, young, star-forming galaxies (SFGs), with active star formation rate (SFR). On a colour-magnitude diagram, well-separated bimodal distributions are identified as red-sequence QGs and blue-cloud SFGs \citep{Chester_1964, Baldry_2004}. As a galaxy ages and exhausts its cold gas, it transits from an SFG to a QG, terminating its star-formation activity in a process known as galaxy quenching, thereby traversing the area between the two populations on the colour-magnitude diagram, also called green valley (GV, \citealt{Salim_2014,Schawinski_2014}).

As a transitional population moving from SFGs to QGs, recently quenched galaxies (RQGs) are ideal for studying the quenching process \citep{Wu_2020}. The RQGs are still under quenching or have recently completed it \citep{Newberry_1990, Quai_2019,Mao_2022}. They have ceased most of their star-formation activity, yet still contain a young stellar population. This nature makes them valuable for probing the physical processes involved in quenching. RQGs have quenched their star formation in the past tens of millions to a few billion years \citep{Couch_1987, Poggianti_2009}. Therefore, their stellar population is dominated by A-type stars that have ages between $0.1$ and $1\,\mathrm{Gyr}$. 
This particular stellar composition adds strong Balmer absorption lines to the passive galaxy spectrum \citep{Dressler_1983,Pablo_2023, Pablo_2023b}. Various methods have been used in the literature to select RQGs, including rest-frame colours, as used in \citet{Belli_2019} and \citet{Mao_2022}, star-formation history (SFH) timescales \citep{Q1-SP044}, and spectral characteristics \citep{Citro_2017, Quai_2018}. 
The RQG population is also referred to as post-starburst galaxies (PSBs). However, the starburst phase is not essential to the quenching process that creates these stellar populations \citep{Poggianti_2009}. Therefore, in this work, we refer to this population as RQGs. 
However, we note that the rejuvenation process, which temporarily reactivates star formation in quenched galaxies \citep{Mancini_2019, MartinNavarro_22}, can affect the identification of RQGs. The galaxies in rejuvenation can be classified as RQGs, yet they are transitioning from QGs to SFGs, contrary to the expected direction of evolution. 

At low redshifts, galaxy quenching is primarily traced by two parameters: stellar mass and environment, which correlate strongly with the red fraction of galaxies \citep{Peng_2010, Darvish_2016}. 
Quenching mechanisms are classified into mass-dependent (internal) and environment-dependent (external, see section\,10.6 in \citealt{Cimatti_2019}) types.
In contrast, the high redshift scenario is more complex, featuring a greater interaction between stellar mass and environmental influences  and their evolution with redshift \citep{DeLucia_2012, Hatamnia_2026}.

Among stellar-mass-driven quenching mechanisms, the most extensively studied is active galactic nucleus (AGN) feedback \citep{Best_2005, Hopkins_2005, Belli_2024}. AGN outflows \citep{Debuhr_2012,Combes_2017} and heating \citep{Croton_2006,Man_2018,Zinger_2020} suppress star formation. In low-mass galaxies, supernova feedback is another key mass-quenching mechanism, where cold gas is expelled by high-energy ejections and super winds \citep{Dekel_1986, Dekel_2003}. Morphological quenching \citep{Martig_2009} is also a mass quenching mechanism, although the parameter directly responsible is the disc's dynamical stability rather than stellar mass itself: the stabilisation of galactic discs prevents gas fragmentation and the subsequent collapse necessary for star formation.

Environmental quenching mechanisms involve interactions between galaxies and their surrounding environment. A widely established mechanism of environmental quenching is strangulation, also known as starvation \citep{Larson_1980}. The virial shocks from the halo heat the cool gas, stopping the gas supply to galaxies \citep{Bekki_2002, Man_2018}. Galaxy mergers can be another mechanism of environmental quenching. In particular, major mergers can trigger starbursts, which are followed by rapid consumption of all remaining cool gas \citep{Gabor_2010, Man_2018}. Other environmental quenching mechanisms include galaxy harassment that occurs in high-speed encounters \citep{Moore_1996, Moore_1998, Moore_1999, Boselli_2006}, ram pressure stripping that occurs during moving through the intracluster medium \citep{Gunn_1972, Roediger_2014, Boselli_2022}, and the tidal effect where the stellar component is stripped by the gravitational potential of the cluster \citep{Gnedin_2003}.

Given the multiple mechanisms that can lead to quenching, it is essential to identify the parameters that can distinguish between them. One critical parameter is the quenching timescale, which varies between the mechanisms. Most of these timescales are obtained from simulations \citep{Wright_2019, Walters_2022}.
Among mass quenching mechanisms, supernova feedback rapidly quenches star formation in $\sim0.1\,{\rm Gyr}$ \citep{Ceverino_2009}. The AGN feedback timescale remains a topic of debate. The gas outflow driven by quasars can quench star formation at million-year-level timescales \citep{Smethurst_2021}, while the radio-mode AGN feedback is found to take up to and beyond $1\,\mathrm{Gyr}$ to establish a balance between cooling and heating to reach a low-SFR phase \citep{Fabian_2012}. \citet{Hirschmann_2017} simulated the SFH of galaxies with AGN and found that the quenching timescale can have a wide range, from a few hundred Myr to a few Gyr. For environmental quenching mechanisms, strangulation is a long-term process that lasts a few billion years \citep{Boselli_2006, Peng_2015}. Merger-driven quenching occurs with a median $1.5\,\mathrm{Gyr}$ delay time between the merger event and the start of quenching, and the timescale varies over a wide range \citep{Rodriguez_2019}. Tidal effects quench galaxy star formation over a wide range of timescales, but even fast cases require $2$--$6\,\mathrm{Gyr}$ \citep{Lokas_2020}. In the extreme case, ram pressure stripping is a rapid type of quenching, with a timescale as short as $\sim0.2\,\mathrm{Gyr}$ \citep{Steinhauser_2016}. 
Thus, the quenching timescale is essential for constraining the physical mechanisms involved in quenching. We also need to note that in many cases multiple quenching mechanisms are involved. 

To transform SFGs to QGs, not only quenching of star formation but also morphological transformation is required \citep{Quilley_2022}. The QGs are mainly with spheroidal morphologies, while the SFGs are mostly disc-dominated. The morphological transformation of galaxies is generally thought to be coupled with the quenching process \citep{Bundy_2010, Liu_2021, Dimauro_2022}, although it is not clear whether it occurrs simultaneously or on longer timescales or follows the quenching \citep{Pozzetti_2010}. Thus, morphological studies of galaxies can further constrain the possible physical mechanisms of quenching. 
Mechanisms like AGN feedback can increase the S\'ersic index of massive galaxies by suppressing disc growth \citep{Dubois_2013}, and can produce lower S\'ersic index of Milky Way--mass galaxies by suppressing bar growth \citep{Irodotou_2022}.
Tidal effects \citep{Moore_1999, Gnedin_2003} strip disc galaxies, thereby shrinking their sizes and increasing their S\'ersic indices, leaving only the bulge component. By studying the morphological parameters (i.e., effective radius, S\'ersic index), we can put further limitations on quenching scenarios \citep{Q1-SP044}.

Deep extragalactic surveys using optical to near-infrared ground-based telescopes (e.g., Subaru, Gemini, Keck, VLT) and space telescopes -- e.g., \HST, {\it James Webb} Space Telescope (JWST), \Euclid\ -- enable reconstruction of galaxy formation and evolution. The SFR density peaks at cosmic noon $z \sim 2\mbox{--}3$ and declines at lower redshifts \citep{Madau_Pozzetti_Dickinson_1998, Madau_2014, Khostovan_2015}, related to the quenching of individual galaxies. Therefore, $z<3.0$ is a key cosmic epoch in quenching studies.

The \Euclid mission \citep{EuclidSkyOverview} provides both photometric and spectroscopic data, covering a wide area of the sky, enabling statistical studies of RQGs and quenching in both photometric and spectroscopic ways. 
Already with the first Euclid Quick Data Release (Q1; \citealt{Q1-TP001}), detailed works on galaxy physical properties \citep{Q1-SP031} 
and morphology \citep{Q1-SP040} are examined, enabling a detailed study of quenching through RQGs.

In this paper, we use data from Q1 to classify and investigate SFG, QG, and RQG populations, both photometrically and spectroscopically, and to reconstruct the evolutionary path of galaxies through these transitional phases. 
We describe sample selection (Sect.\,\ref{sc:Data}) from \Euclid data,  and rest-frame colour diagrams used to classify galaxy populations (Sect.\,\ref{sc:Method}). We analyse the fraction of each population (Sect.\,\ref{sc:quenchingproperties}), and their properties (Sect.\,\ref{sc:galproperties}), discussing possible quenching scenarios (Sect.\,\ref{sc:discussion}). In Sect.\,\ref{sc:spec}, we present the spectroscopic subsample and stacked spectra for each population.

In this work, a flat $\Lambda$CDM cosmology model with $H_0 = 70\,\kmsMpc$, $h_{70}=H_0/(70\,\kmsMpc)=1$, $\Omega_{\rm m} = 0.3$ is adopted. All magnitudes are in AB photometric system \citep{ABmagnitude}.

\section{\label{sc:Data}Data}
In this work, we use Q1 data release \citep{Q1cite}. 
This release consists of observations of the three Euclid Deep Fields (EDFs) over $\sim63.1\, \deg^2$. It includes photometric data from the Visible Camera \citep[VIS;][]{Q1-TP002}, and photometric plus slitless spectroscopic data from Near Infrared Spectrometer and Photometer \citep[NISP;][]{Q1-TP003} instruments, and photometric external data \citep{Q1-TP004}. 
The visible filter \IE reaches a $5\,\sigma$ point-source sensitivity of $24.5\,\mathrm{mag}$ \citep{EuclidSkyVIS}, and the NIR filters \YE, \JE, and \HE each reach $24\,\mathrm{mag}$ \citep{EuclidSkyNISP}. Further, Q1 provides already slitless spectroscopic data using NISP, extracted for the bright subsample of galaxies with $\HE < 22.5$, validating approximately three million spectra for sources observed, with four dithers, in the red-grism mode \citep{EuclidSkyNISP,Q1-TP006, Q1-TP007}.\footnote{{The red-grism cover the wavelength range $\num{12100}\,\AA$ to $\num{18900}\,\AA$.}}

In this paper, we use, for the galaxies properties, such as photometric redshift, stellar mass, SFR, and rest-frame colours, the catalogue derived in \citet{Q1-SP031} that combines the Infrared Array Camera (IRAC, \citealt{Fazio_2004}) data with the \Euclid MER catalogue \citep{Q1-TP004} to have a more robust estimate of photometric redshifts and physical properties. \citet{Q1-SP031} obtains the photometric redshifts and physical properties (i.e., stellar mass, SFR, rest-frame colours, SFH timescale) with the nearest-neighbours algorithm (\texttt{NNPZ}, \citealt{Desprez-EP10, EP-Enia, Q1-TP005}), which finds 80 nearest neighbours in the template space (i.e., magnitude and colour), from a reference sample constructed from stellar population synthesis (SPS), assuming a delayed exponentially declining SFH and the \citet{Kroupa_2001} initial mass function \citep[IMF, then converted to][]{Chabrier_2003}, and derives physical properties from their weighted distribution (in this work, median properties).
The stellar mass and SFR derived from \texttt{NNPZ} depends on $h_{70}^{-2}$.
All galaxy properties, including the rest-frame colours, have been derived from multiband photometry after correction for local reddening, depending on the position in the sky \citep{Q1-TP001,Q1-TP005,Q1-SP031}. 

The sample is cleaned by applying the same criteria of \citet{Q1-SP031}:
\begin{itemize}
    \item \texttt{SPURIOUS$\_$FLAG} $= 0$;
    \item \texttt{DET\_QUALITY\_FLAG} $< 4$;
    \item \texttt{MUMAX\_MINUS\_MAG} $> -2.6$;
    \item \texttt{PHZ$\_$QSO$\_$PROB} $< 0.86$;
    \item \texttt{PHZ$\_$STAR$\_$PROB} $< 0.10$;
    \item $q > 0.05$;
    \item $0.01\,{a} <$ \texttt{$R_{\mathrm e}$} $< 2\,{a}$;
    \item ${a} > 13$ or ${a} < 20$ or $R_{\mathrm e} < 0.15$ or $R_{\mathrm e} > 0.5$.
\end{itemize}
The aim is to remove spurious sources, using the photometry quality flags derived in \citet{Q1-TP004} -- \texttt{SPURIOUS$\_$FLAG} and \texttt{DET$\_$QUALITY$\_$FLAG} -- and point-like objects, using \texttt{MUMAX$\_$MINUS$\_$MAG} and classification probabilities (\texttt{PHZ$\_$QSO$\_$PROB, PHZ$\_$STAR$\_$PROB}), derived in \citet{Q1-TP005}. 
We further remove diffraction spikes and cosmic-rays, as discussed in \citet{Q1-SP031}, also applying morphological criteria to discard unsuccessful fits and spurious sources as discussed in detail in \citet{Q1-SP047} and \citet{Q1-SP040}.
Specifically, we use the Sérsic axis ratio $q$, the Sérsic radius $R_\mathrm{e}$, and the isophote semi-major axis $a$, all in \IE-band pixel units. 
We then apply the total magnitude Q1 depth at $5\sigma$ in the \HE band $\HE < 24$ \citep{Q1-TP005}, and a mask summarised as follows. Our mask uses information from the {\rm VMPZ} survey characteristic maps \citep{Q1-TP001} and external information from HyperLeda \citep{Makarov_2014} and {\it Gaia} \citep{GAIA}. We require at least three standard exposures in \IE and \HE, and remove regions of high noise in the ground-based observations. This latter cut involves removing the area at declinations lower than $\ang{63.9}$ from EDF-N to work with a more homogeneous data set. To remove bright objects and their surrounding regions, we derive simple scalings for elliptical apertures based on the magnitudes and axis ratios of galaxies (from HyperLeda) and stars (from {\it Gaia}). Next, we perform more extensive masking of bright star diffraction spikes, again based on each star's magnitude and the position angle of the observations. Finally, we use the flag maps to identify optical ghosts in the \IE images, and fit enlarged bounding ellipses around the affected pixels to remove any sources impacted by their presence.

\section{\label{sc:Method}Photometric classification}
In this section, we first confirm the reliability of photometric redshifts and introduce and apply the methods of rest-frame colour-colour diagrams for classifying different galaxy populations. 

\subsection{\label{sc:zvalidation}Redshift validation}
\citet{Q1-SP031} used a subsample with external spectroscopic redshift to check the quality of the photometric redshift estimation. 
Specifically external redshift samples from DESI \citep{DESI_DR2_2025}, SDSS \citep{SDSS_DR16, SDSS_DR17, SDSS_DR18}, 2MRS \citep{Huchra_2012}, PRIMUS \citep{Coil_2011,Cool_2013}, OzDES \citep{Lidman_2020}, 3dHST \citep{Brammer_2012, Momcheva_2016}, 2dF-GRS \citep{Colless_2001}, 2dFLenS \citep{2dflens}, 6dFGS \citep{Jones_2009}, MOSDEF \citep{Kriek_2015}, JADES \citep{DEugenio_2024,Eisenstein_2025}, VVDS \citep{LeFevre_2013}, Quaia \citep{quaia}, MUSE Wide \citep{MUSEwide}, MUSE-HUDF \citep{MUSEHUDF}, VANDELS \citep{VANDELS}, HETDEX \citep{HETDEX1,HETDEX2}. 
To quantify the quality of $z_{\rm phot}$ estimation, \citet{Q1-SP031} derived the normalised median absolute deviation (NMAD, \citealt{Hoaglin_1983}), the catastrophic outlier fraction $f_{\rm out}$ and the bias \citep[see definitions in][]{Q1-SP031}. For the subsample with matched external $z_{\rm spec}$, $\sigma_{\rm NMAD}$ is $0.039$, $f_{\rm out}$ is $12.7\%$, and the bias is $+0.01$. At low redshift, the outlier fraction increases \citep{Q1-SP031}, therefore, galaxies at $z<0.2$ are excluded from the analysis of the RQGs' properties in this work.

\subsection{\label{sc:ccdiagram}Rest-frame $UVJ$ diagram}
In this work, we use the rest-frame colour-colour diagram ($U-V$) versus ($V-J$) diagram (hereafter $UVJ$) based on the photometric redshifts to separate QGs, SFGs, fast- and slow-quenching recently-quenched galaxies (fRQGs and sRQGs), and compare with other methods.
The $UVJ$ diagram was first introduced by \citet{Williams_2009} to distinguish between passive and star-forming galaxies. \citet{Mao_2022} introduced a new selection based on an extension of the $UVJ$ diagram by \citet{Schreiber_2018}
with boundaries based on model evolution tracks of exponentially declined SFHs
that further classifies fast and slow-RQGs. In that work, a constant SFH followed by an exponential decline in SFR was assumed in a galaxy with \citet{Chabrier_2003} IMF and solar metallicity. The evolutionary pathways of galaxies with e-folding SFH timescales $\tau = 0.1\,\mathrm{Gyr}$ and $\tau = 1\,\mathrm{Gyr}$ were used to set the boundaries between fRQG and sRQG on the $UVJ$ diagram (see details in section\,3.2 in \citealt{Mao_2022}). With this additional $UVJ$ selection, we can examine each stage of quenching and investigate the fast and slow quenching routes.
\citet{Mao_2022} note that a galaxy with a longer quenching timescale stays longer in the RQG region on the $UVJ$ diagram (section\,3.2 in \citealt{Mao_2022}). A typical fRQG with $0.1\,\mathrm{Gyr}$ SFH timescale will be in the corresponding RQG region for $\sim 0.43\,\mathrm{Gyr}$, and a typical sRQG with $1\,\mathrm{Gyr}$ SFH timescale will be in the RQG region for $\sim 1.5\,\mathrm{Gyr}$.

The rest-frame $U$, $V$ \citep{Johnson_2006}, and 2MASS $J$ \citep{Huchra_2012} filters used in \citet{Mao_2022}, which are the same as \texttt{EAZY} rest-frame filters in COSMOS2020 \citep{Weaver_2022} are not identical to the Johnson $U$, $V$ \citep{JohnsonFilters}, and \JE \citep{EuclidSkyNISP} filters used in this work. Therefore, a shift in the boundaries is necessary to align them with the \Euclid data set. 
The offset between the COSMOS and the \Euclid data set is discussed in Appendix \ref{appd:boundary shift}, along with the updated boundaries. 

\begin{figure}[htbp!]
\centering
\includegraphics[angle=0,width=1.0\hsize]{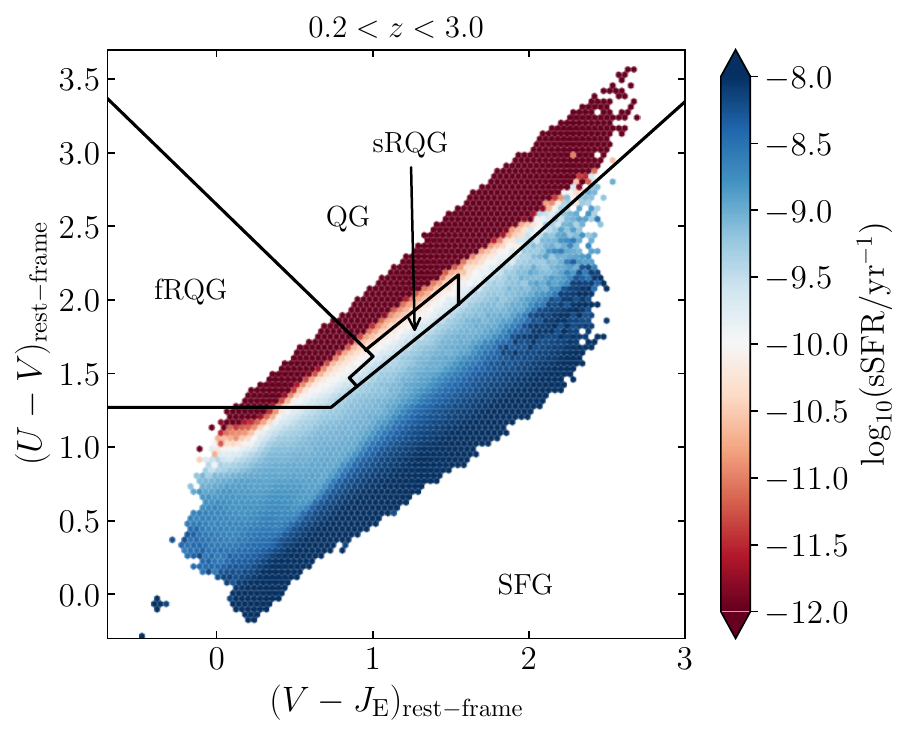}
\caption{Distributions of QGs, SFGs, fRQGs, and sRQGs on the rest-frame $UVJ$ diagram at $0.2 < z < 3.0$, using hex bins and colour-coded by median sSFR in each cell, from red (low sSFR) to blue (high sSFR). The sSFR values outside the colourbar range are assigned the colours at the upper and lower ends.} 
\label{fig:uvj}
\end{figure}

In Fig.\,\ref{fig:uvj}, we show the selected QGs, SFGs, and RQGs, separated by the new boundaries. The plot is colour-coded by specific SFR (sSFR), showing the expected distributions: low-sSFR QGs and high-sSFR SFGs, with intermediate-sSFR galaxies located near the boundary between QGs and SFGs, where we expect to find both slow- and fast-quenching populations. 
The evolution with redshift in the colour-colour distribution is shown in Appendix \ref{appd:boundary shift}, and reflects the evolution of the different populations, as we discuss in Sect.\,\ref{sc:quenchingproperties}.
In Appendix\,\ref{apdx:comparison}, we compare the $UVJ$ classification with the rest-frame $\mathrm{NUV} - r$ versus $r-J$ diagram classification in \citet{Ilbert_2013}. 
While QGs and SFGs are mostly located in their corresponding regions, RQGs lie near the boundary between passive and star-forming regions.

We analyse the extent to which the galaxy classification is affected by photometric redshift uncertainties. We derive the physical properties as stellar mass, SFR, rest-frame colours, and the classification at $z_{\rm spec}$ for the spectroscopic subsample. Specifically, we run the \texttt{NNPZ} algorithm, extracting nearest neighbours from the reference sample within the window $z_{\rm spec} \pm 0.01$. Apart from $\sim 12.9\%$ redshift outliers showing large differences between $z_{\rm spec}$ and $z_{\rm phot}$ (Sect.\,\ref{sc:zvalidation}), that contaminate the sample, overall for $z_{\rm phot}>0.2$, we find that most QGs and SFGs ($>90\%$) retain their original classification. In contrast, only $\sim 54\%$ of RQGs do, showing  significant overlap with both QGs and SFGs, even though the total number of RQGs remains mostly unchanged (Appendix\,\ref{apdx:photz-specz-classification}). For a transitional population like RQGs, the colour-colour boundaries are inherently blurred by nature: the $UVJ$ diagram is a projection of a continuous physical process, and objects near the boundary between RQGs, SFGs, and QGs are genuinely intermediate, and the overlap and mixing reflect the transitional nature of RQGs. Nevertheless, using the spectroscopic subsample in the range $0.2<z<1.5$, we verify that the impact on the results obtained with the photometric sample is not substantial. Indeed, the spectroscopic subsample of RQGs shows relative properties similar to those of the photometric selected sample. This is to ensure that our results are robust.

In the following analysis, we utilise SFGs, QGs, fRQGs, and sRQGs, which are classified according to the $UVJ$ diagram. At $0.2<z<3$, we identify \num{5618011} SFGs, \num{420101} QGs, \num{133598} sRQGs, and \num{339407} fRQGs, within the Q1 area of $63.1\,\mathrm{deg}^2$.

\section{\label{sc:quenchingproperties}Quenching properties}
\begin{figure}[htbp!]
\centering
\includegraphics[angle=0,width=1.0\hsize]{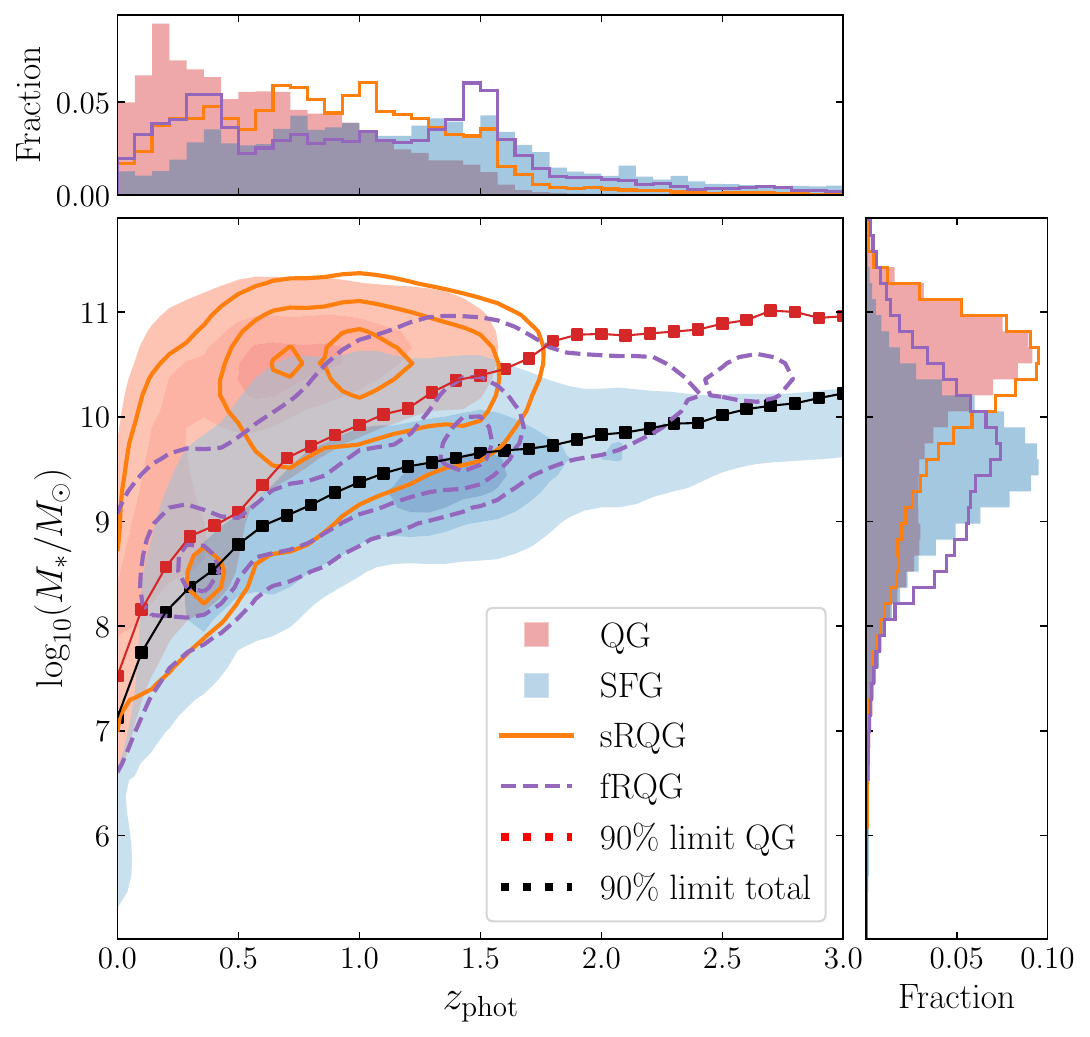}
\caption{Stellar mass versus redshift relations of the four populations, with a $y$-axis side panel showing the stellar mass distributions and an $x$-axis side panel showing the photometric redshift distributions. The sample is limited to 
$\HE <24$. On the main panel, the red and blue filled contours contain $10\%$, $50\%$, and $90\%$ of the total QGs and SFGs, respectively. The dashed purple and solid orange contours show the same percentiles for the total fRQGs and sRQGs. The black and red lines indicate the $90\%$ mass completeness of all and QGs galaxies, respectively. 
The same colours have been used on the side panels for the histograms.}
\label{fig:mass-z}
\end{figure}

In this section, we investigate how quenching properties depend on stellar mass and redshift.
We first show in Fig.\,\ref{fig:mass-z} the relation between stellar mass and redshift for QG, SFG, fRQG, and sRQG, along with the distribution of stellar mass and redshift for each population. 
Following the method in \citet{Pozzetti_2010} we derive the $90\%$ mass completeness limit, due to the magnitude limit of $\HE<24$, showing in Fig.\,\ref{fig:mass-z} the limit for the total sample and the more conservative one for QGs.
From this figure, we note that QGs, as expected, lie in the massive tail of the mass distribution, while SFGs dominate at low masses.  
Furthermore, 90\% of massive QGs are below $z\sim1.5$, as expected given their decrease in density observed in previous surveys \citep{Ilbert_2013,Davidzon_2017,Weaver_2023}.
For the RQGs, we note that sRQGs exhibit a stellar mass distribution similar to that of QGs, whereas fRQGs are more similar to SFGs. 
We also note that both RQG populations are present up to $z=3$ (upper panel of Fig. \ref{fig:mass-z}), even if $90\%$ of sRQGs are at $z<1.7$, while 90\% of fRQGs are at $z<2.5$, suggesting that both quenching mechanisms occur throughout most of the cosmic epoch below $z<3$, but with different importance. 

In particular, the fast-quenching population peaks at two redshifts (low-$z$ and high-$z$) and is always at lower masses than sRQGs, whereas the slow-quenching population peaks at intermediate redshifts and systematically at higher masses. 
We note that a peak at $z\sim1.5$ is also present in SFGs and in the full redshift distribution. Its presence could be due to some bias effect in \texttt{NNPZ} photometric-redshift solution (see figure\,17 in \citealt{Q1-TP005}). We verify that the two peaks of fRQGs remain even after normalising to the excess in the total distribution at those redshifts. 
The two distinct peaks in the fRQG population (at low-$z$ with low-mass and at high-$z$ with high-mass) suggest that not all fast-quenching processes can be attributed to the same physical mechanisms. However, we note that the peaks are below the $90\%$ mass completeness derived from the QG sample, and therefore excluded from the sample to study the relative fraction of RQGs.

In the following sections, we discuss the relative fractions and ratios of the four populations: QG, SFG, fRQG, and sRQG, and their dependence on stellar mass and redshift. Following \citet{Mao_2022}, we define quenching efficiency (QE) as the ratio of RQG to the SFG population (i.e., $n_{\rm RQG}/n_{\rm SFG}$) and quenching stage (QS) as the ratio of RQG to the QG population (i.e., $n_{\rm RQG}/n_{\rm QG}$). 
In the quenching process, SFGs transform into RQGs and then into QGs. A high QE value indicates that more SF galaxies are undergoing this transformation and the quenching is more efficient. When QS value is high, a large fraction of RQGs relative to QGs are present, suggesting that quenching is still happening or is in its early stages. Conversely, a low QS value indicates that the quenching process is nearly complete, with most RQGs having evolved into QGs, reflecting a later stage of quenching.

We estimate the classification uncertainties through 1000 Monte Carlo (MC) realisations, perturbing the rest-frame $U$, $V$, and $\JE$ magnitudes, redshift and stellar mass according to the uncertainties from \texttt{NNPZ} (Appendix \ref{apdx:MC}). For all the properties in this section, the errors are the combination of  Poisson and classification uncertainties, being the latter on average from $\sim 20\%$ to a factor of 4 larger than Poisson ones.
Additionally, this MC analysis, which was performed across mass and redshift bins generates mixing classification matrices that, when inverted \citep{Starck_2002}, confirmed patterns in fractions, QE, and QS with stellar mass and redshift, consistent with the observed data,
as reported in Appendix \ref{apdx:MC} and \ref{apdx:QEQS}. Drawing errors that neglect positive covariance from shared \texttt{NNPZ} derivations can inflate mixing, which should be treated with caution as an upper limit. Our analysis shows that mixing has a moderate impact without compromising data interpretation and confirms the transitional nature of RQG populations.

We stress that the fractions derived here do not account for the different times required to cross the RQG region on the $UVJ$ diagram, which depends on the quenching timescale \citep{Mao_2022}, which is of the order of $\sim1.5$ Gyr for sRQGs and even shorter than $0.5$ Gyr for fRQGs. This results in a higher chance of finding an sRQG than an fRQG. Therefore, the fractions of fast and slow RQGs do not directly reflect the relative importance of fast and slow quenching mechanisms \citep{Wild_2016, Belli_2019, Park_2023}. In principle, from these fractions, in particular from the quenching efficiency, we could derive the quenching timescale for each population of RQG, taking into account the time required to cross the RQG region in the $UVJ$ diagram and the typical lifetime of an SF galaxy (equation\,6 in \citealt{Quai_2018} and the references therein).

For all the analysis in this section, we apply a stellar mass cut at $\log_{10}(M_\ast/M_{\odot}) < 11.25$, as some extremely high stellar mass values result from unreliable photometric redshift estimates (figures\,4 and 5 in \citealt{Q1-SP031}).
We repeat our analysis without stellar mass cuts, and the general trends remain unchanged. Furthermore, in each redshift bin, we apply the conservative limit of $90\%$ mass completeness for QGs at the corresponding redshift (Fig.\,\ref{fig:mass-z}), due to the magnitude limit of $\HE<24$ (Sect.\,\ref{sc:Data}).
In our analysis, we do not include the Eddington bias or cosmic variance in our error budget. As discussed in \citet{Weaver_2022}, the Eddington bias tends to overestimate the fraction of massive objects due to flux measurement errors. While stellar mass uncertainties are smaller for the most massive and brighter galaxies, catastrophic errors in determining photometric redshifts may lead to significant overestimations of stellar mass. In our study 
this risk has been mitigated imposing the upper limit on stellar mass at $\log_{10}(M_\ast/M_{\odot}) < 11.25$.
Further, we estimate that the effect of cosmic variance is less than 1\%, as it is inversely proportional to the total sampled volume. Indeed, Q1 samples an area approximately 50 times larger than COSMOS, where it is $<30\%$ at $z<2$ and $\log_{10}(M_\ast/M_{\odot}) <11.5$ (figure\,3 in \citealt{Weaver_2023}).

\subsection{\label{sc:populationfraction}Fractions of the different galaxy populations} 
\begin{figure}[htbp!]
\centering
\includegraphics[angle=0,width=1.0\hsize]{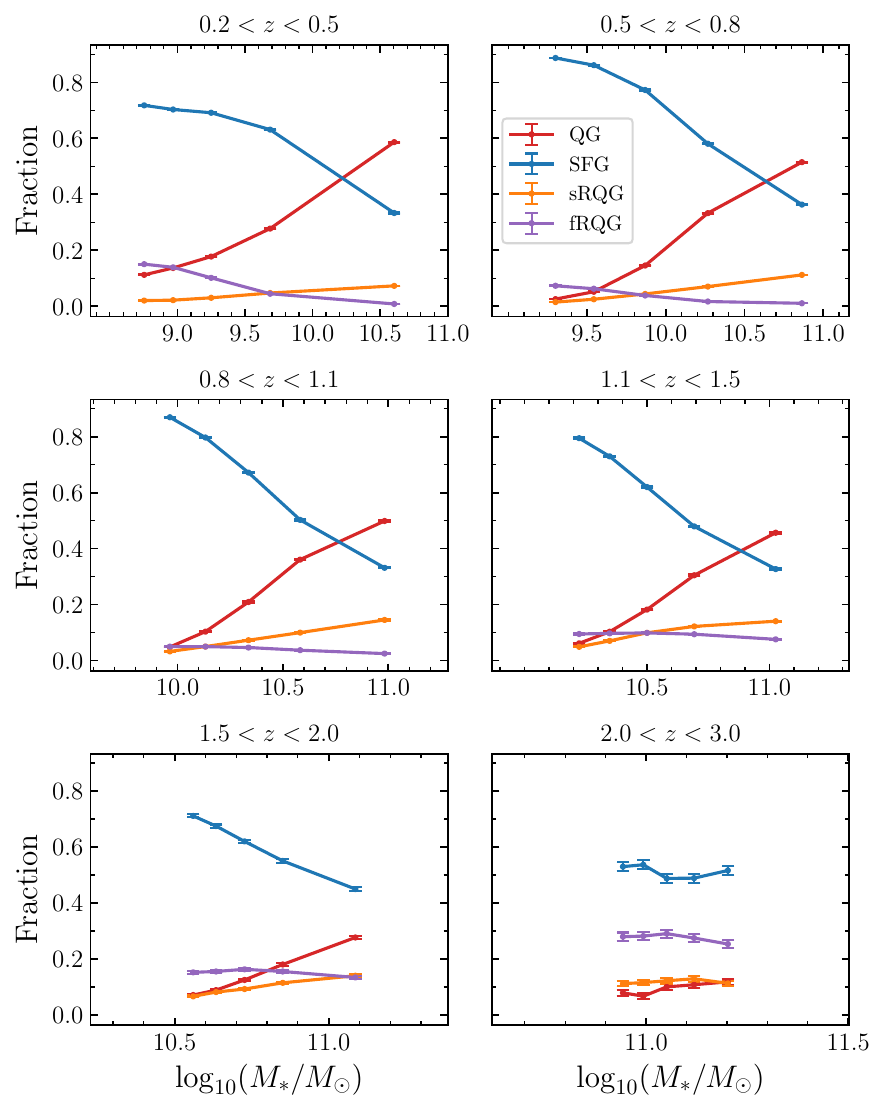}
\caption{Fractions of QGs, SFGs, fRQGs, and sRQGs across six redshift bins, with a stellar mass cut at $\log_{10}(M_\ast/M_{\odot}) < 11.25$. Different line colours show various populations. 
In each redshift bin, the full sample is divided into five stellar mass percentiles, each containing the same number of galaxies. Note that we plot in each panel only the mass range above the corresponding QG completeness limit.}
\label{fig:populationfraction}
\end{figure}
In this section, we discuss the fraction of each population relative to the total as a function of stellar mass and redshift.
As shown in Fig.\,\ref{fig:populationfraction}, as expected, the fraction of SFGs consistently decreases with stellar mass, while the fraction of QGs increases. The fraction of sRQGs also increases with stellar mass, demonstrating that slow quenching is more dominant 
for massive galaxies, even if it never exceeds $15\%$. The fraction of fRQGs, instead, exhibits a more complex behaviour. It decreases with stellar mass at low redshift, whereas at higher redshifts, the dependence becomes weaker, and it never exceeds $30\%$. We note that these are the fractions of galaxies that are still in or have just finished quenching, which is a relatively short period of time. Therefore, they represent only lower limits for the {total} fraction of galaxies that will eventually pass through this phase.
 
As Fig.\,\ref{fig:populationfraction} shows, at $z>1.5$, SFGs dominate in all stellar-mass bins, indicating that the main galaxy activity remains star formation. 
At $z<1.5$, the fraction of QGs increases progressively with cosmic time from higher to lower masses, exceeding, for the most massive, that of SFGs, indicating that quenching is mostly complete for massive galaxies.
We further note that the stellar mass at which the SFG and QG populations have similar fractions ($\sim40\%$) increases with redshift, from $\log_{10}(M_\ast/M_{\odot}) \sim 10.25$ at $0.2<z<0.5$ to $\log_{10}(M_\ast/M_{\odot}) \sim 10.9$ at $1.1<z<1.5$. This trend is the same as found in previous studies \citep{Pozzetti_2010, Bolzonella_2010}, confirming that quenching is a downsizing mechanism \citep{Neistein_2006}.

From this analysis, we infer that fast quenching dominates in earlier epochs ($z>1.5$) at masses above $\log_{10}(M_\ast/M_{\odot}) > 10.5$. At lower redshift, it remains the main mechanism of quenching only for less massive galaxies, likely driven by a different process.
For massive galaxies at lower redshift, slow quenching becomes the primary mechanism. Our data interpretation does not change if we consider the fractions derived by inverting MC mixing classification matrices, as reported in  Appendix\,\ref{apdx:MC} and shown in Fig.\,\ref{fig:fraction_corrected}.

\begin{figure}[htbp!]
\centering
\includegraphics[angle=0,width=1.0\hsize]{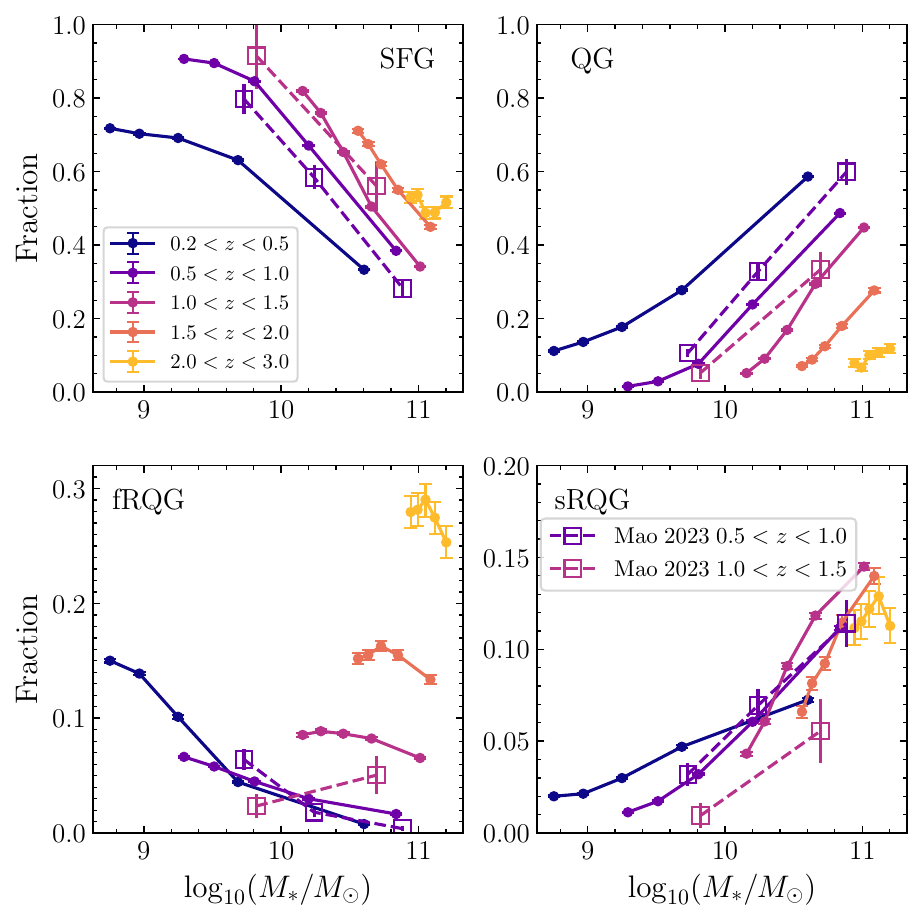}
\caption{Fractions of QG, SFG, fRQG, and sRQG as a function of stellar mass across five redshift bins in the range $0.2<z<3$, with a stellar mass cut at $\log_{10}(M_\ast/M_{\odot}) < 11.25$. A mass completeness cut at the upper redshift limit is applied for each redshift bin. From dark purple to light yellow line, the redshift increases. At each redshift, the full sample is divided into five stellar mass bins to contain a similar number of galaxies.
The filled circles are from Q1 data, and the empty squares are from \citet{mao_thesis} from COSMOS2020 data with members in X-ray detected galaxy groups and clusters in \citet{Gozaliasl_2019}. Note that fraction axes in panels have different scales.}
\label{fig:fraction-all-z}
\end{figure}
Figure\,\ref{fig:fraction-all-z} shows the fractions of each population and their stellar mass dependence across five redshift bins.
While the QG fraction decreases with redshift, the SFG and fRQG fractions increase. This behaviour indicates that star-formation activity generally declines in the Universe after cosmic noon. The fraction of sRQG has a different redshift dependence for massive and less massive ends. For low mass and intermediate mass galaxies, the fraction of sRQG decreases with redshift, while it increases at the massive end.
This indicates that slow quenching primarily occurs in massive galaxies at high redshift. At lower redshift, as most massive galaxies have already quenched, slow quenching becomes more prominent in less massive galaxies.

We compare the fraction of populations with a sample in \citet{mao_thesis}, which is selected from the COSMOS2020 catalogue as galaxy members of the X-ray--detected clusters or groups in \citet{Gozaliasl_2019} at $0.5<z<1.5$. We note that the fractions of SFGs and QGs generally agree with our results. However, there is a clear difference at $1<z<1.5$ for fRQGs and sRQGs. The sRQG fraction at the massive end and the fRQG fraction at the low-mass end in the \citet{mao_thesis} sample is smaller than in the Q1 sample. These differences may be related to the particular environment 
of the X-ray–selected clusters and groups. 
A possible environmental dependence of fast and slow quenching at $1<z<1.5$ may also be present.

\subsection{\label{sc:qe}Quenching efficiency}
In this section, we discuss the quenching efficiency (QE, i.e., $n_{\rm RQG}/n_{\rm SFG}$) of transforming SFGs into RQGs. We recall that when the QE value is high, relatively more SFGs are going through quenching process, hence quenching is more efficient, and vice versa.

\begin{figure}[htbp!]
\centering
\includegraphics[angle=0,width=1.0\hsize]{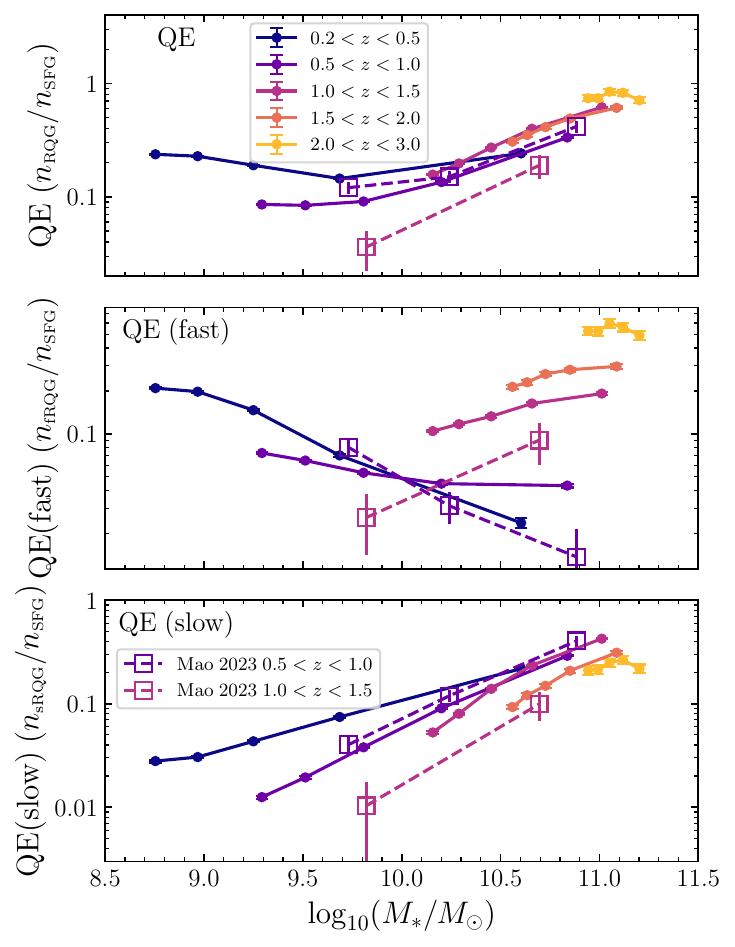}
\caption{QE as a function of stellar mass across five redshift bins in the range $0.2<z<3$, with a stellar mass cut at $\log_{10}(M_\ast/M_{\odot}) < 11.25$. Symbols and colours as in Fig.\,\ref{fig:fraction-all-z}. Note that QE axes in three panels have different scales.}
\label{fig:qe-all-z}
\end{figure}
In Fig.\,\ref{fig:qe-all-z}, we show the QE's mass dependence evolution with redshift. 
The total QE shows little or no evolution at $0.2<z<2.0$, remaining below $60\%$, but rises to $80\%$ at $2<z<3$, indicating an intense quenching process around cosmic noon. However, the contributions of fRQG and sRQG show a distinct trend. At $0.2<z<0.5$ QE is high for both high-mass and low-mass ends due to the contribution of slow and fast quenching, respectively. For high-mass galaxies with $\log_{10}(M_\ast/M_{\odot})>10$, the fast-quenching efficiency increases with stellar mass and keeps increasing with redshift up to about $30\%$ at $z=2$ and $60\%$ at $z>2$. 
For less massive galaxies with $\log_{10}(M_\ast/M_{\odot}) <10$, the fast-quenching efficiency decreases with stellar mass and shows little redshift evolution. Again, this suggests different fast-quenching mechanisms, one of which preferably occurs in massive galaxies at high redshift, while the other occurs in low-mass galaxies at low redshift.
In all redshift bins, the slow-quenching efficiency increases with stellar mass, linking it to mass quenching mechanisms. It exhibits a clear redshift evolution, decreasing with redshift for all stellar masses. However, the difference is not as pronounced as for fast quenching. The contribution to the QE mass dependence in each redshift bin is reported in Appendix\,\ref{apdx:QEQS} and shown in Fig.\,\ref{fig:qe}, together with the quantities derived inverting the MC mixing classification matrices, which show that the trends are only moderately influenced.  The QE is higher for sRQG than for fRQG at $z<1.5$ for massive galaxies and vice versa for fRQG at $z<1.5$ for less massive galaxies. At high-$z$ fRQGs' QE always dominates above the mass limit sampled.

We compare our results with \citet{mao_thesis}, where QE is analysed for X-ray-detected cluster or group members in \citet{Gozaliasl_2019}. The trends agree well at $0.5<z<1$, but at higher redshifts, the QE is higher in the \Euclid sample than in the members of X-ray clusters or groups, suggesting an environmental dependence of quenching at $1<z<1.5$. Highly efficient quenching appears to occur outside X-ray clusters or groups, which could occur either in the field, in unrelaxed clusters, or in protoclusters. The early quenching in proto-clusters \citep{Andrews_2025, Shi_2021} can partly explain this difference. 

\subsection{\label{sc:qs}Quenching stage}
In this section, we discuss the quenching stage (QS, i.e., $n_{\rm RQG}/n_{\rm QG}$) to determine if the quenching process for turning RQGs into QGs is in an early or late stage. When the QS value is high, the quenching process is ongoing, while when the QS value is low, the quenching is almost complete, with most RQGs having already evolved into QGs and the quenching process is overall terminated, with no more quenching on-going.
\begin{figure}[htbp!]
\centering
\includegraphics[angle=0,width=1.0\hsize]{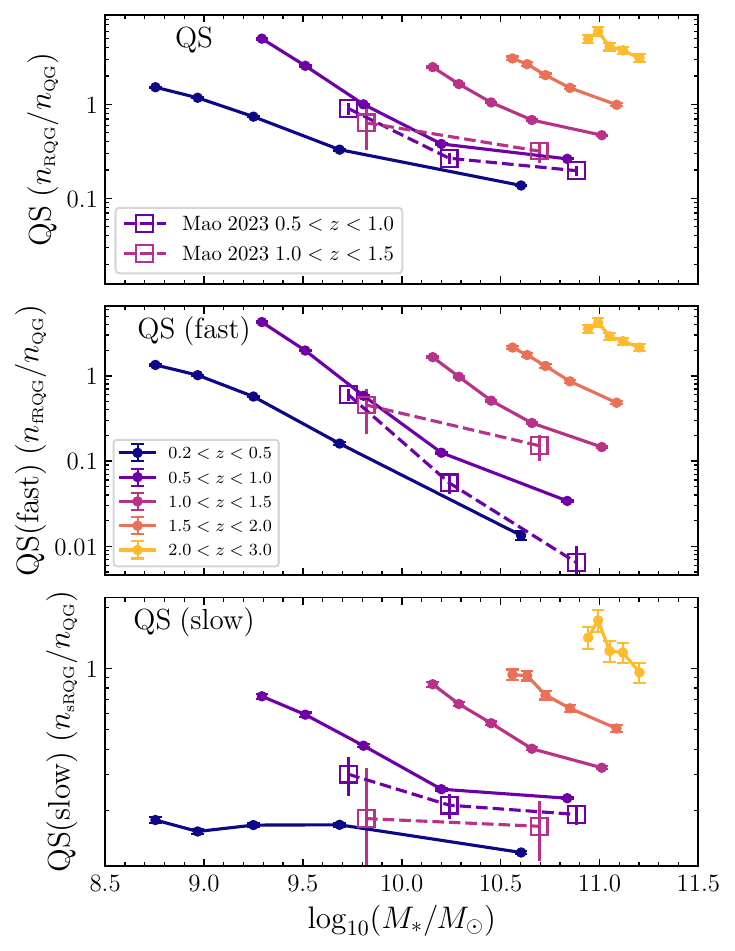} 
\caption{As in Fig.\,\ref{fig:fraction-all-z} but now for QS.
}
\label{fig:qs-all-z}
\end{figure}

In Fig.\,\ref{fig:qs-all-z}, we show the QS's mass dependence evolution with redshift, applying a mass completeness cut. 
Our analysis reveals that for all RQGs QS values evolve within the sampled redshift interval, regardless of stellar mass, showing an earlier quenching stage at higher redshift, consistent with the peak of star formation during cosmic noon.
The QS decreases with stellar mass, supporting the downsizing scenario where massive galaxies quench earlier. 
The contribution of fast and slow RQGs to QS's mass dependence in each redshift bin, along with the result of empirical deconvolution obtained by inverting MC mixing classification matrices, thus demonstrating that the trends are not affected, is reported in Appendix\,\ref{apdx:QEQS} and shown in Fig.\,\ref{fig:qs}. QS of fRQGs is always higher at low-mass and vice versa for sRQGs at high-mass at $z<1.5$, while fRQGs QS is similar or higher than for sRQG at $1.5<z<3$ and high-masses.

The results in Fig.\,\ref{fig:qs-all-z} are compared with \citet{mao_thesis}, where the analysis of QS is based on the member galaxies of X-ray clusters and groups. The differences found for sRQG at $1<z<1.5$, could also indicate environmental influences on the quenching process, across different mass ranges and redshifts.

\subsection{\label{sc:GAEA}Comparison with the GAEA model}
\begin{figure}[htbp!]
\centering
\includegraphics[angle=0,width=1.0\hsize]{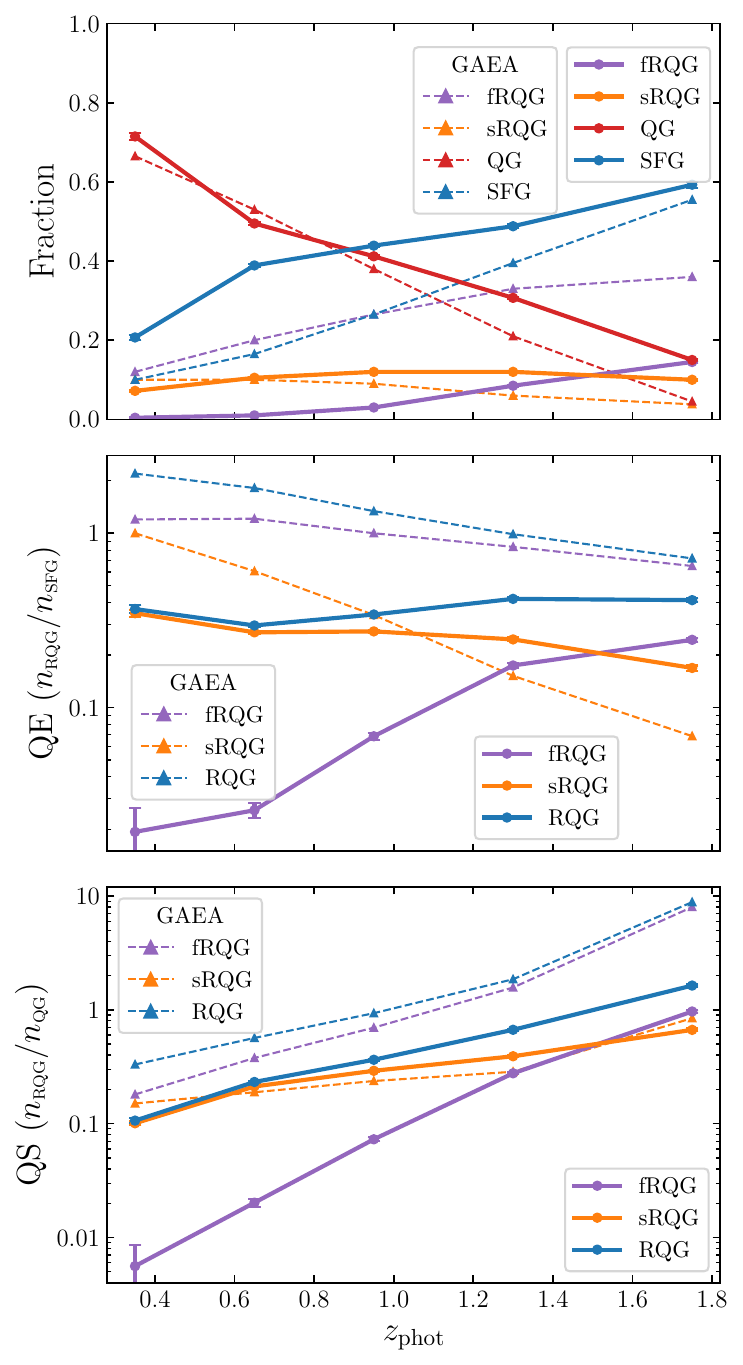}
\caption{Fractions ({\it top panel}), QE ({\it middle panel}), QS ({\it bottom panel}) of QG, SFG, fRQG, and sRQG as a function of redshift (corresponding to the five redshift bins at $z<2.0$ in Fig.\,\ref{fig:populationfraction}). 
Mass cut is applied at $\log_{10}(M_\ast/M_{\odot}) > 10.8$, which corresponds to the $90\%$ mass completeness of QG at $z\sim2$.
}
\label{fig:fraction-QE-QS-z}
\end{figure}

In Fig.\,\ref{fig:fraction-QE-QS-z}, we summarise the redshift evolution of the fraction, QE, and QS for a sample with $\log_{10}(M_\ast/M_{\odot}) > 10.8$, mass-complete up to $z<2$. In particular, we find that at these masses the fraction of SFGs decreases with cosmic time, while the QG population increases. 
The fraction of fRQGs dominates over sRQGs at early epochs and decreases towards the local universe as the fraction of sRQGs increases.
Furthermore, we find that the QE and QS of fRQGs also decrease with cosmic time, indicating that fast quenching in this mass range is completed at $z > 1$. For sRQGs, instead, while their QE increases, their QS decreases with cosmic time.

Below $z\sim 1.5$, the relative importance of fRQGs and sRQGs swap in fraction, QE, and QS.  In future studies, we plan to use these fractions as a starting point, to predict the growth rate of the quiescent population and compare it with the observed growth rate of the quiescent numerical densities, thereby contributing to our understanding of the various quenching mechanisms. 

We compare these quantities with the semi-analytical model GAlaxy Evolution and Assembly (GAEA, \citealt{GAEA-2014}, \citealt{GAEA-2016}, \citealt{GAEA-2017}, \citealt{GAEA-2020}) latest model rendition \citep{GAEA}, using the same rest-frame colour-colour criteria used with the \Euclid data (Sect.\,\ref{sc:ccdiagram}). We find that the general trends of fractions at $z < 2.0$ are well reproduced in GAEA. 
However, while QGs and sRQGs are well reproduced, within the range $10\mbox{--}30\%$ up to $z<1$ and up to factor $2.5\mbox{--}3$ at $z\sim2$,  GAEA strongly overestimates the fraction, QE and QS for fRQGs, from a factor $2.5$ up to a factor $100$, depending on the redshifts, and underestimates the fraction of SFGs up to a factor $2.5$.
We do not consider uncertainties of galaxy properties in the GAEA model for this comparison. However, we tested that adding simplified, Gaussian random uncertainties, to the colours, the comparison does not improve significantly. 
The motivation behind these differences lies beyond the scope of this article and requires further investigation of the SFHs and quenching path of the different populations in GAEA. 

\section{\label{sc:galproperties}Galaxy properties}
In this section, we examine the physical and morphological properties of QG, SFG, fRQG, and sRQG, and investigate the differences between these various populations. For the analysis of the relations among different properties, we restrict the sample to galaxies at $0.2 < z < 1.5$ to { limit} the impact of redshift evolution. Including a broader redshift range could blend trends at different epochs, therefore obscuring the intrinsic relations.

The physical properties discussed in this section are obtained with the \texttt{NNPZ} algorithm (Sect.\,\ref{sc:Data}). The morphological properties are estimated in \citet{Q1-SP047}, and the parameters (i.e., S\'ersic index, effective radius) are obtained by fitting a S\'ersic profile in \citet{Q1-SP040}.
We remind that the overlap among the classes (Appendix \ref{apdx:MC}) could moderately influence the distribution of galaxy properties, while also reflecting the transient nature of the RQG population.

\subsection{\label{sc:pp}Physical properties}
This section discusses the physical properties of galaxies, including redshift, stellar mass, SFR, and SFH timescale $\tau$, which is the e-folding timescale built from an exponentially delayed SFH.

In Fig.\,\ref{fig:mass-z}, we have already illustrated the relation between stellar mass and redshift and their distributions, for each populations.
It is evident that RQGs hold intermediate stellar masses, with fRQGs resembling less massive SFGs and sRQGs resembling the most massive QGs.

\begin{figure}[htbp!]
\centering
\includegraphics[angle=0,width=1.0\hsize]{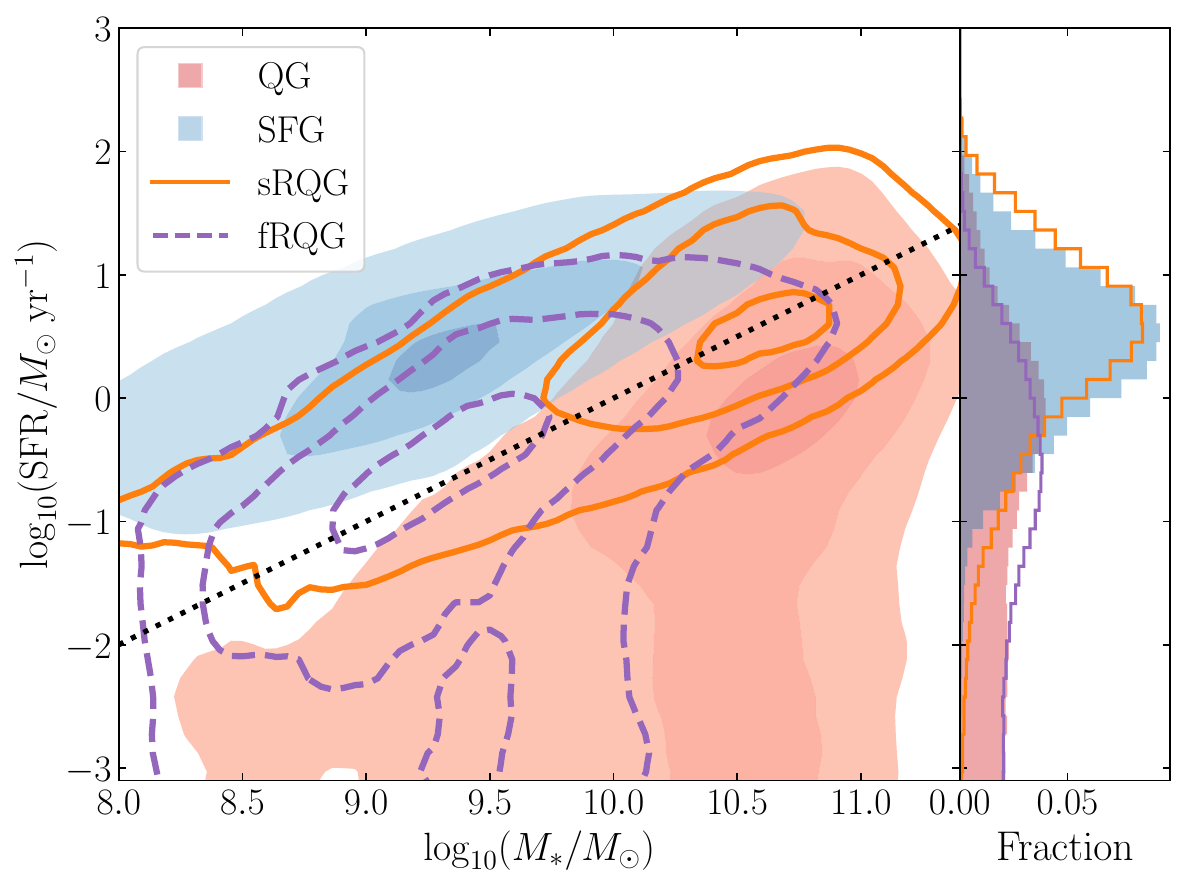}
\caption{SFR versus stellar mass relation of the four populations in the range $0.2<z<1.5$, with the SFR distribution shown as a side panel. Contours and colours as in Fig.\,\ref{fig:mass-z}. The dotted line marks sSFR of $10^{-10}\,\mathrm{yr}^{-1}$.
}
\label{fig:SFR-m}
\end{figure}
Figure\,\ref{fig:SFR-m} illustrated the relation between SFR and stellar mass, 
together with the SFR distribution for each population. The QG population reaches the lowest SFR and, on average, an sSFR lower than $10^{-10}\,\mathrm{yr}^{-1}$, as expected.
The SFG population falls within the regime of intermediate to low stellar mass
and intermediate to high star formation rates, reaching up to very high sSFR, with $90\%$ of them having $\mathrm{sSFR} > 10^{-10}\,\mathrm{yr}^{-1}$, and lies on the well-defined star formation main sequence (see \citealt{Q1-SP031} for its evolution in redshift). 
The fRQG and sRQG populations show an  intermediate SFR to stellar mass relation, with the densest $10\%$ of both fRQGs and sRQGs having similar sSFR, around $10^{-10}\,\mathrm{yr}^{-1}$, with the fRQGs showing an extended tail towards the lower SFR and sSFR.
For comparison, the spectroscopic subsamples show similar relative properties but are limited to the brightest and more massive populations (see Appendix \ref{apdx:photz-specz-classification} and Fig.\,\ref{fig:spec-sample-sfr-mass}).

\begin{figure}[htbp!]
\centering
\includegraphics[angle=0,width=1.0\hsize]{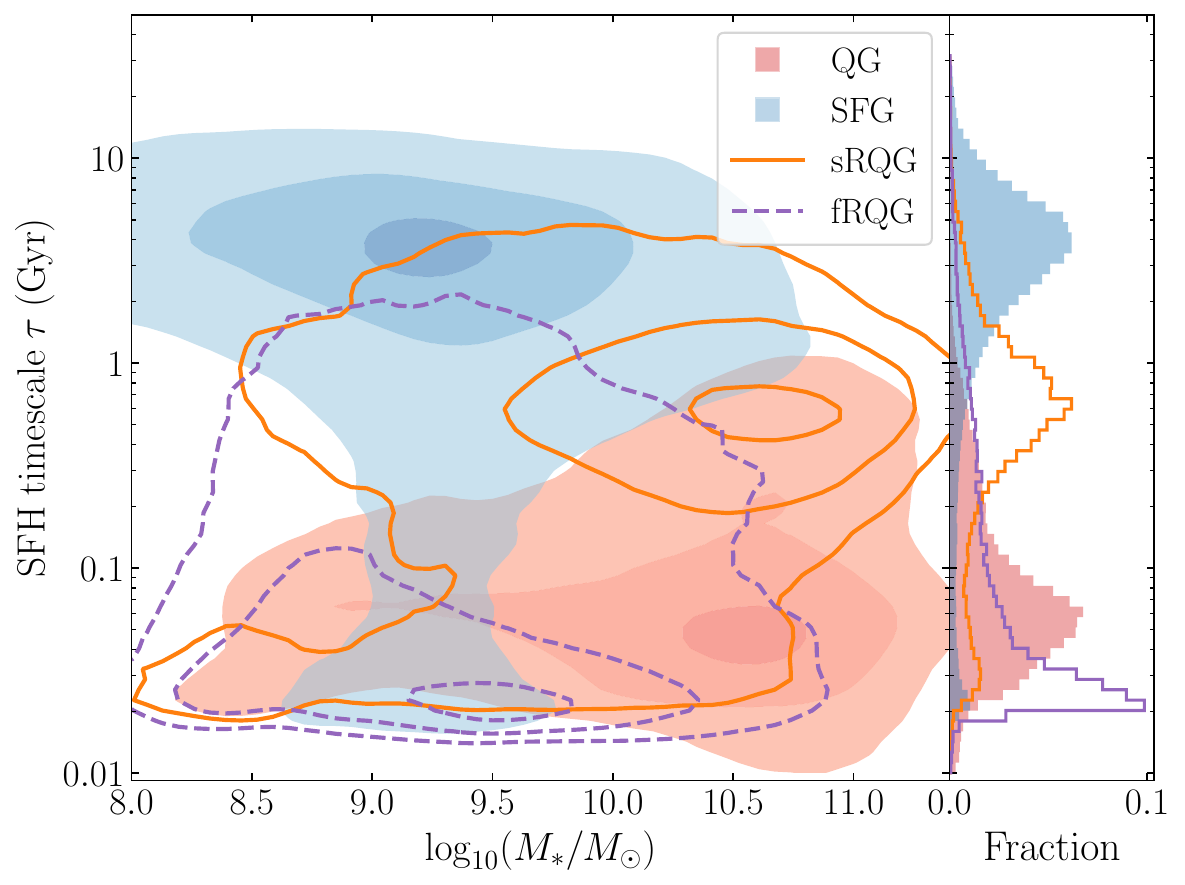}
\caption{SFH timescale versus stellar mass relation of the four populations in the range $0.2<z<1.5$, with SFH timescale distribution shown as a side panel. Contours and colours as in Fig.\,\ref{fig:mass-z}.}
\label{fig:tau-m}
\end{figure}
Figure\,\ref{fig:tau-m} shows the relation between the SFH timescale $\tau$ and the stellar mass, and the distribution of $\tau$ values for each galaxy population. SFGs show the most extended $\tau$ distribution, peaking at $\tau \sim 4\,\mathrm{Gyr}$, since their star-formation activity has not yet started to decline. The sRQGs also have a wide range of $\tau$, but peaking at $\tau \sim 0.7\,\mathrm{Gyr}$, confirming that they are slowly quenching their star-formation activity. The fRQGs also cover a wide range but reach the shortest $\tau$, peaking at $\tau \sim 0.02\,\mathrm{Gyr}$. Finally, QGs have $\tau$ intermediate between those of sRQGs and fRQGs, peaking at $\tau \sim 0.06\,\mathrm{Gyr}$, with tails in both directions up to $1\, \mathrm{Gyr}$ and even lower than $0.02\, \mathrm{Gyr}$, which is consistent with the fact that QG are evolved from both types of RQGs.
\begin{figure}[htbp!]
\centering
\includegraphics[angle=0,width=1.0\hsize]{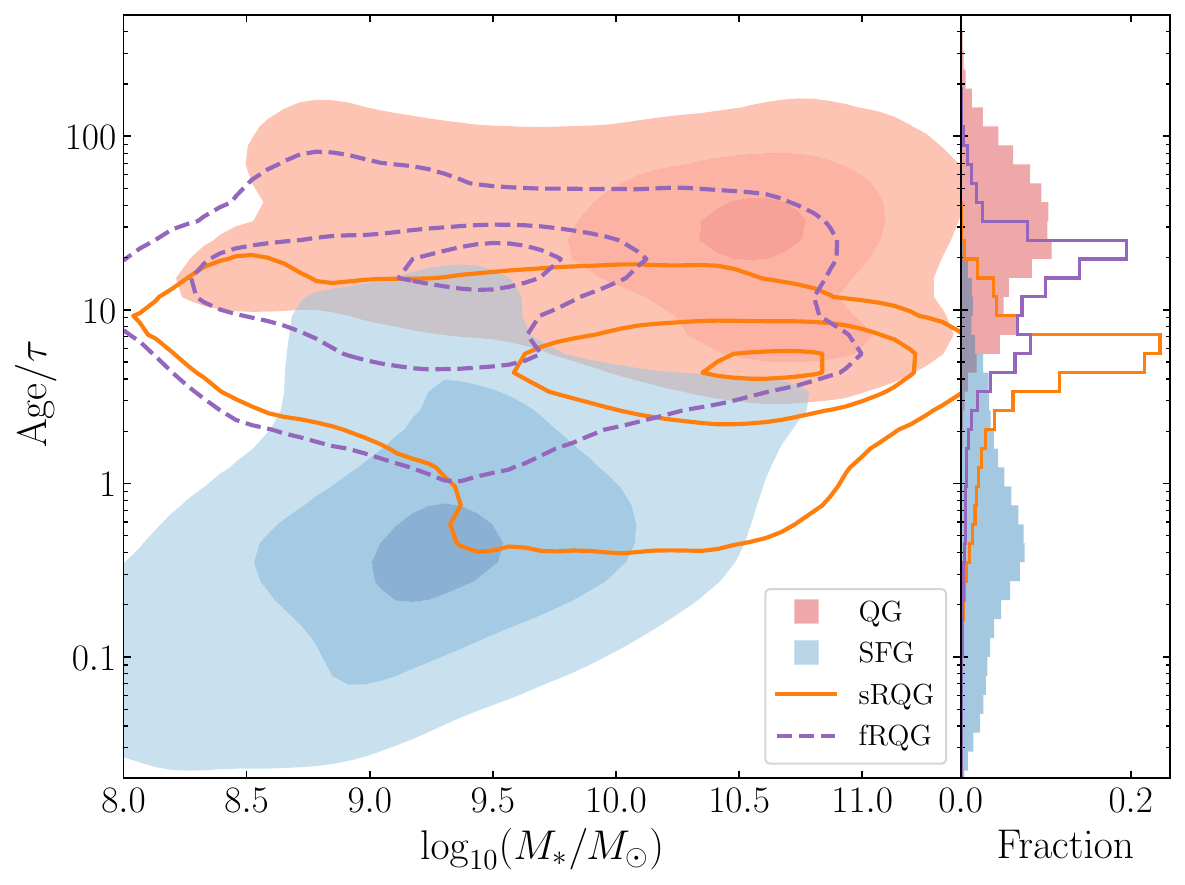}
\caption{Age/$\tau$ versus stellar mass relation in the range $0.2<z<1.5$ of the four populations, with age/$\tau$ distribution as a side panel. Contours and colours as in Fig.\,\ref{fig:mass-z}. 
}
\label{fig:phase-mass}
\end{figure}

The age/$\tau$ parameter indicates whether the galaxy is in the early or late phase of its evolution. Figure\,\ref{fig:phase-mass} shows its distributions for the four populations. SFGs are in an earlier phase of evolution, whereas QGs are in a late phase. We find that RQGs are in an evolutionary phase between SFGs and QGs, confirming an intermediate evolutionary link between SFGs and QGs.

\subsection{\label{sc:mp}Morphological properties}
This section discusses the morphological properties of galaxies, analysing the S\'ersic index and radius measured for each galaxy using VIS images \citep{EuclidSkyVIS,Q1-TP004}.

\begin{figure}[htbp!]
\centering
\includegraphics[angle=0,width=1.0\hsize]{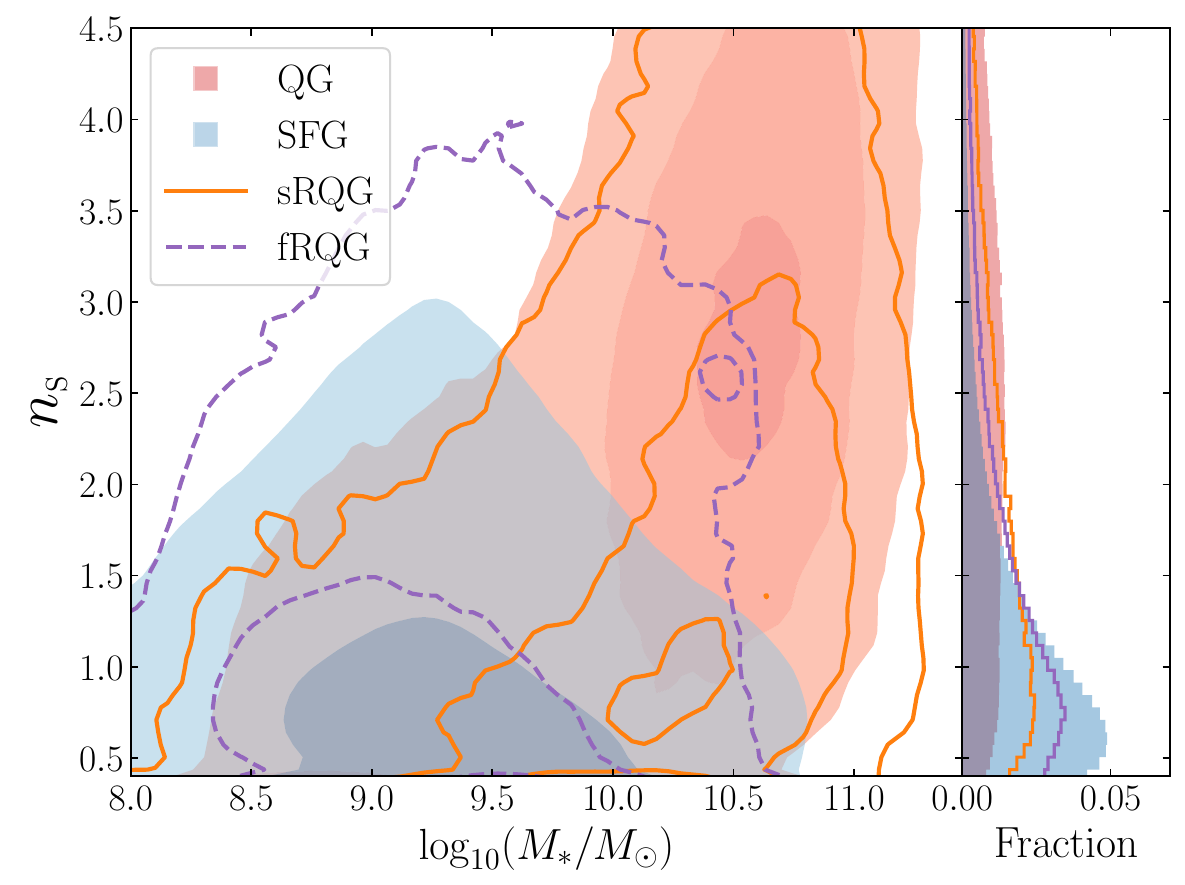}
\caption{S\'ersic index versus stellar mass relation of the four populations, with S\'ersic index distribution as a side panel. The sample is cut at $0.2<z<1.5$. Contours and colours as in Fig.\,\ref{fig:mass-z}. 
}
\label{fig:n-mass}
\end{figure}
Figure\,\ref{fig:n-mass} shows the relation between the S\'ersic index and the stellar mass of the four populations. 
We find that SFGs and fRQGs have similar $n_\sfont{S}$ distributions, characterised by low stellar mass and S\'ersic index values, the latter indicating weakly concentrated light profiles. In contrast, QGs and sRQGs have an $n_\sfont{S}$ distribution more extended towards higher values, indicating more concentrated light profiles.
SFGs and fRQGs peak at $n_\sfont{S}<1$, with over $70\%$ of SFGs and fRQGs having $n_\sfont{S}<1.5$ and $n_\sfont{S}<1.7$, respectively. QGs and sRQGs have more extended distributions, with $55\%$ of sRQGs having  $n_\sfont{S}>1.5$, being therefore already transformed to more concentrated light profiles. Finally, the QGs peak between $2<n_\sfont{S}<3.5$, in particular for high-mass QGs.
Therefore, in terms of the S\'ersic index, our analysis suggests that fRQGs have not yet undergone the morphological transformation already experienced by a large fraction of sRQGs.

\begin{figure}[htbp!]
\centering
\includegraphics[angle=0,width=1.0\hsize]{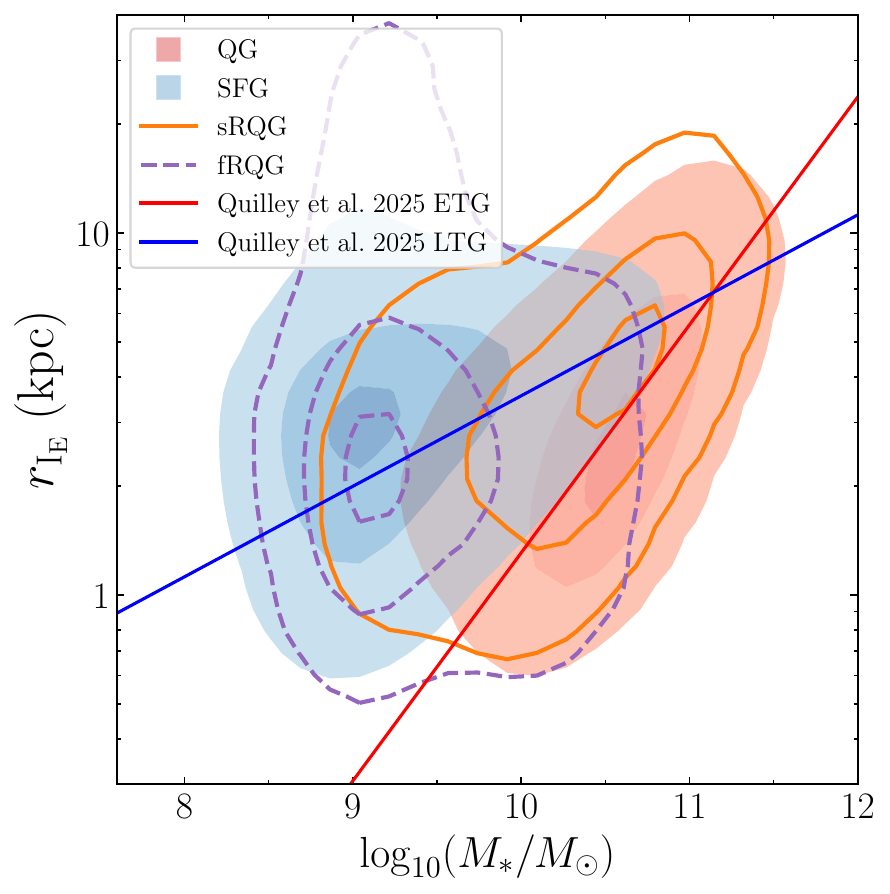}
\caption{\IE-band effective radius versus stellar mass relation of the four populations. The sample is cut with $0.8<z<1.0$.
Contours and colours as in Fig.\,\ref{fig:mass-z}.
The solid lines show the radius-mass relations fitted at $0.8 < z < 1.0$ by \citet{Q1-SP040} for early-type galaxies (red line) and late-type galaxies (blue line). 
}
\label{fig:r-m}
\end{figure}
However, using the S\'ersic index alone is insufficient to distinguish differences between populations. Therefore, we also looked into the radius-mass relation. Figure\,\ref{fig:r-m} shows the relation between the \IE-band effective radius and the stellar mass for each population. To compare with the size-mass for early- and late-type galaxies measured by \citet{Q1-SP040}, we restrict the sample to $0.8<z<1.0$. The slope of the size-mass relation is different for SFGs and QGs. For SFGs, the slope agrees with that found for late-type galaxies (LTGs), and for QGs, it agrees with that for early-type galaxies (ETGs) in \citet{Q1-SP040}. 
Visual inspection indicates that sRQGs have a relation that is  similar, but with slightly larger radii and flatter slope, to that of QGs, generally agreeing with the ETG size-mass relation.  
The fRQGs also appear to have a distribution in size-mass, similar to that of SFGs, 
and consistent with the LTG size-mass relation. Visually inspecting the size-mass relation in different redshift bins, we find a similar evolutionary trend to \citet{Q1-SP040}, finding that the size of SFGs and ETGs increases progressively with decreasing redshift, while the slope is steeper at higher redshift for both SFGs and QGs as found by \citet{vanderWel_2014}. 

We also investigated the relation between asymmetry, Gini coefficient, and the stellar mass, but there is no significant difference between the distributions of the four populations (Appendix \ref{apdx:CAS}). In particular, there is no clear evidence for more irregular morphologies for RQGs compared to SFGs and QGs,
excluding merger as a primary mechanism. Furthermore, from the concentration analysis we find that fRQGs are similar to SFGs and less concentrated than sQRGs, confirming the results from the S\'ersic index (Fig. \ref{fig:n-mass}).

\begin{figure}[htbp!]
\centering
\includegraphics[angle=0,width=1.0\hsize]{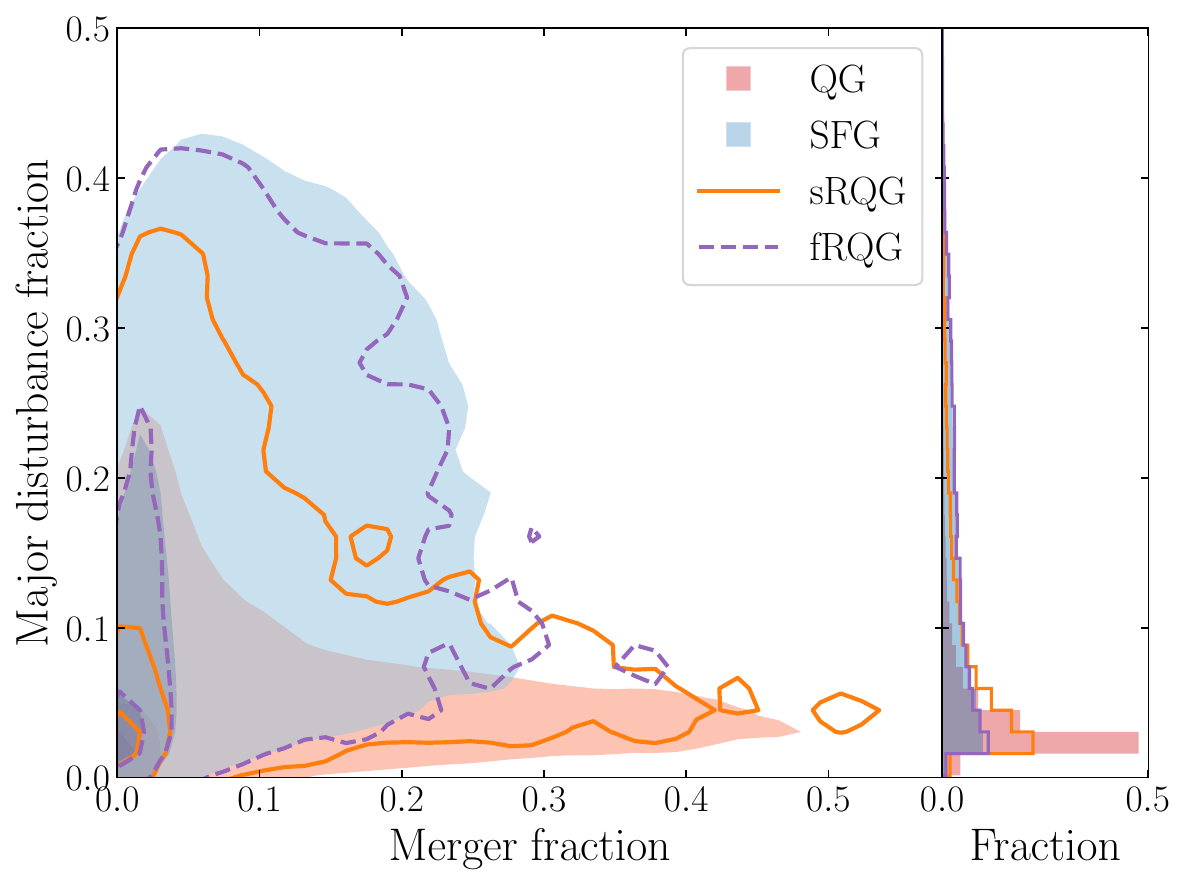}
\caption{Merger fraction and major distrubance fraction in QG, SFG, fRQG, and sRQG for a bright subsample from \citealt{Q1-SP047} at $0.2<z<1.5$. Contours and colours as in Fig.\,\ref{fig:mass-z}.}
\label{fig:merger}
\end{figure}

\begin{figure}[htbp!]
\centering
\includegraphics[angle=0,width=1.0\hsize]{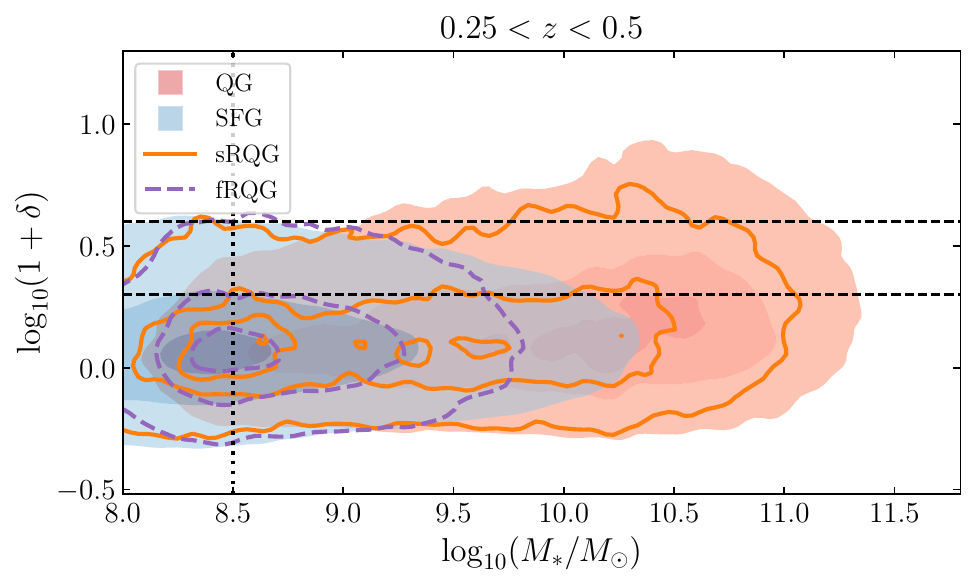}
\includegraphics[angle=0,width=1.0\hsize]{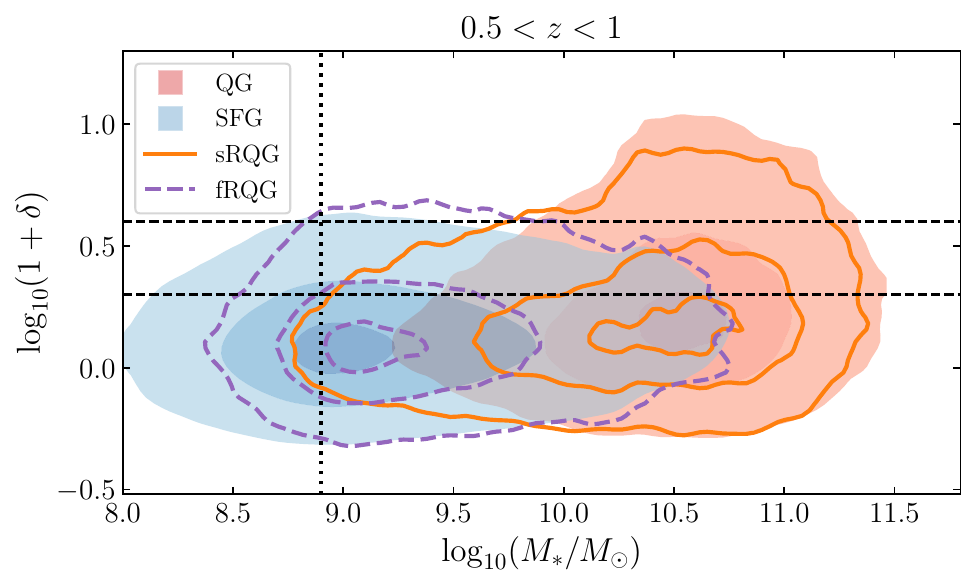}
\includegraphics[angle=0,width=1.0\hsize]{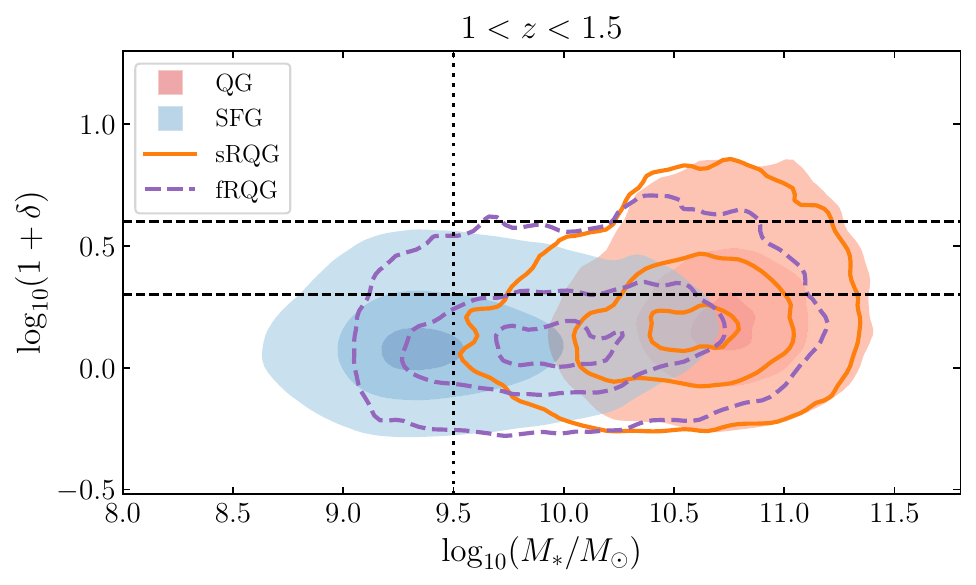}
\caption{Overdensity versus stellar mass relation of the four populations in three redshift bins, $0.25<z<0.5$, $0.5<z<1$, and $1<z<1.5$. Contours and colours as in Fig.\,\ref{fig:mass-z}. The vertical dotted line gives the 90\% mass limit at the high-redshift edge of each bin. The horizontal dashed lines give the $1\sigma$ and $2\sigma$ thresholds above median in overdensity as in \citet{Q1-SP069}. }
\label{fig:env-mass}
\end{figure}

We also use the mergers identified by \citet{Q1-SP047} in the bright subsample [i.e., $\IE < 20.5$ and ${\rm segmentation\_area}>200$ pixels ($\sim 2\,\mathrm{arcsec}^2$), or ${\rm segmentation\_area}>\num{1200}$ pixels ($\sim 12\,\mathrm{arcsec}^2$) and examine the merger probabilities in the four populations. In Fig.\,\ref{fig:merger}, the merger fraction indicates the possibility that a single galaxy is undergoing a merger, while the major disturbance fraction is the probability that a galaxy shows a tidal disturbance feature or has undergone a major merger in the past \citep{Q1-SP047}. In general, we find that merger and disturbance features are present only in a small fraction of the analysed sample, confirming a minor contribution of mergers to galaxy evolution also in the bright galaxies. In particular, the distribution of fRQGs is similar to that of SFGs. 
In contrast, the sRQG population, similarly to QG, has a tail with a higher probability of a merger in progress, whereas the probability of major disturbances or major mergers in the past is lower than in fRQGs and SFGs, but higher than in QGs. This sample is biased toward bright, massive galaxies, which tend to have a rich merger history. 

In conclusion, for morphological properties, we find the majority of RQGs have regular morphologies, ruling out merging as the main quenching method, with fRQGs being more similar to SFGs, whereas sRQGs  more similar to QGs.
This analysis suggests that fRQGs have not yet undergone morphological transformation, while sRQGs have already transformed to a more bulge-dominated morphology, similar to QGs. This result provides an indication of the possible quenching mechanisms (Sect.\,\ref{sc:discussion}).

\subsection{\label{sc:env}Environmental properties}

After analysing the properties and morphologies of the various populations, we investigate the environments in which these four populations exist.

Figure\,\ref{fig:env-mass} shows the density of the four populations in the stellar mass-overdensity panel. The environment is estimated up to $z\sim1$ in \citet{Q1-SP069} as the overdensity in redshift slices using $\Sigma_5$ estimator and derive the density contrast $\log_{10}(1+\delta)=\log_{10}(1+\frac{\Sigma_5-\bar{\Sigma}_5}{\bar{\Sigma}_5})$. We further extend the overdensity estimate up to $z=1.5$, even if we are aware that the uncertainties will be larger. The uncertainties mainly come from the pdz of the galaxies and are taken into account through a Monte Carlo integration. We define environments below the $({\rm median}+\sigma)$ of $\log_{10}(1+\delta)$\footnote{$\sigma$ being the standard deviation of the distribution estimated through sigma-clipping} as low-density, and those above $({\rm median}+2\sigma)$ as high-density.
At $z<0.5$ most of the populations reside in low density environment at all masses, but there is evidence in the high-mass regime of a tail extended to higher-density environments for QGs, where their fraction increases compared to lower-density environment (see Fig.\,\ref{fig:Fraction-mass-env} in Appendix\,\ref{apdx:env-fraction} for fraction plots). 
At $\log_{10} (M_\ast/M_\odot)<9.5$, fRQGs, sRQGs, and SFGs cover similar environments and extend to even lower density compared to QGs. At $z>0.5$, the tail in the high-density environment includes not only high-mass QGs, but also high-mass sRQGs and an intermediate-mass population of fRQGs, whose fraction increases with redshift (Fig.\,\ref{fig:Fraction-mass-env}).
At $z > 1$, the situation is similar to the $0.5<z<1$ regime, confirming that RQGs possibly also exist in high-density environments at high masses. Our results indicate that the environment can plays a role in both fast and slow quenching processes at $z>0.5$, suggesting an earlier evolution in high-density regions. This result can be related to JWST works finding early quenched galaxies in overdense environments \citep{Jin_2024,Baker_2026}.

\section{Quenching scenario\label{sc:discussion}} 

The analysis we perform on the photometrically selected RQG populations allows us to derive their average properties, morphologies, and environments, shedding light on possible quenching mechanisms. The robustness of our approach is demonstrated by our ability to extract a clean transitional signal despite the rigid selection criteria and potential mixing from SFGs and QGs.

\begin{table*}[htbp!]
\centering
\caption{Summary of fast quenching (fRQGs) and slow quenching (sRQGs) properties.}
\smallskip
\label{table:summary}
\smallskip
\renewcommand{\arraystretch}{1.5}
\begin{tabular}{c@{\hskip 1em}|>{\centering\arraybackslash}m{6.2cm}|>{\centering\arraybackslash}m{6.2cm}}
  \hline
  & Fast quenching fRQG & Slow quenching sRQG \rule{0pt}{2.5ex}\\
  \hline
  \hline
Selection 
& \multicolumn{2}{c}{Classified by \Euclid rest-frame $UVJ$ diagram plus SFH model track split} \\
\hline
sSFR & $\sim10^{-10}\,\mathrm{yr}^{-1}$, between SFG and QG & $\sim10^{-10}\,\mathrm{yr}^{-1}$, between SFG and QG \\
\hline
Stellar mass & $\sim10^8$--$10^{10}\,M_\ast/M_\odot$, lower than sRQG at fixed $z$ & $\sim10^8$--$10^{11}\,M_\ast/M_\odot$, higher than fRQG at fixed $z$\\
\hline
SFH timescale $\tau$ & $<0.1\,\mathrm{Gyr}$ & $\sim 1\,\mathrm{Gyr}$ \\
\hline
Age/$\tau$ & $\sim20$, between SFG and QG & $\sim5$, between SFG and QG \\
\hline
S\'ersic index $n_\sfont{S}$ & Smaller than sRQG, similar to SFG & Larger than fRQG, similar to QG \\
\hline
Size-mass relation & Similar to SFG & Similar to QG \\
\hline
Environment & Prefer low-density environments at $0.25<z<0.5$, marginally also in dense environment at $0.5<z<1.5$ & In low-density environments at $0.25<z<0.5$, in all environments at $0.5<z<1.5$ \\
\hline
Fraction & Decreases with mass at $z<1$, increases with mass at $z>1$, decreases with redshift & Increases with mass, decreases with redshift for $\log_{10} (M_\ast/M_\odot)<10.5$ galaxies, increases with redshift for $\log_{10} (M_\ast/M_\odot)>10.5$ galaxies \\
\hline
Quenching efficiency (QE) & Decreases with mass at $z<1$, increases with mass at $z>1$, increases with redshift & Increases with mass, decreases with redshift \\
\hline
Quenching stage (QS) & Later at massive end, later at low-$z$ & Later at massive end, later at low-$z$ \\
\hline
Merger & Higher probability of past major merger & Higher probability of ongoing merger \\
\hline
\multirow{2}{*}{Proposed quenching mechanisms} 
  & High-mass: radiatively efficient AGN feedback 
  & High-mass: merger, morphological quenching, tidal effect, jet-mode AGN \\
  \cline{2-3}
  & Low-mass: ram pressure stripping, supernovae 
  & Low-mass: strangulation, morphological quenching \\
\hline
\end{tabular}
\smallskip
\end{table*}

In Table\,\ref{table:summary}, we summarise different properties of the two populations of RQGs and the two different quenching paths.  
Our analysis suggests that various phases and mechanisms should be in place to quench star formation in SFGs and transform them into QGs.
Based on our analysis, we propose a galaxy-quenching scenario that starts before cosmic noon in high-mass galaxies and proceeds to lower masses. 
First, in early epochs, fast quenching  dominates at $z>1.2$ in galaxies with $\log_{10}(M_\ast/M_{\odot}) > 10.5$. The typical quenching timescale is about $0.1\,\mathrm{Gyr}$ and is always shorter than $1\,\mathrm{Gyr}$, and the morphological transformation is not yet complete. A possible mechanism at high masses could be radiatively efficient AGN feedback of quasars. At lower redshifts, fast quenching decreases in importance even if it continues to act at $0.5<z<1$, progressively at lower masses among galaxies with $\log_{10}(M_\ast/M_{\odot}) <10$, residing marginally also in high-density environments. This result suggests that environmental mechanisms, such as the ram pressure stripping, may be at play. This is consistent with the fact that the anchoring forces keeping the gas on the disc of perturbed galaxies strongly decreases with stellar mass \citep{Boselli_2022}. At even lower redshift, fast-quenching becomes important at lower masses in various environments, suggesting possible supernova feedback or back-splash galaxies ejected from denser environments \citep{Hirschmann_2014}. 
On the other hand, slow mechanisms dominate quenching at high masses at $z<1.2$ and become progressively more significant at lower masses as the redshift decreases. In this case, the typical quenching timescale is about $1\,\mathrm{Gyr}$ or longer, and galaxies have already or are simultaneously transforming their morphologies into more bulge-dominated morphologies, similar to QGs. We also find evidence of past mergers in sRQGs. 
Mechanisms such as mergers, strangulation, jet-mode AGN, and morphological quenching can lead to morphological transformation, particularly in high-density environments. In Fig.\,\ref{fig:env-mass}, a high fraction of slow quenching occurs in low-density environments at $z<0.5$, suggesting mechanisms such as morphological quenching in low-mass galaxies and jet-mode AGN feedback in massive galaxies. The lack of RQGs at $z<0.5$ in high-density environments at the highest masses, combined with the fact that both fast and slow quenching predominantly occur in the field at lower masses, suggests that quenching in high-density environments occurs {soon} after cosmic noon and is almost complete by $z\sim0.5$.

However, we remind the reader that the identified fRQGs and sRQGs may be contaminated by rejuvenated galaxies moving from the QG region back to the RQG and even the SFG regions. 
In principle, one might expect a galaxy undergoing rejuvenation to exhibit a patchy morphology due to regions of star formation, which could
affect its uniformity and Gini coefficient.
There are no significant differences in smoothness or the Gini coefficient between the four populations, suggesting that rejuvenation does not significantly affect RQGs. This is consistent with the discussion in \citet{wu_2021}, where the authors found that the rejuvenation fraction in the TNG100 simulation was $\sim 2\%$. We also need to note that the patchiness related to star-forming regions may be difficult to detect in galaxies at $0.5<z<1.5$, due to their small angular resolution. 
The effect of rejuvenation for fast/slow quenching pathways needs further investigation. 

\section{\label{sc:spec}Spectroscopy of QGs and RQGs}
In this section, we use the Q1 spectra from NISP \citep{Q1-TP007, EuclidSkyNISP}, to analyse the spectroscopic properties of QG and RQG populations. Given the average low signal to noise ratio (S/N) of single \Euclid spectra, we derive stacked spectra, using the spectroscopic subsamples of QGs, fRQGs, sRQGs and SFGs with matched spectroscopic redshifts from external surveys as mentioned in Sect.\,\ref{sc:Data}. 
Even if this spectroscopic subsample is  not statistically representative due to the biases introduced by the different selections of the various spectroscopic surveys, our objective is to assess the reliability of our photometric selection and estimate the main differences among the various populations. 

\begin{figure}[htbp]
\centering
\includegraphics[angle=0,width=1.0\hsize]{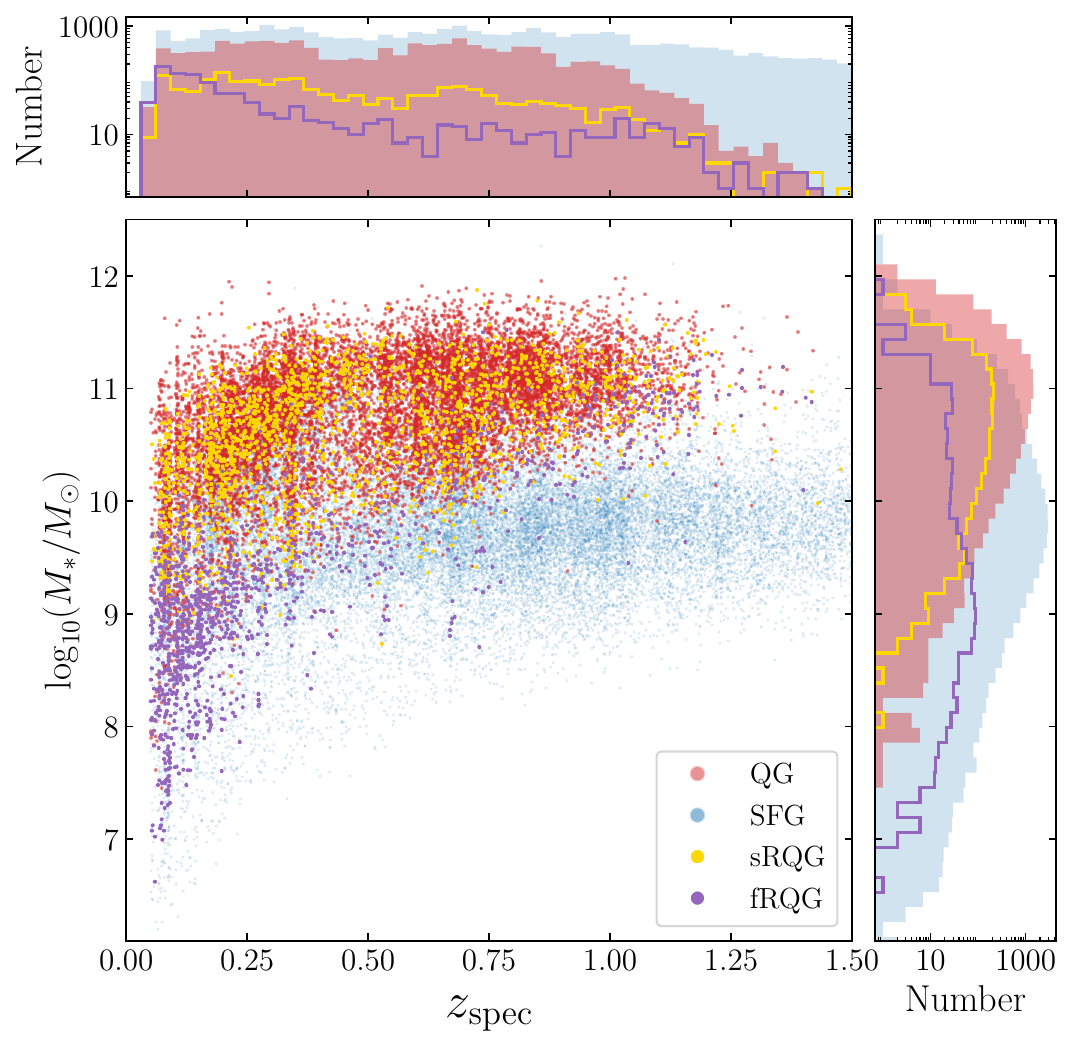}
\caption{Stellar mass versus redshift relation of galaxies for the spectroscopic subsample. 
}
\label{fig:spec-sample-mass-z}
\end{figure}

As reported in Sect.\,\ref{sc:Data}, for the galaxies in this spectroscopic sample, we derive the physical properties as stellar mass, SFR, and rest-frame colours, at $z_{\rm spec}$, using \texttt{NNPZ} algorithm, and, therefore a new galaxy classification (Appendix\,\ref{apdx:photz-specz-classification}). Figure\,\ref{fig:spec-sample-mass-z} shows the redshift and stellar mass distribution of this sample.
In total we have about 45 thousands galaxies with external spectroscopic redshift, with  \num{12190} QGs, \num{2081} sRQGs, and \num{1138} fRQGs, and \num{29451}  SFGs, photometrically classified, covering the redshift range $0<z_{\rm spec}<1.5$. Beyond this redshift limit, the number of QGs and RQGs with reliable external $z_{\rm spec}$ is too small to reach a sufficiently high S/N for stacking spectra analysis.
In this sample we were able to check the fraction of AGN in the subsample of spectra observed in DESI DR1, composed of \num{6652} QGs, 876 sRQGs, and 504 fRQGs and \num{16859} SFGs. 
We look into these galaxies in a catalogue cross-matching DESI AGN/QSO catalogue (Juneau et al. in prep.) with \Euclid sample (as mentioned in \citealt{Q1-SP027}), and find that only a few per cent ($< 5\%$) are identified as AGN/QSO in this catalogue. However, the main DESI selection can bias these fractions, and correcting for such bias is beyond the aim of this paper.

\subsection{\label{sc:stackfeature}Stacked spectra and templates}
In this section, we present the stacked spectra of each population.
In Appendix\,\ref{sc:stackmethod} we introduce the stacking method and the stacked spectra in different redshift and stellar mass bins.
Here, we construct the global templates for each population, combining the stacked spectra at $z<1.5$.
These templates can be successfully used to check the reliability of our photometric selection of QGs and RQGs and for the identification of spectral features and compared to spectral population synthesis models, as well as used to estimate redshift in single objects within the \Euclid framework.
The large range of redshift covered does not allow a lambda range common to all spectra; therefore, we combined the spectra stacked in redshift bins. We first combine two stacked spectra at a time, renormalising in a common rest-frame wavelength range with no significant emission or absorption lines. 
Then, we calculate the weighted average flux of the two stacked spectra, assigning a weight to each pixel based on the number of spectra used in each stacking. This is mathematically equivalent to a mean stack of all spectra.  
We also propagate the uncertainty of the two stacked spectra and save the number of spectra used in each pixel. The combination begins with the two lowest redshift bins, and then we repeat this process by combining the output spectrum with the stacked spectrum in the next redshift bin, continuing until all stacked spectra from the redshift bins are included.
We also explore the possibility of using a different method to derive a global stack \citep{Francis_1991}. However, as there is no common lambda range available for all spectra, we are forced to use a running common range, which is more susceptible to the presence of individual noisy spectra, making the final model less stable for a given set of stacking parameters.
Nevertheless, we verify that this new global method produces similar templates.

\begin{figure}[htbp!]
\centering
\includegraphics[angle=0,width=1.0\hsize]{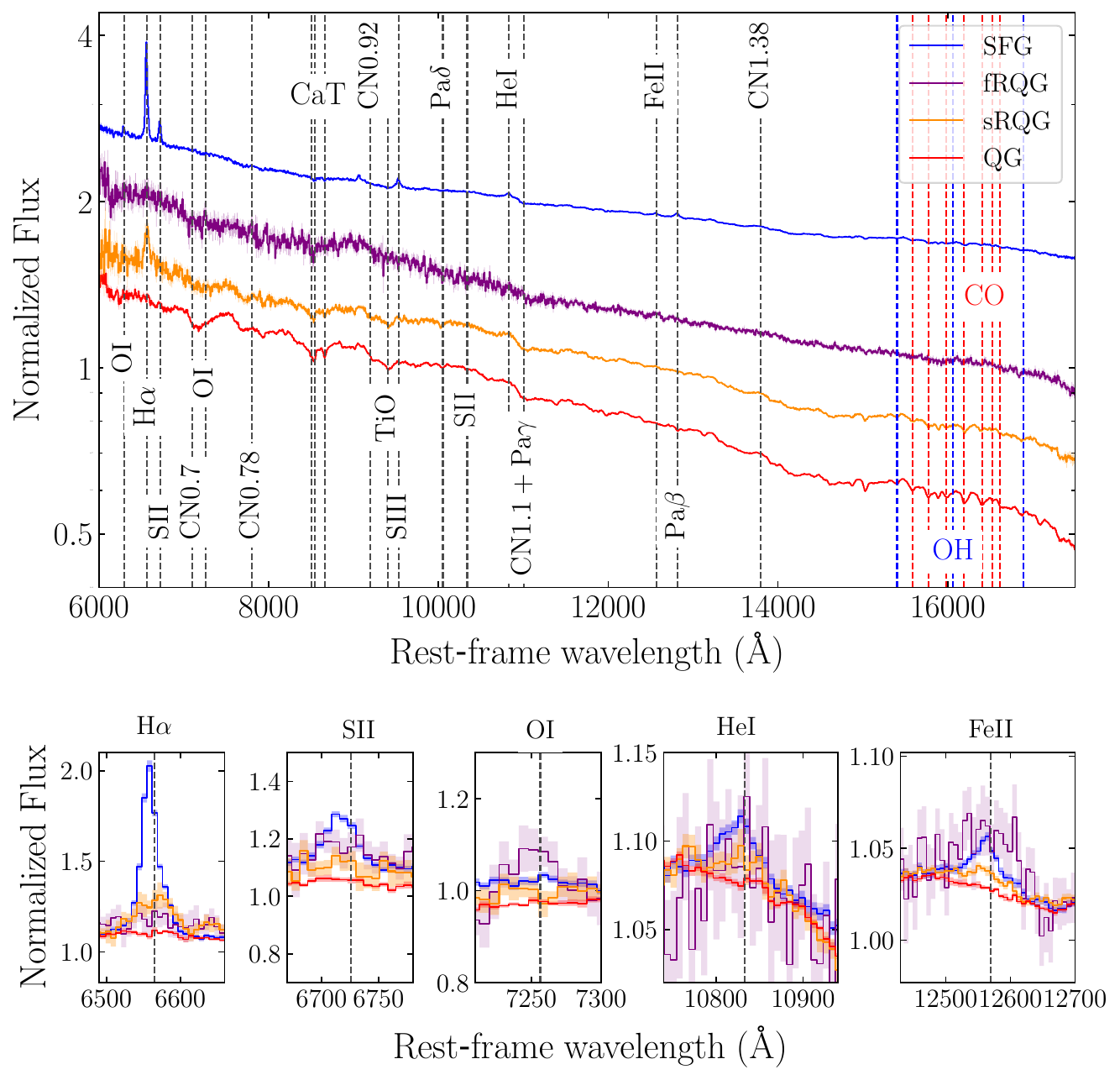}
\caption{Comparison between template stacked spectra of SFGs, QGs, fRQGs, and sRQGs. Some major features are marked. All red vertical lines indicate CO lines, and blue lines indicate OH lines, with a zoom-in to spectral features that highlight the difference between populations.}
\label{fig:stack-in-populations}
\end{figure}
Figure\,\ref{fig:stack-in-populations} shows the template spectra for the QG, fRQG, and sRQG populations from optical to NIR wavelengths at $z<1.5$. For comparison, we also show the template spectrum of SFGs.
Given the high number of stacked spectra, we find that the templates show an average high S/N and a low mean dispersion, with values ranging from S/N $\sim500$ down to 
$50$, with a dispersion from 14\% to 30\%, for QG, sRQG, to  fRQGs templates, respectively. 
From Fig.\,\ref{fig:stack-in-populations} we note, in the NIR range, dominated by low-redshift sources, all three stacked spectra of QGs and RQGs appear quite similar, and very different from SFGs templates, despite higher noise in the fRQG spectrum due to its smaller sample size. The QG spectrum shows more prominent absorption features and sharper continuum breaks, consistent with an old stellar population, as we further discuss in Sect.\,\ref{sc:stack-vs-tmpl}.
Their differences mainly emerge in the optical range, where the contributions are primarily from higher-$z$ sources.
In particular, sRQG spectrum shows clearly a faint \ha \ emission,   while no \ha \ emission appears in the QG spectrum. 
We measure emission lines using \texttt{slinefit}\footnote{The \texttt{slinefit} code is publicly available at \url{https://github.com/cschreib/slinefit}.} \citep{Schreiber_2018, Talia_2023}. The stacked SFG spectrum shows strong, high equivalent width (EW), \ha \ emission (EW $\sim 28\,\AA$, S/N $\sim33$), 
while sRQG and fRQG spectra show only weak \ha \ emission, with a clear detection for the sRQG (EW $\sim8.5\,\AA$ and S/N $\sim7$) and no detection (EW $\sim 1.4\,\AA$ and S/N $\sim1.5$) for the fRQG. 
The emission in sRQGs can be related to the residual star-formation activity or AGNs hosted by RQGs at $z<1.5$. 
We also note that faint (EW $< 2\,\AA$ and low S/N $\sim3$) \ion{S}{ii} emission lines are present in both RQG spectra.
In fRQGs, we also note that there are hints of very faint \ion{O}{i}, \ion{He}{i}, and \ion{Fe} {ii} emission lines, with $1<$ EW $<4\,\AA$ but with S/N $\lesssim 3$, which could be related to AGN activities. These lines are indeed expected, due to the presence of some AGNs in the DESI sample (Appendix\,\ref{sc:stackmethod}).

In the QG template spectrum, some NIR absorption features that we note (e.g., \ion{Ca}{ii} triplet) have previously been used in the literature by \citet{Conroy_2012} for the IMF estimation. Therefore, in the future, using more complete and homogeneous samples, it will be possible to utilise these features to investigate IMF at different redshifts and masses.

In this study, we measure a series of Lick indices and spectral breaks using \texttt{pyLick} \citep{Borghi2022a}, adding NIR indices as defined by \citet{Riffel_2019}. We find that sRQG shows weak $\mathrm{Pa}\,\delta$, and possible $\mathrm{Pa}\,\beta$ absorption, which confirms its recently quenched nature. We further discuss the measurements in Sect.\,\ref{sc:stack-vs-tmpl}.

\subsection{\label{sc:stack-vs-tmpl}Comparison with models and empirical templates}
We compare the stacked templates for QGs and RQGs with E-MILES stellar population models \citep{Vazdekis_2016, Koleva_2012, LaBarbera_2017} at various metallicities ($0.5$--$2\,Z_\odot$) and with the empirical templates by \citet{Mannucci_2001}, produced from spectra observed with TIRGO LonGSp \citep{Vanzi_1997}.

\begin{figure}[htbp!]
\centering
\includegraphics[angle=0,width=1.0\hsize]{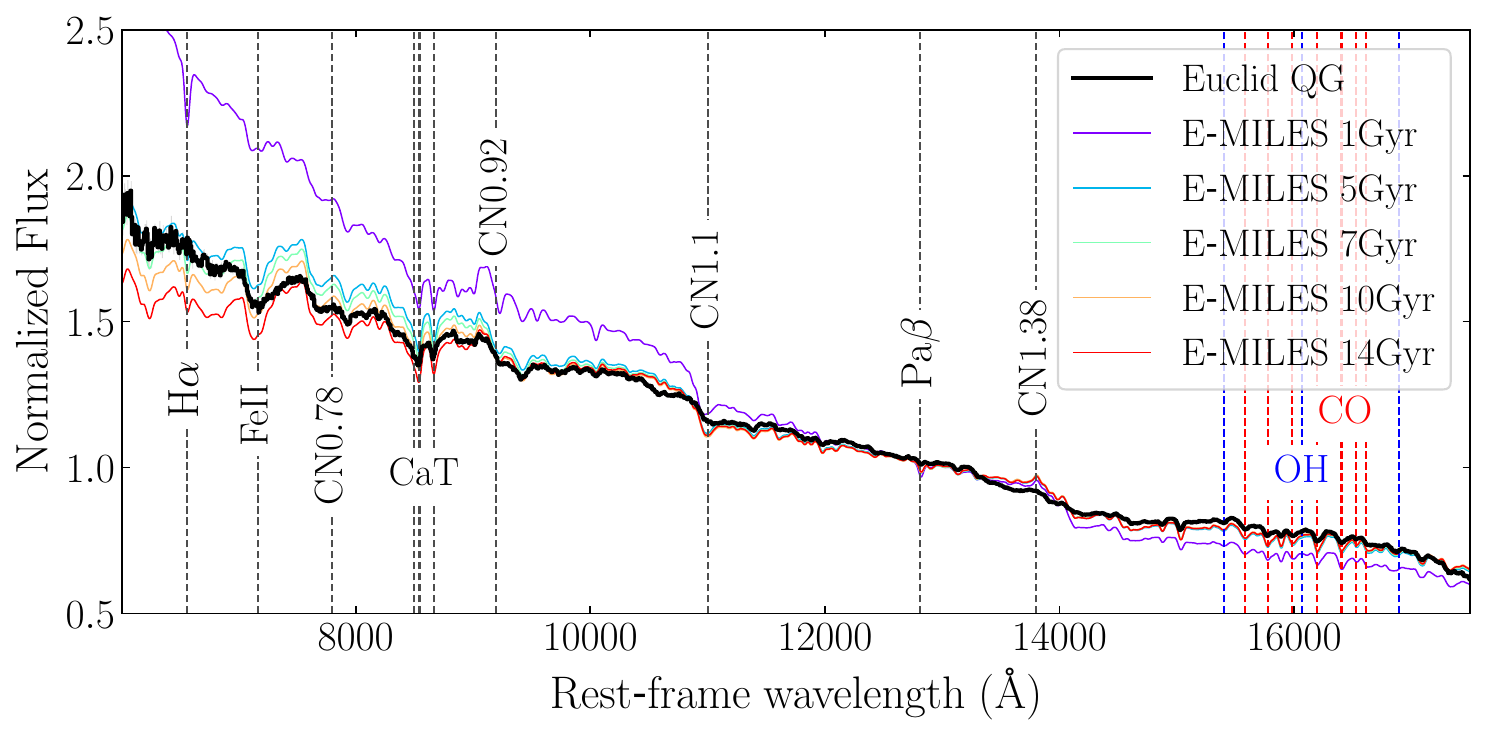}
\caption{E-MILES models at solar metallicity with various ages compared to the stacked QG template. All red vertical lines mark CO lines, and blue lines mark OH lines.}
\label{fig:stack-vs-tmpl-full}
\end{figure}

In Fig.\,\ref{fig:stack-vs-tmpl-full}, we compare the stacked QG spectrum to E-MILES models, downgraded to similar \Euclid resolution, with $Z_\odot$ and various stellar ages. At long wavelengths ($\lambda > \num{11000}\,\AA$), the stacked template agrees with models with ages $5$--$14\,\mathrm{Gyr}$. At shorter wavelengths, the stack agrees instead with models having gradually younger ages of $5$--$7\,\mathrm{Gyr}$. This effect could be because at shorter wavelength the template is dominated by higher-$z$ bins ($z>0.75$, see Fig.\,\ref{fig:stack-in-z-bins}), when the Universe was younger.

From this comparison, we note that our template presents most of the absorption features present in the models, as well as some bumps that we attribute to CN. However, these CN bumps are always more prominent in the models compared to the \Euclid templates. Our results could indicate differences in SPS models compared to those of \Euclid galaxies, which require further investigation using a complete sample from the next \Euclid release.

\begin{figure*}[htbp!]
\centering
\includegraphics[angle=0,width=1.0\hsize]{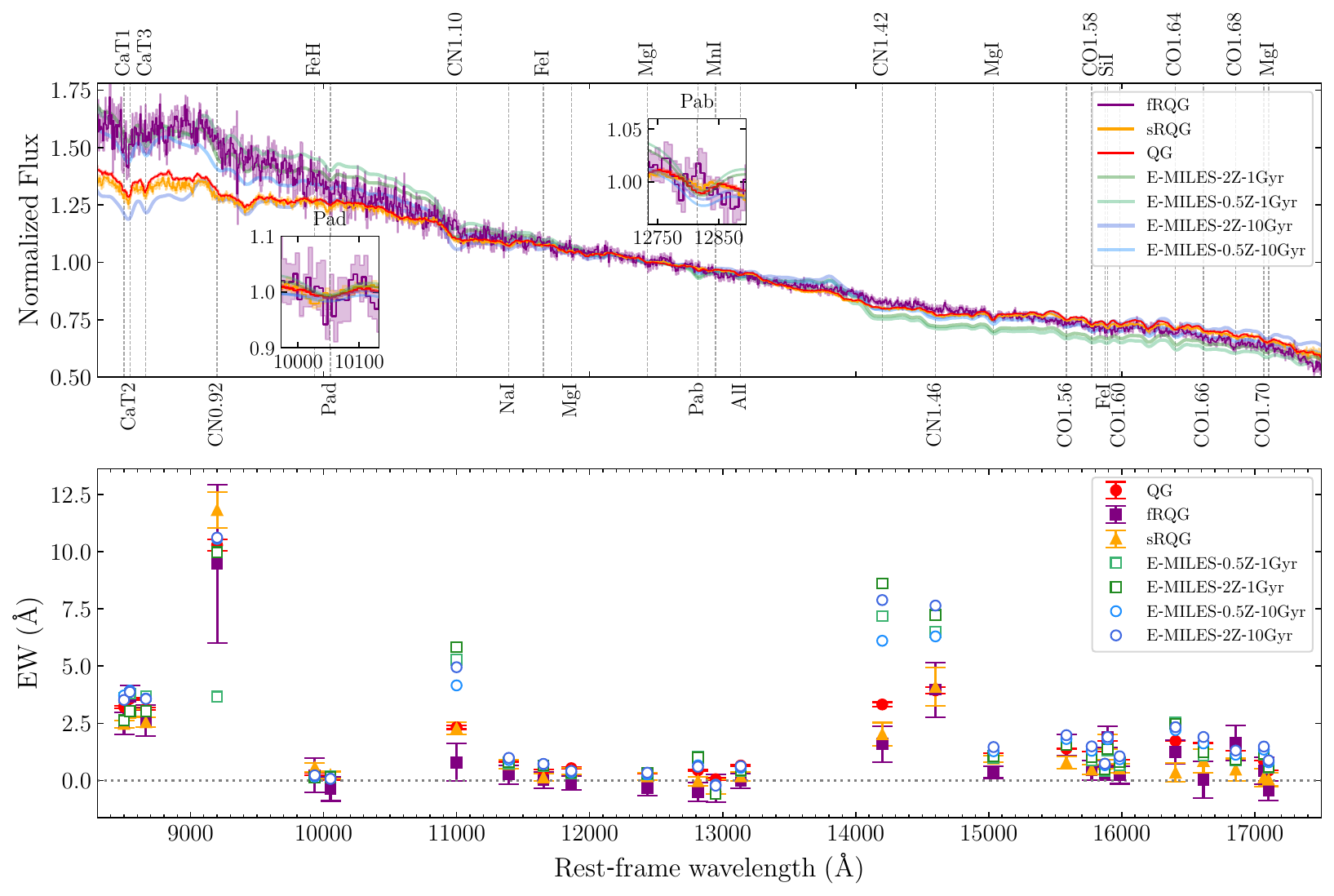}
\caption{Spectral templates ({\it upper panel}) and indices ({\it bottom panel}) of QG, fRQG, and sRQG compared to  E-MILES models for various ages and metallicities. The two small panels inside the upper panel are zoom-in ranges of $\mathrm{Pa}\,\delta$ and $\mathrm{Pa}\,\beta$ features. 
}
\label{fig:index_compare}
\end{figure*}

To quantify possible differences in absorption features between our templates and models, we then measure the spectral indices of the E-MILES model with ages of $1\,\mathrm{Gyr}$ and $10\,\mathrm{Gyr}$, and metallicity values of $0.5\,Z_\odot$ and $2\,Z_\odot$, downgrading their resolution to match the same resolution as our templates. 
In Fig.\,\ref{fig:index_compare}, we compare the measurements with the indices measured from stacked QG, sRQG, and fRQG spectra. Despite the CN break features, other model features generally agree with the \Euclid stacked spectra measurements. In particular, we show and measure Paschen lines, finding $\mathrm{Pa}\,\beta$ and $\mathrm{Pa}\,\delta$ in absorption for QGs and sRQGs consistent with models with ages greater than $1\,\mathrm{Gyr}$, while for fRQGs the S/N is too low to allow comparison with models.
We notice in the E-MILES models that some line strengths evolve with stellar age and metallicity. With a more complete sample, it will be possible to analyse these trends from the stacked \Euclid spectra and put constraints on stellar populations. 
We further investigated the optical spectra of the DESI cross-matched subsample, to analyse the Balmer absorption features of each populations. The optical spectra were stacked with \texttt{SpectraPyle} (Appendix \ref{apdx:DESI_stack}).
We find that QGs and RQGs stacked spectra show evident $\mathrm{H}\,\delta$ absorption and strong $4000$\,\AA\ break, being $\mathrm{H}\,\delta$ stronger and  ${\rm Dn4000}$ smaller in fRQGs. 
All measurements, reported in Appendix \ref{apdx:DESI_stack}, with ${\rm S/N}>50$ are consistent with what expected for RQGs and QGs (see figure\,A.2 in \citealt{Mao_2022}).

\begin{figure}[htbp]
\centering
\includegraphics[angle=0,width=1.0\hsize]{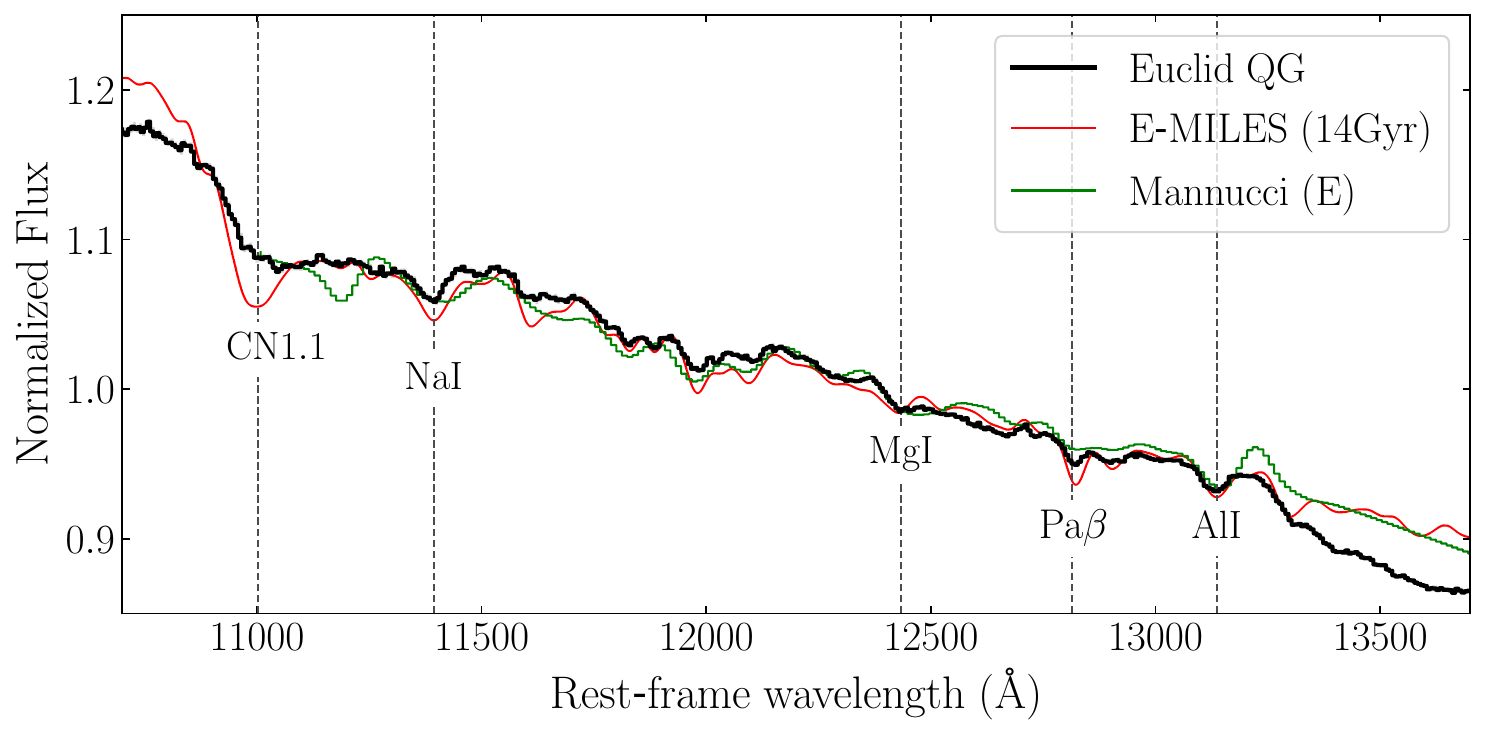}
\includegraphics[angle=0,width=1.0\hsize]{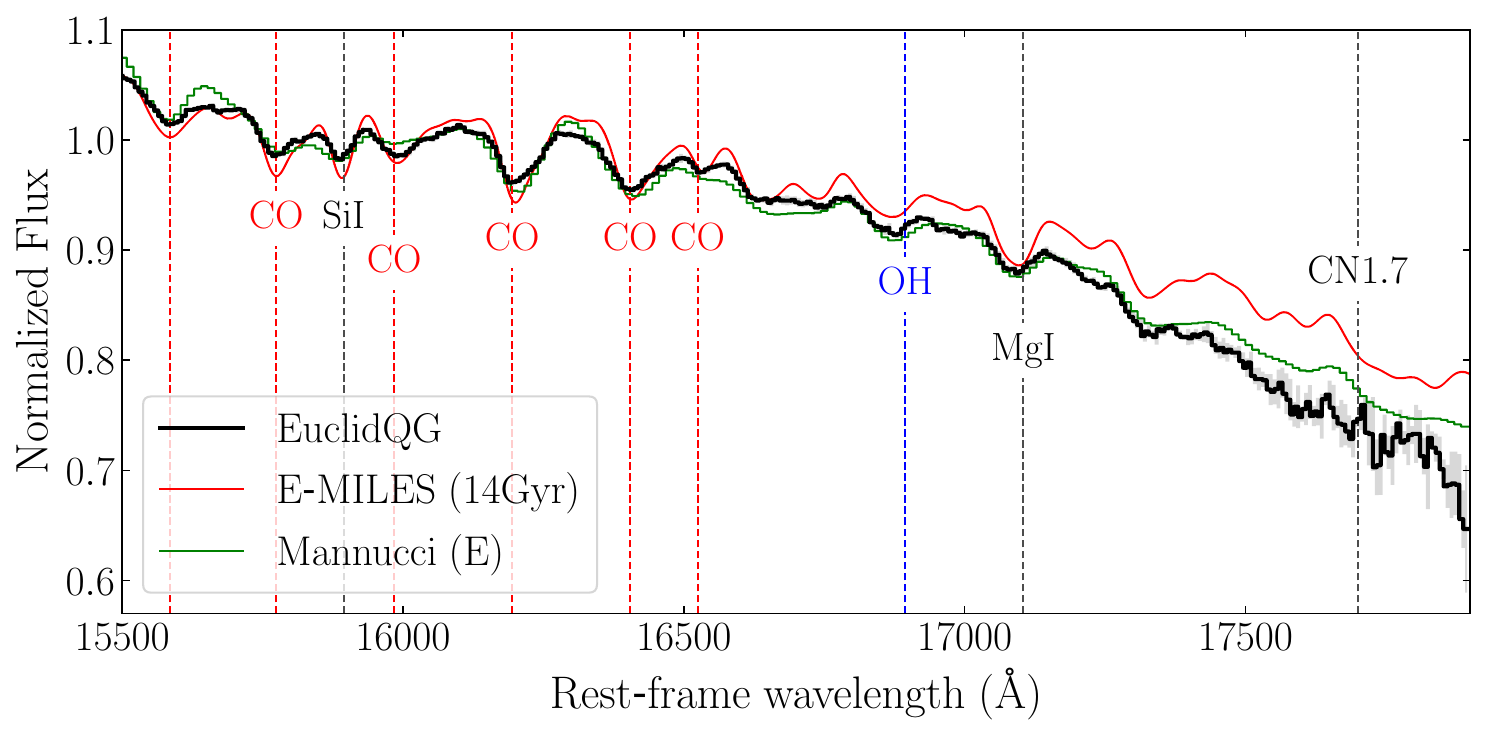}
\caption{Stacked QG spectrum compared to E-MILES model with $14\,\mathrm{Gyr}$ stellar age and to Mannucci template for elliptical galaxies \citep{Mannucci_2001}. The top panel shows a comparison in the $J$ band, and the bottom panel shows a comparison in the $H$ band. In the $J$ band, the Mannucci template begins at a wavelength of \num{11000}\,\AA.}
\label{fig:stack-vs-z-tmpl-NIR}
\end{figure}

In Fig.\,\ref{fig:stack-vs-z-tmpl-NIR}, we compare the stacked QG spectrum with the elliptical galaxy template from \citet{Mannucci_2001} and with the E-MILES model at $Z_\odot$ and an age of $14\,\mathrm{Gyr}$ in the $J$ and $H$ bands. Our template agrees well with both bands. In the $J$ band, the agreement with the E-MILES model is better, while there are some differences with the \citet{Mannucci_2001} elliptical template. 
In the $H$ band, the agreement with the Mannucci template is better, especially at wavelengths longer than \num{16500}\,\AA, while the E-MILES models present more prominent features.
These agreements show that the stacked \Euclid QG template can reconstruct the continuum shape and absorption features with relatively good precision. This template has been provided to the \Euclid consortium team and has been successfully used for the identification and redshift estimation of passive galaxies in the \Euclid pipeline in DR1 (Euclid Collaboration: Le Brun et al., in prep.). 

\section{Summary and perspectives}
In this work, we identify \num{420101} QGs, \num{5618011} SFGs, \num{339407} fRQGs, and \num{133598} sRQGs at $0.2<z<3$ from Q1, using the rest-frame $UVJ$ diagram. Our approach, originally introduced by \citet{Mao_2022}, has been adapted to the \Euclid filter set. 

We investigate the quenching process through the fraction of populations, QE, and QS. Then we compare the physical properties (i.e., stellar mass, SFR, sSFR, $\tau$, and age/$\tau$), the S\'ersic indices, the mass-size relations, and the environments of the four populations. From these analysis, we find the following results:
\begin{itemize}
\item
The fast-quenching mechanisms are different at high-$z$ and low-$z$. 
The fast-quenching process begins in massive galaxies following the cosmic noon, then moves to lower-mass galaxies at low redshift. 
\item
The slow quenching begins at high redshift in massive galaxies. Although slow quenching continues to occur in massive galaxies, it also appears in less massive galaxies at lower redshift. 
\item  
Intense quenching processes occur around cosmic noon in massive galaxies, then downsizing to lower-mass galaxies.
\item In contrast, the GAEA models for massive galaxies significantly overestimate the fraction, QE and QS of fRQGs, whilst underestimating the fraction of SFGs.
\item
The stellar mass, SFR, sSFR, and age/$\tau$ of RQGs lie between those of SFGs and QGs. 
Regarding the SFH timescale, fRQGs typically have short timescales $\tau < 0.05\,\mathrm{Gyr}$, while sRQGs have long timescales $\tau \sim 1\,\mathrm{Gyr}$. 
\item
The fRQGs have morphologies similar to SFGs, while sRQGs have morphologies more similar to QGs. This indicates that the morphology transformation occurs simultaneously or earlier than slow quenching, but later than fast quenching. 
\item
We examine the environment of the four populations up to $z\sim1.5$ and find that sRQGs have more evident tails in dense environments at $z>0.5$, and marginally also fRQGs, while they mainly exist in lower-density regions at $z<0.5$, indicating that quenching in high-density environments ends earlier than in fields.
\end{itemize}

Combining results from the fraction of RQGs and their properties, we conclude that a series of complex processes cause quenching, and we propose the following scenario. 
Fast quenching operates in massive galaxies 
likely driven by quasars shortly after cosmic noon, while in lower redshift, it occurs in lower-mass galaxies likely due to ram pressure stripping in a dense environment or by supernovae in the field. Slow quenching begins in massive galaxies during cosmic noon and 
could be driven by mergers, strangulation, tidal effects, jet-mode AGN, and morphological quenching across various environments. At $z<0.5$, slow quenching mainly occurs in the field, likely driven by jet-mode AGN in massive galaxies or morphological quenching in lower mass galaxies.

We also carry out spectroscopic analysis stacking $\sim 45$ thousand Q1 spectra with reliable $z_{\rm spec}$ provided by various external surveys.  Among them, $\sim 25$ thousand targets are observed in DESI DR1.
\begin{itemize}
\item 
We derive high S/N and homogeneous stacked templates for the populations of QGs, sRQGS, fRQGS, and SFGs. The templates of QGs and RQGs appear quite similar, with no prominent emission lines, confirming the passive nature for QGs and the recently quenched nature for RQGs, in contrast with strong emission lines detected in the SFGs template.
\item
A weak residual \ha \ emission line is found in the sRQG template, indicating a remaining star-formation activity or an AGN contribution.
\item
Despite the incompleteness of the current spectroscopic stacking sample, the template of QG agrees well with E-MILES models with stellar population older than $5\,\mathrm{Gyr}$, even if the observed CN bumps are always less prominent. It also agrees with empirical templates of elliptical galaxies from \citet{Mannucci_2001}.
\item
The stacked DESI spectra of DESI cross-matched objects show strong Balmer absorption features for both fRQG and sRQG stacks. All QG, fRQG, and sRQG stacks show strong Dn4000, in contrast with SFG stacked spectrum.  These features confirm the corresponding nature of our photometrically selected sample.
\end{itemize}
Future \Euclid data releases, DR1 for instance, will provide spectra across deeper and wider fields. This will allow us to increase significantly the current sample size and perform spectral analysis in redshift and mass bins. DR1 will allow us to have better estimation of $z_{\rm spec}$. By using $z_{\rm spec}$ measured from \Euclid spectra, we will have greater control over the selection function, and, consequently, a better understanding of the stellar population in each galaxy type.

\begin{acknowledgements}
ZhMa, LuPo acknowledge support from Italian Ministry of University and Research (Premiale MITIC 2017). ZhMa, LuPo, AnEn, SaQu, MaTa, MiBo, FaGe, LoGa, GaDe, EmDa, ChDE acknowledge support from the ELSA project. ``ELSA: Euclid Legacy Science Advanced analysis tools'' (Grant Agreement no. \num{101135203}) is funded by the European Union. Views and opinions expressed are however those of the author(s) only and do not necessarily reflect those of the European Union or Innovate UK. Neither the European Union nor the granting authority can be held responsible for them. UK participation is funded through the UK HORIZON guarantee scheme under Innovate UK grant \num{10093177}.
S.B. is supported by ERC grant \num{101076080}.
AnEn, MiBo acknowledge support from the INAF MiniGrant 2023 ``ADIEU: Anomaly Detections In EUclid''.
TK acknowledges financial support from JSPS KAKENHI Grant Numbers 24H00002 (Specially Promoted Research by T. Kodama et al.), 22K21349 (International Leading Research by S. Miyazaki et al.), and JSPS Core-to-Core Program (JPJSCCA20210003; M. Yoshida et al.).

In preparation for this work, we used the following codes for Python: \texttt{Numpy} \citep{numpy}, \texttt{Scipy} \citep{scipy}, \texttt{Pandas} \citep{mckinney-proc-scipy-2010}, \texttt{NNPZ}\ \citep{Tanaka_2018}.
  \AckQone
  \AckDatalabs
  \AckEC 
\end{acknowledgements}

%
%

\bibliography{Euclid,bib,DR1}

\begin{appendix}
\section{\label{appd:boundary shift}Adaption of colour-colour diagrams to \Euclid filters}

The $U$, $V$ \citep{Johnson_2006}, and 2MASS $J$ \citep{Huchra_2012} filters used in \citet{Mao_2022} are different from $U$, $V$ \citep{JohnsonFilters}, and \JE filters used in this work. 
The differences between \citet{Johnson_2006} and \citet{JohnsonFilters} $U$ and $V$ filters are discussed in \citet{Johnson_2006}.
\begin{figure}[htbp!]
\centering
\includegraphics[angle=0,width=1.0\hsize]{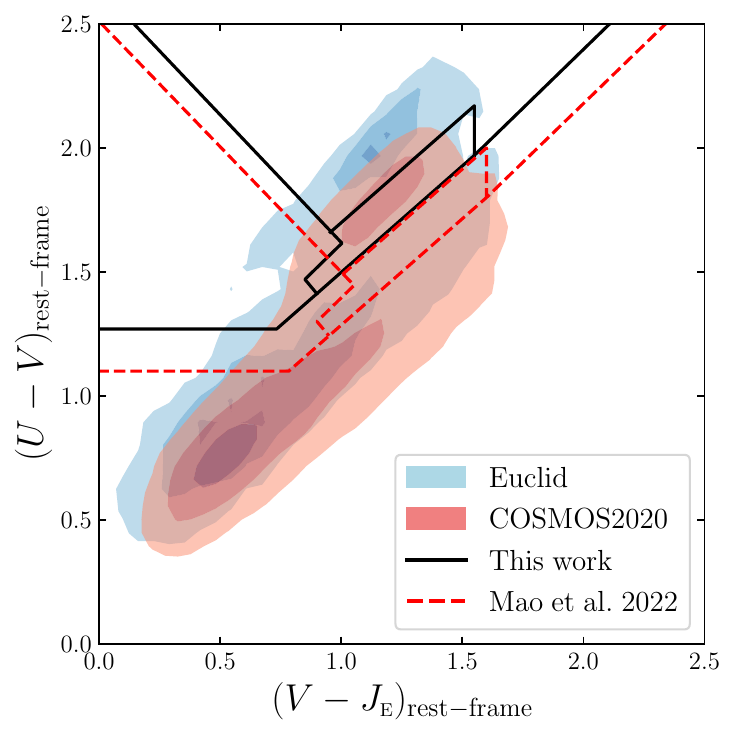}
\caption{Offset between rest-frame $UVJ$ distribution of COSMOS2020 and Q1 massive galaxies with $\log_{10}(M_\ast/M_\odot) > 9$ at $0.6<z<0.7$. Red-filled contours are the distribution of Q1 galaxies. Three contours include $10\%$, $50\%$, and $90\%$ of the total sample. Blue-filled contours show the distribution of COSMOS2020 galaxies within the same percentage.
Red dashed lines are original boundaries used in \citet{Mao_2022} and black solid lines are boundaries we apply to separate QG, SFG, fRQG, and sRQG in Sect.\,\ref{sc:ccdiagram}.}
\label{fig:uvj-boundary}
\end{figure}
Figure\,\ref{fig:uvj-boundary} shows the distribution on the $UVJ$ diagram of COSMOS2020 and \Euclid galaxies with $\log_{10}(M_\ast/M_\odot) > 9$ at $0.6<z<0.7$, being less uncertain in the rest-frame colours. 
Comparing the peaks of the density contours to the COSMOS2020 sample, we find that the \Euclid galaxies are redder in $(U-V)$ colour and bluer in $(V-J)$ colour. 
The boundaries are shifted accordingly and reported here.
The SFGs are separated from other populations by boundaries
\begin{equation} \label{eq:SFG}
\begin{aligned}
& U - V < 1.27
& \text{OR}\quad U - V < 0.857\,(V - \JE) + 0.641 .
\end{aligned}
\end{equation}
Apart from these two boundaries, the QG population is separated from fRQG and sRQG by three other boundaries
\begin{equation} \label{eq:QG}
\begin{aligned}
& V-\JE > 1.55 \quad \text{OR} \\
& \left\{
\begin{array}{l}
    U-V>0.857\,(V-\JE)+0.841,\\
    U-V>-1.165\,(V-\JE)+2.784.
\end{array}
\right.
\end{aligned}
\end{equation}
Finally, between fRQGs and sRQGs, there are two other boundaries
\begin{equation} \label{eq:RQG}
\left\{
\begin{array}{lr}
    U-V<0.857\,(V-\JE)+0.741, & \\
    U-V>-1.165\,(V-\JE)+2.46.
\end{array}
\right.
\end{equation}

\begin{figure*}[htbp!]
\centering
\includegraphics[angle=0,width=1.0\hsize]{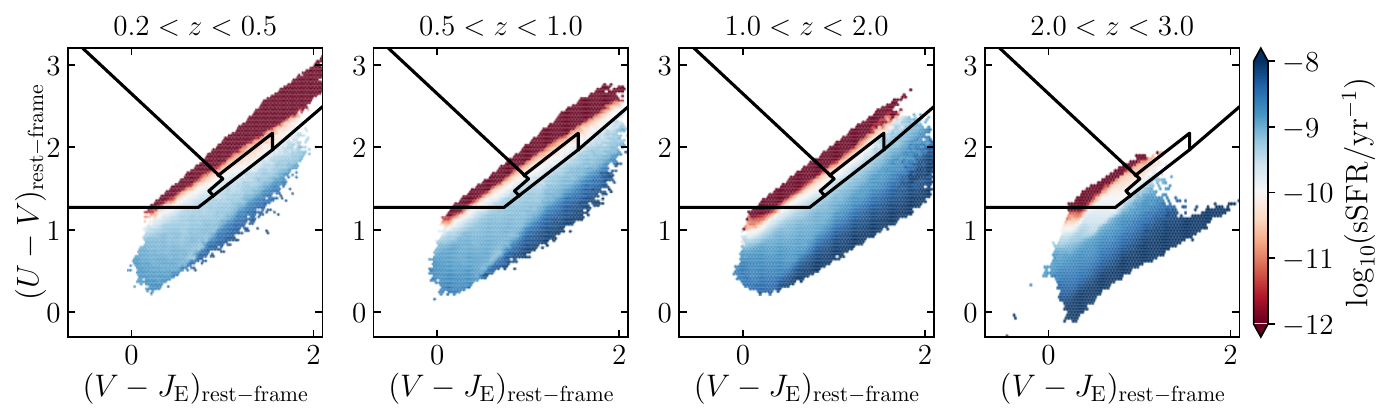}
\caption{Distribution of $UVJ$ classified galaxies in four redshift bins between $0.2<z<3.0$. The plot is colour-coded by sSFR from red (low sSFR) to blue (high sSFR).
}
\label{fig:uvj-z}
\end{figure*}

Here, we further show that the $UVJ$ boundaries do not change with redshift.
The distributions of galaxies on $UVJ$ diagram are shown in Fig.\,\ref{fig:uvj-z} in four redshift bins.
After applying the $90\%$ QG mass-completeness limit in each redshift interval (Fig.\,\ref{fig:mass-z}), we examined the distribution of the fitted sSFRs of the RQG populations. We use the peak of the sSFR distribution.
Overall, we find a value $\log_{10}(\mathrm{sSFR}/\mathrm{yr}^{-1})\sim -10 $, with a variation by less than $0.2$ dex, and 
no evidence for a smooth systematic evolution of the  peak of the sSFR distribution that would require a redshift-dependent $UVJ$ boundary.

\section{\label{apdx:comparison}Comparison of two classifications}
\begin{figure*}[htbp!]
\centering
\includegraphics[angle=0,width=1.0\hsize]{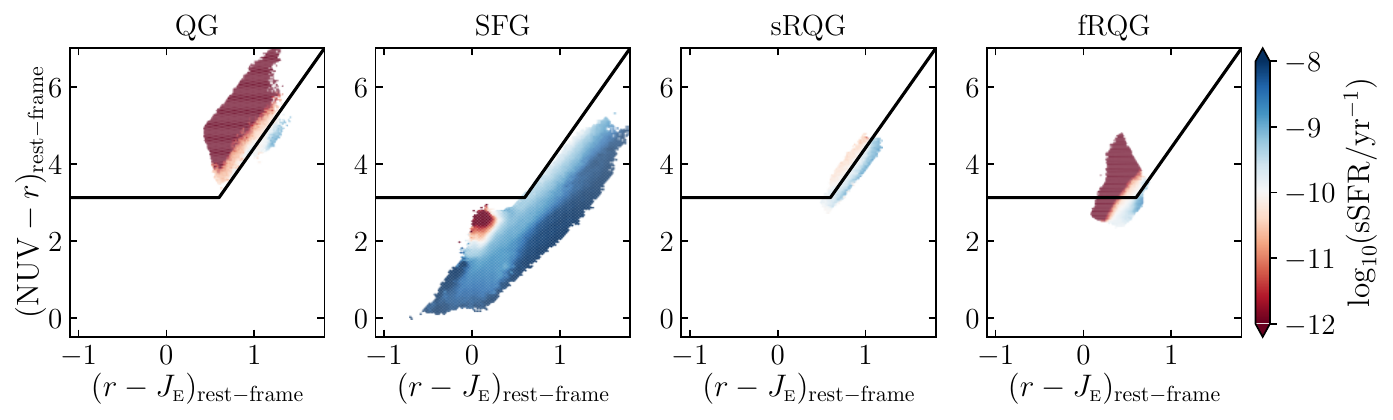}
\caption{Distribution of $UVJ$ classified galaxies on $\mathrm{NUV}RJ$ diagram at $0.2<z<3.0$. The plot is colour-coded by sSFR from red (low sSFR) to blue (high sSFR).}
\label{fig:uvjonNUVRJ}
\end{figure*}
The rest-frame ($\mathrm{NUV} - r$) versus ($r-J$) diagram (hereafter NUV$rJ$ digram) introduced in \citet{Ilbert_2013} is widely used in separating passive and star-forming galaxies. We compare the $UVJ$ classification to the NUV$rJ$ classification, shifting the original boundaries redder by 0.2 mag in ($V-\JE$), due to differences with \Euclid filter \JE. Each population classified by $UVJ$ diagram is shown on the NUV$rJ$ diagram in Fig.\,\ref{fig:uvjonNUVRJ}.

Among the $UVJ$ classified galaxies, $\sim96\%$ of QGs fall into the passive region, and $\sim99\%$ of SFGs fall into the star-forming region. The RQGs either fall within the passive region or near the boundary in the star-forming region, with $\sim43\%$ of fRQGs, and $\sim35\%$ of sRQGs falling into passive regions. 
This shows that the NUVrJ diagram also has the potential to classify galaxies with different levels of star formation activity. With an appropriate scheme, it may further separate sRQGs from fRQGs.

\section{\label{apdx:photz-specz-classification}Classification and properties  for the spectroscopic subsample} 
\begin{figure*}[htbp!]
\centering
\includegraphics[angle=0,width=1.0\hsize]{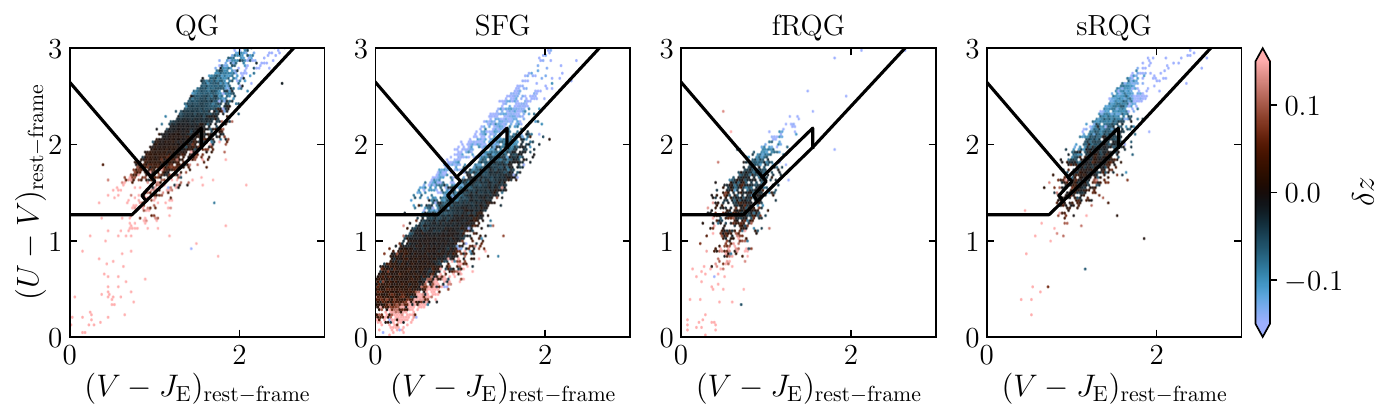}
\caption{$UVJ$ diagram colour-coded by the relative redshift difference $\delta z = (z_{\rm phot}-z_{\rm spec})/z_{\rm spec}$. The galaxies are classified with colours estimated at $z_{\rm spec}$, and the plot shows their positions on the $UVJ$ diagram when their colours are estimated at $z_{\rm phot}$.} 
\label{fig:reclassify}
\end{figure*}
For the spectroscopic  sample  analysed in Sect.\,\ref{sc:spec}, we estimate again the colours of all galaxies at the fixed external spectroscopic redshift.
Figure\,\ref{fig:reclassify} shows how the classification of galaxies changes.
It shows where the populations classified at $z_{\rm spec}$ originally are when the colours are estimated at $z_{\rm phot}$. The plot is colour-coded by the relative redshift difference
$\delta z := (z_{\rm phot}-z_{\rm spec})/z_{\rm spec}$.

\begin{figure}[htbp!]
\centering
\includegraphics[angle=0,width=1.0\hsize]{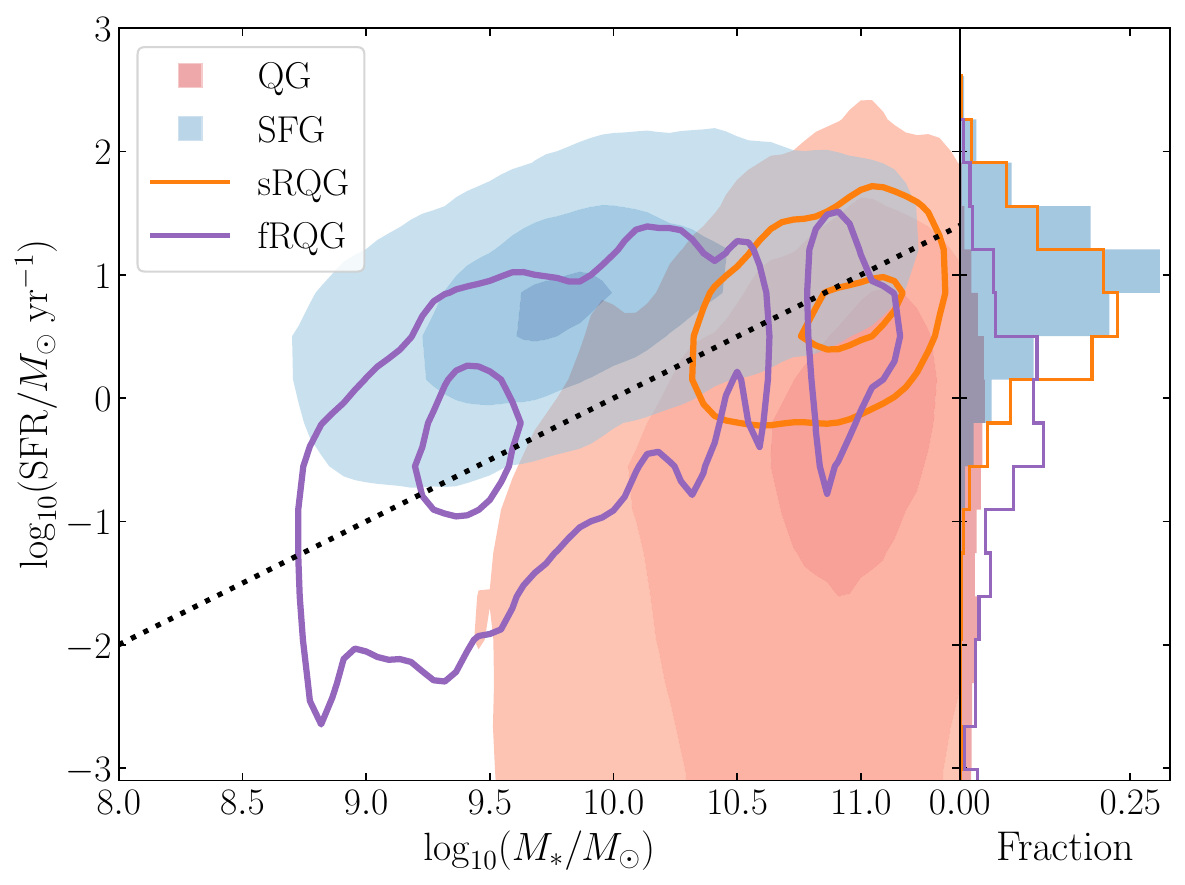}
\caption{SFR versus stellar mass relation of the four populations' spectroscopic subsample in the range $0.2<z<1.5$, with the SFR distribution shown as a side panel. Contours and colours as in Fig.\,\ref{fig:mass-z}. The dotted line marks sSFR of $10^{-10}\,\mathrm{yr}^{-1}$.}
\label{fig:spec-sample-sfr-mass}
\end{figure}

The higher precision of spectroscopic redshift, in particular at $z_{\rm phot}<0.2$, leads to fluctuations in galaxy classifications, but the inflow of new objects is nearly balanced by their outflow, maintaining a stable overall number in each class, $(102.9\pm1.5)$\% for QG, $(98.4\pm0.7)$\% for SFG,  $(110.3\pm7.2)$\% and $(101.3\pm4.1)$\% for fRQG and sRQG).
In particular, we find that $(94\pm 1)\%$ QGs and $(96\pm 1)\%$ SFGs retain their original classification, whereas only $(54\pm3)\%$ and $(53\pm 2)\%$ for fRQGs and sRQGs, respectively.
The significant overlap of RQGs with both QGs and SFGs reflects a mixed population and the transitional nature of these evolutionary phases.
As an illustration of the low impact of these systematics we show the SFR-mass distribution of QG, SFG, fRQG, and sRQG in $z<1.5$ spectroscopic sample in Fig.\,\ref{fig:spec-sample-sfr-mass}. Compared with Fig.\,\ref{fig:SFR-m}, the relative locations between the four populations appear consistent, with RQGs centred at sSFR of $\sim 10^{-10}\,\mathrm{yr}^{-1}$, but only with higher stellar mass, as spectroscopic surveys are limited to the brightest populations.

\section{\label{apdx:MC} Monte Carlo simulations} 
We perform 1000 Monte Carlo realisations by perturbing the rest-frame U, V, and \JE magnitudes, redshift and stellar mass of each galaxy. The uncertainty distributions from \texttt{NNPZ} peak around $\sigma(U)\sim0.12$, $\sigma(V)\sim0.10$, $\sigma(\JE)\sim0.15$, $\sigma(z_{\rm phot})\sim0.04$, $\sigma[\log_{10}(M_\ast/M_{\odot})]\sim0.08$. 
Note that, neglecting the positive covariance from shared \texttt{NNPZ} derivations in drawing errors leads to an artificial increase of colour scatter and a conservative increase in the mixing across narrow colour selections. Overall, our Monte Carlo realisations yield mean class purities of $(91.2 \pm 0.2)\%$, $(94.6 \pm0.1)\%$ for QGs and SFGs, and $(54.4 \pm 0.1)\%$ and $(51.0 \pm 0.2)\%$ for fRQGs and sRQGs -- showing a minimal decrease with redshift  ($<5\%$), and an increase ($<15\%$) with stellar mass and full consistency with the spectroscopic sample (Appendix \ref{apdx:MC}) -- while successfully preserving the relative abundances of each galaxy populations, $(104.0\pm0.2)$\% for QG, $(98.6\pm0.1)$\% for SFG,  $(96.7\pm0.3)$\% for fRQG, and $(102.1\pm0.5)$\% for sRQG.

\begin{figure}[htbp!]
\centering
\includegraphics[angle=0,width=1.0\hsize]{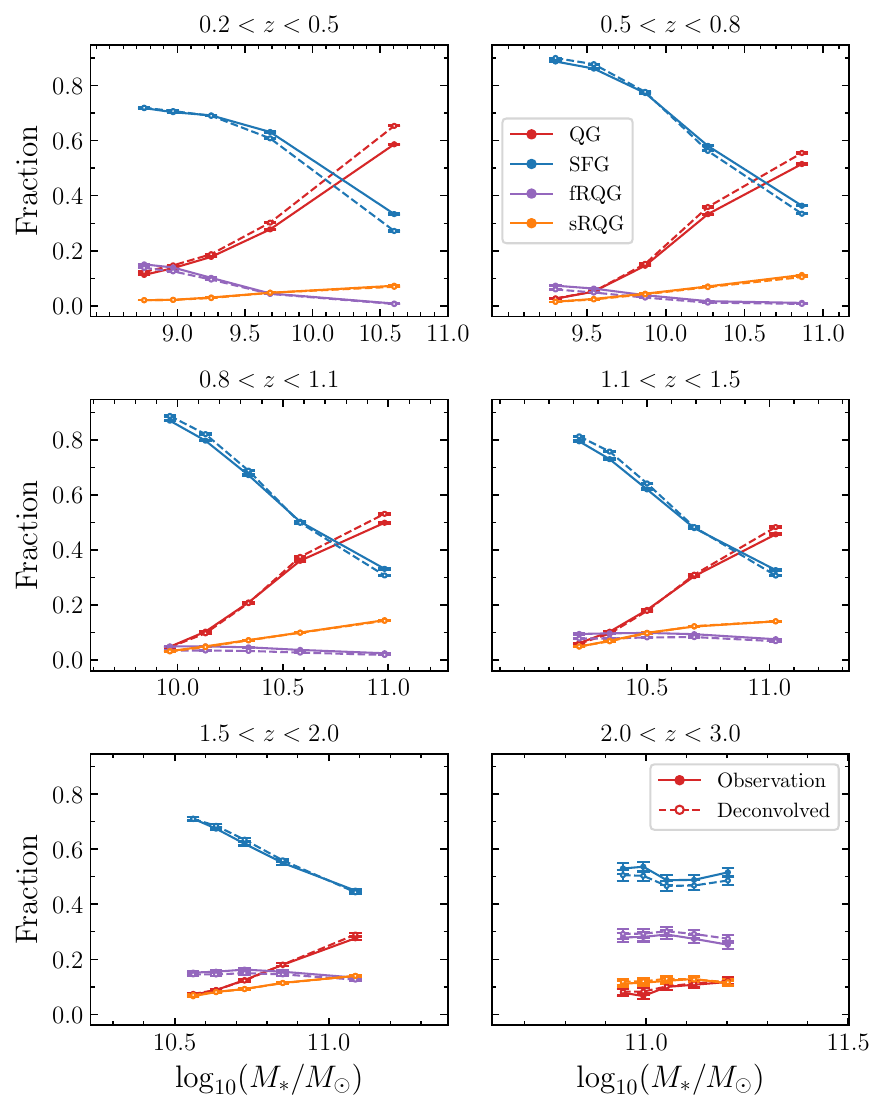}
\caption{{ Fractions of QGs, SFGs, fRQGs, and sRQGs as a function of stellar mass in six redshift bins, with a stellar mass cut at $\log_{10}(M_\ast/M_{\odot}) < 11.25$. 
At each redshift, the full sample is divided into five equipopulated stellar mass bins. Note that we plot in each panel
only the mass range above the corresponding QGs completeness
limit. For comparison,  we show with dashed lines
the deconvolved fractions derived by inverting the MC mixing classification matrices. }}
\label{fig:fraction_corrected}
\end{figure}

For each mass and redshfit bin, we invert the mixing matrix obtained from a MC analysis \citep{Starck_2002} for each class to derive the deconvolved fractions, QE, and QS. 
In Fig.\,\ref{fig:fraction_corrected}, we compare the mass dependences of fractions at different redshifts before and after deconvolution. The impact on QE, and QS is shown in Appendix \ref{apdx:QEQS}. This effect should be interpreted as an upper limit, given that the MC overestimates mixing as a consequence of both the over-smoothing of the data, and the independent sampling of the three bands, redshift and mass.

\section{\label{apdx:QEQS} QE and QS in redshift bins} 
\begin{figure}[htbp!]
\centering
\includegraphics[angle=0,width=1.0\hsize]{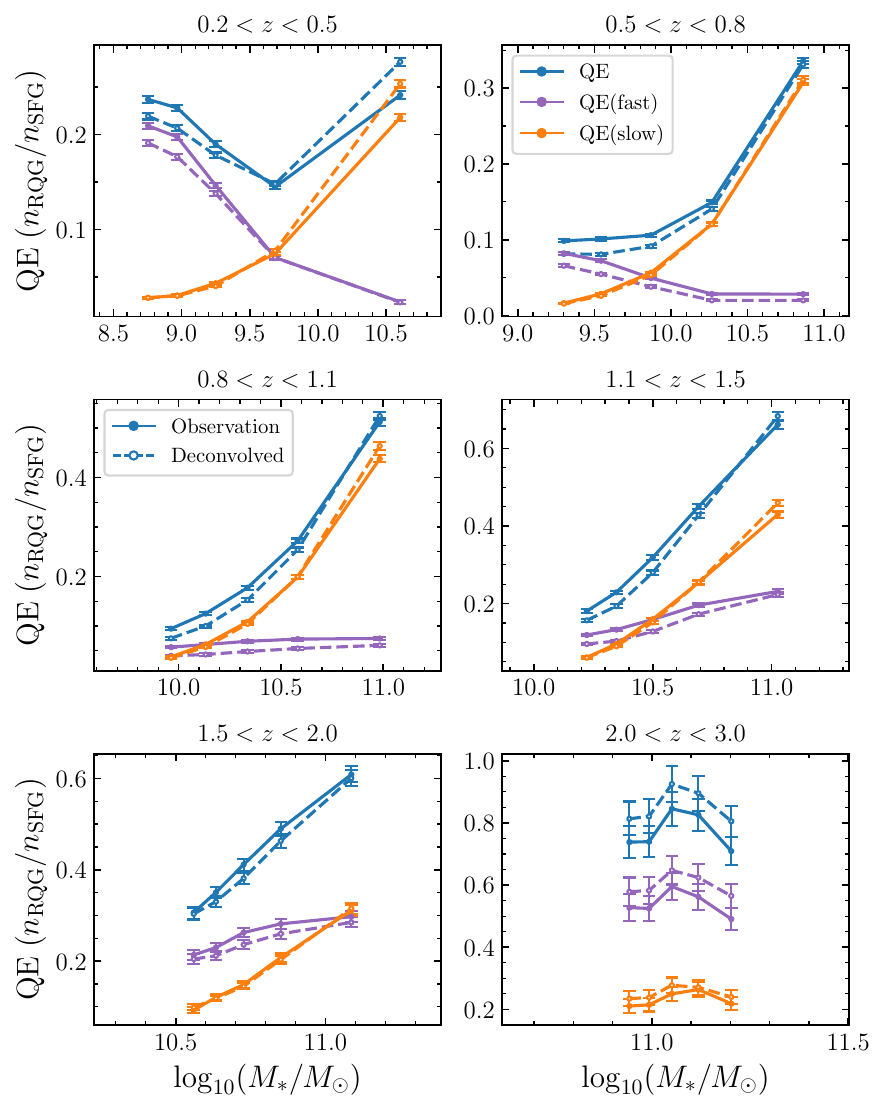}
\caption{Quenching efficiency QE ($n_{\rm RQG}/n_{\rm SFG}$) as a function of stellar mass in six redshift bins, with a stellar mass cut at $\log_{10}(M_\ast/M_{\odot}) < 11.25$. The blue line shows QE for all RQGs, while purple and orange lines the contribution of fast- and slow-quenching RQGs, respectively. 
At each redshift, the full sample is divided into five equipopulated stellar mass bins. Note that we plot in each panel
only the mass range above the corresponding QGs completeness
limit. We show with dashed lines the QE derived by inverting the MC mixing classification matrices. }
\label{fig:qe}
\end{figure}
In Fig.\,\ref{fig:qe}, we investigate the mass dependences of total QE at different redshifts, applying the corresponding {the most conservative mass completeness cut for QGs (Sect.\,\ref{sc:quenchingproperties})}, together with the contribution of fast and slow quenching. The uncertainties are the combinations of classification uncertainties and Poisson errors (Sect.\,\ref{sc:quenchingproperties}).
Slow quenching dominates the most massive galaxies up to $z=1.5$, reaching $40$--$50\%$ at $0.8<z<1.5$, while fast quenching is efficient for low-mass galaxies at low redshifts ($z<0.5$), arriving up to $20\%$, and is efficient for high-mass galaxies at high redshift ($z>1.5$), reaching $50\%$. At $0.5<z<2$, the total QE always increases with stellar mass, indicating that quenching happens earlier for massive galaxies at higher redshifts. The slow-quenching efficiency increases with stellar mass in all redshift bins {below $z<2$}, connecting it to the mass quenching mechanisms. On the other hand, the fast quenching's dependence on stellar mass varies with redshift, {being decreasing with stellar mass at $z<0.8$, flat at $0.8<z<1.1$, and increasing at $1.1<z<1.5$}. 
This suggests that the mechanisms of fast quenching are different for galaxies at low and high redshifts.

\begin{figure}[htbp!]
\centering
\includegraphics[angle=0,width=1.0\hsize]{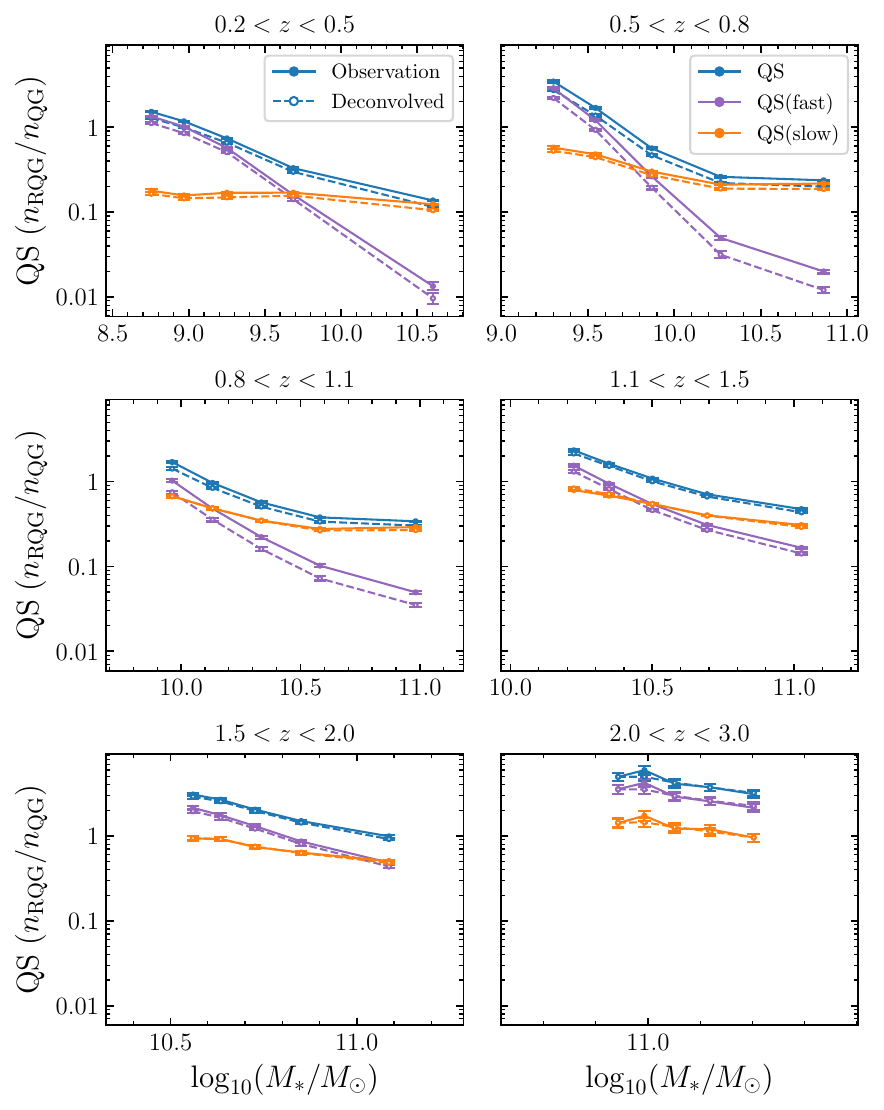}
\caption{Quenching stage ($n_{\rm RQG}/n_{\rm QG}$) as function of stellar mass in five redshift bins, with a stellar mass cut at $\log_{10}(M_\ast/M_{\odot}) < 11.25$.  Colours and binning as in Fig.\,\ref{fig:qe}. For comparison, we show with dashed lines the QS derived by inverting the MC mixing classification matrices. }
\label{fig:qs}
\end{figure}
Figure\,\ref{fig:qs} shows the mass dependence of the total QS, together with the contribution of fast and slow quenching, at different redshifts. 
The total, fast-quenching and slow-quenching QS all decrease with stellar mass at $0.2<z<3$, supporting the downsizing scenario. At $2<z<3$, slow quenching is at a later stage than fast quenching for galaxies of all stellar masses, suggesting that it began earlier. Conversely, at $0.2 < z < 2$, low-mass galaxies experience fast quenching first, while massive galaxies show it later, implying that fast quenching predominantly occurs in massive galaxies at $z > 2$ and in low-mass ones at $0.2 < z < 3$. 

\section{\label{apdx:CAS}Morphological properties}
\begin{figure}[htbp!]
\centering
\includegraphics[angle=0,width=1.0\hsize]{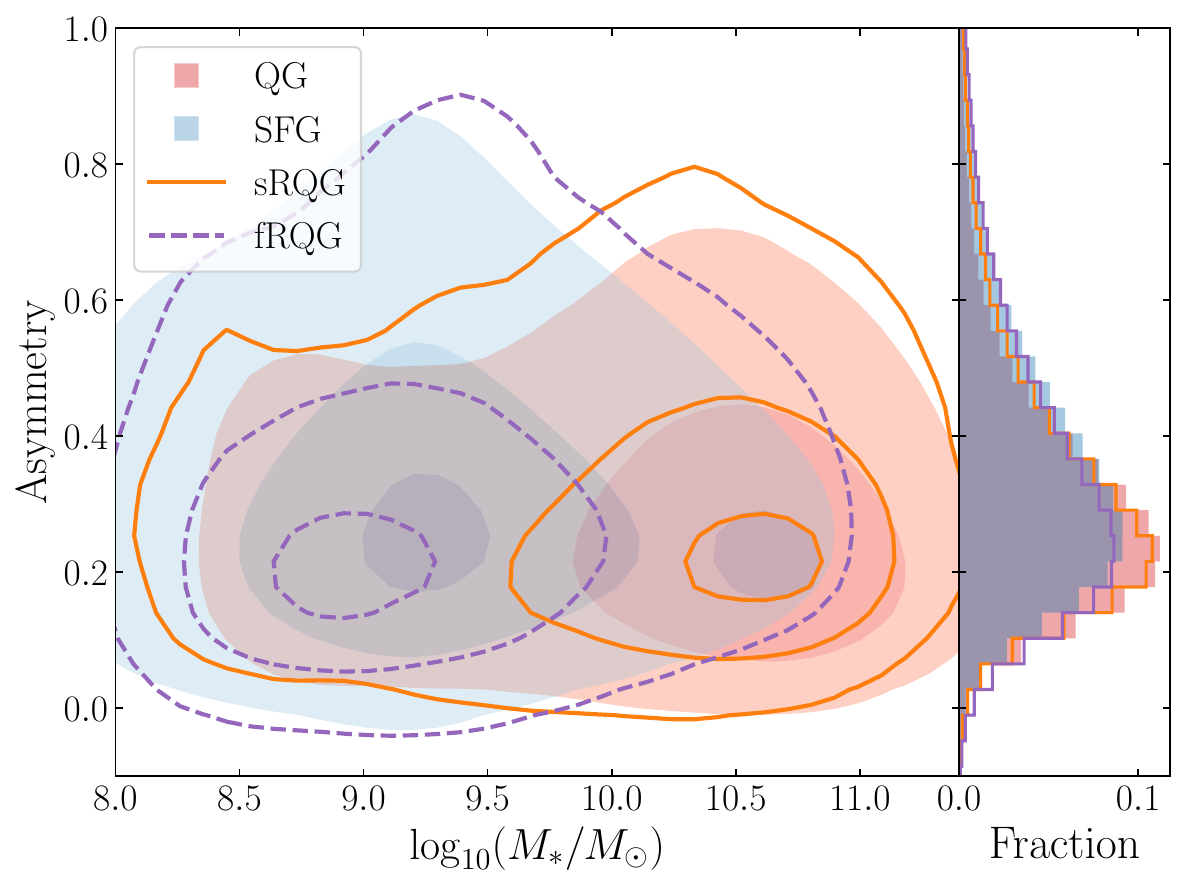}
\caption{Asymmetry versus stellar mass relation of the four populations, with asymmetry distribution as a side panel. The sample is cut with $0.2<z<1.5$. Contours and colours as in Fig.\,\ref{fig:mass-z}.
}
\label{fig:asymmetry-mass}
\end{figure}

\begin{figure}[htbp!]
\centering
\includegraphics[angle=0,width=1.0\hsize]{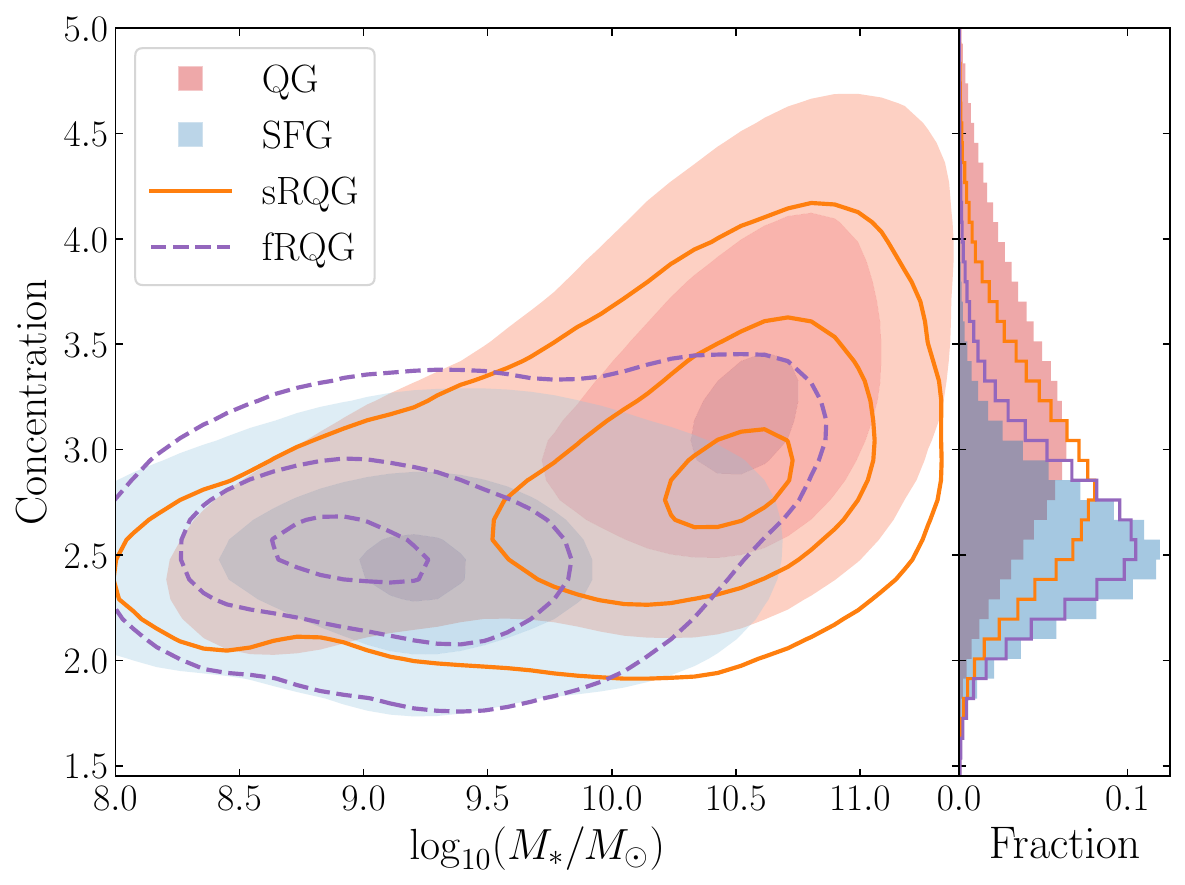}
\caption{Concentration versus stellar mass relation of the four populations. The sample is cut with $0.2<z<1.5$. Contours and colours as in Fig.\,\ref{fig:mass-z}.
}
\label{fig:concentration-mass}
\end{figure}

\begin{figure}[htbp!]
\centering
\includegraphics[angle=0,width=1.0\hsize]{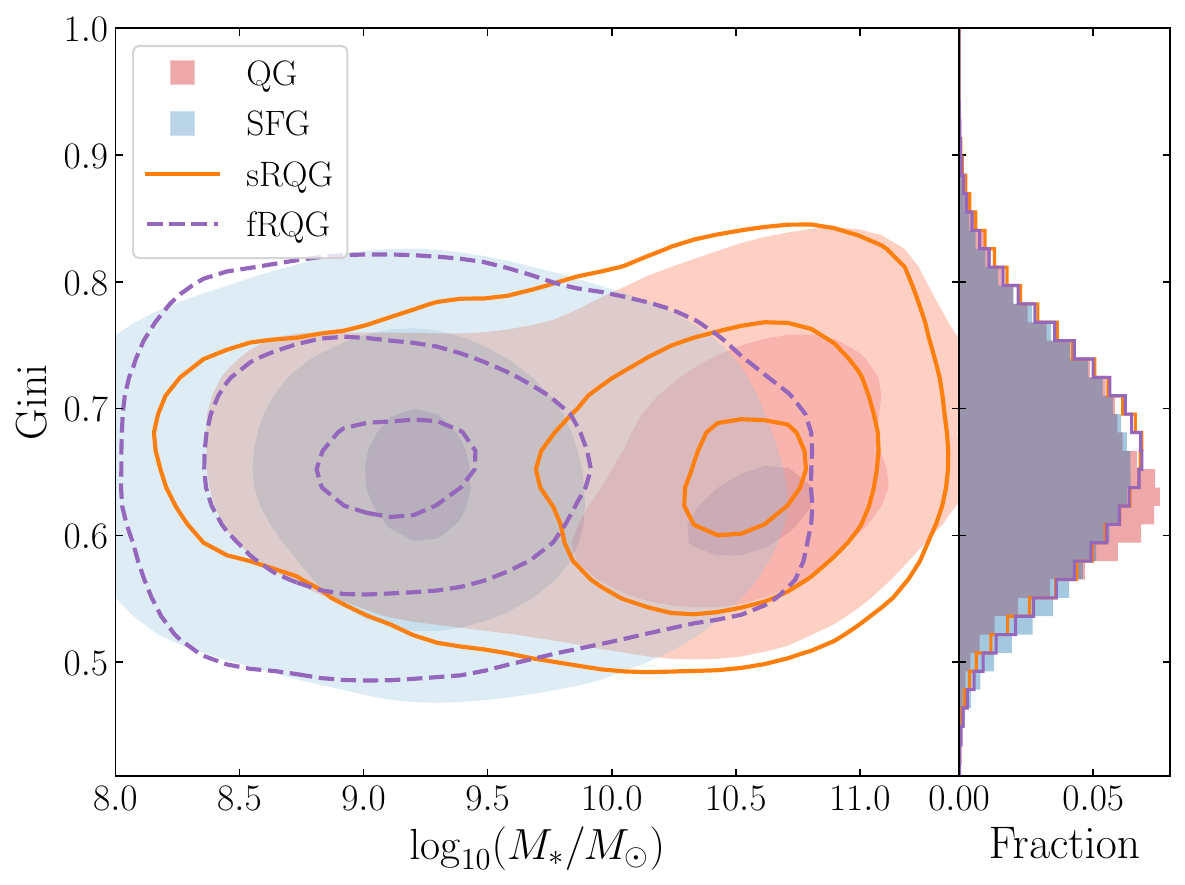}
\caption{Gini coefficient versus stellar mass relation of the four populations. The sample is cut with $0.2<z<1.5$. Contours and colours as in Fig.\,\ref{fig:mass-z}.
}
\label{fig:gini-mass}
\end{figure}

The asymmetry, and Gini coefficient indicate the irregularity of a galaxy and can be used to distinguish whether a galaxy has undergone a merger or other morphological disturbance. 
{The concentration coefficient is a degenerate indicator of the morphological type of a galaxy. The asymmetry, concentration coeeficient and Gini coefficient are measured in \citet{Q1-TP004} in VIS band.} Here we show the relations of asymmetry (Fig.\,\ref{fig:asymmetry-mass}), concentration (Fig.\,\ref{fig:concentration-mass}), and Gini coefficient (Fig.\,\ref{fig:gini-mass}) with stellar mass. The RQGs do not show any significant differences from SFGs and QGs, {in their asymetry and Gini coefficient,} consistent with Sect.\,\ref{sc:discussion}, which indicates that merger is not a major mechanism for quenching. {The concentration coefficient, instead, shows that QG and sRQG are more concentrated than the fRQG and the SF, and confirm the result find using the S\'ersic index (Fig.\,\ref{fig:n-mass}). The higher concentration in QG than SFG at high masses is consistent with \cite{Quilley_2022}. }
\section{\label{apdx:env-fraction}Environmental distribution fractions}
\begin{figure*}[htbp!]
\centering
\includegraphics[angle=0,width=1.0\hsize]{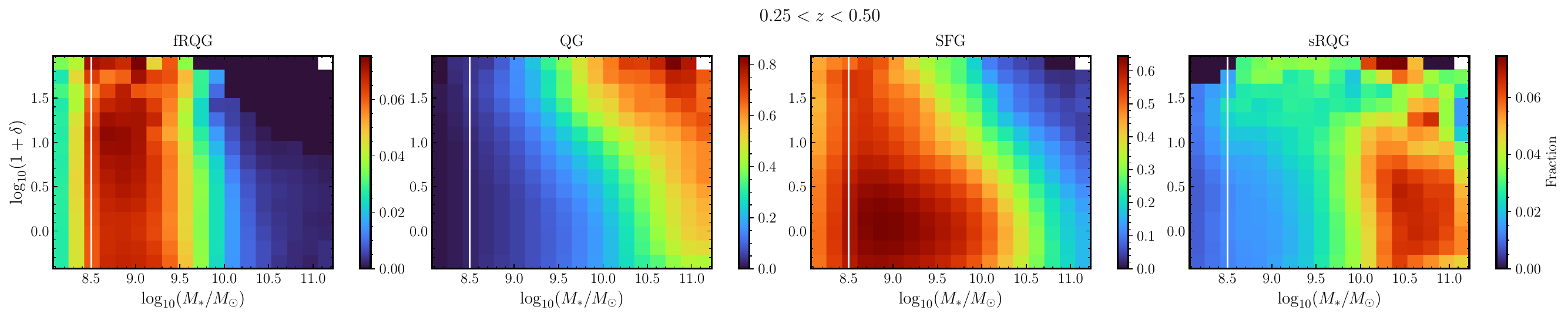}
\includegraphics[angle=0,width=1.0\hsize]{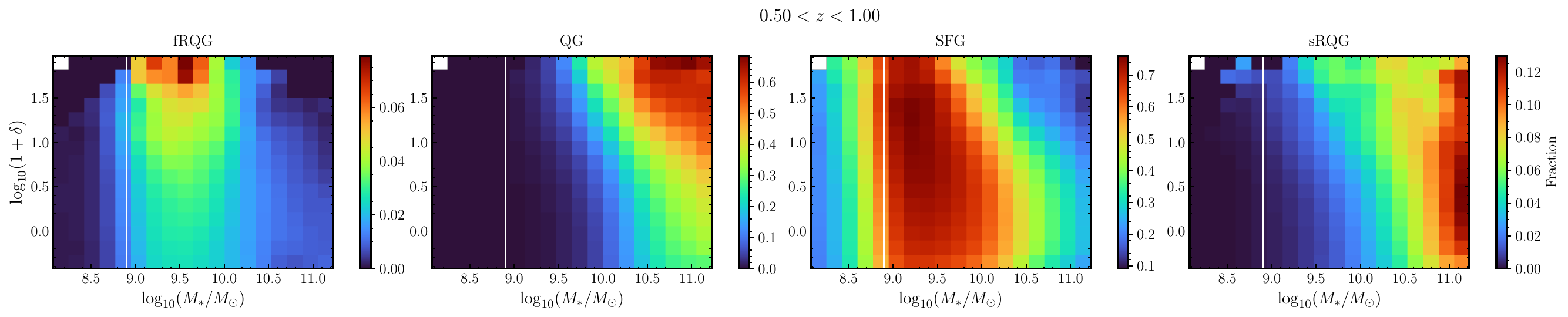}
\includegraphics[angle=0,width=1.0\hsize]{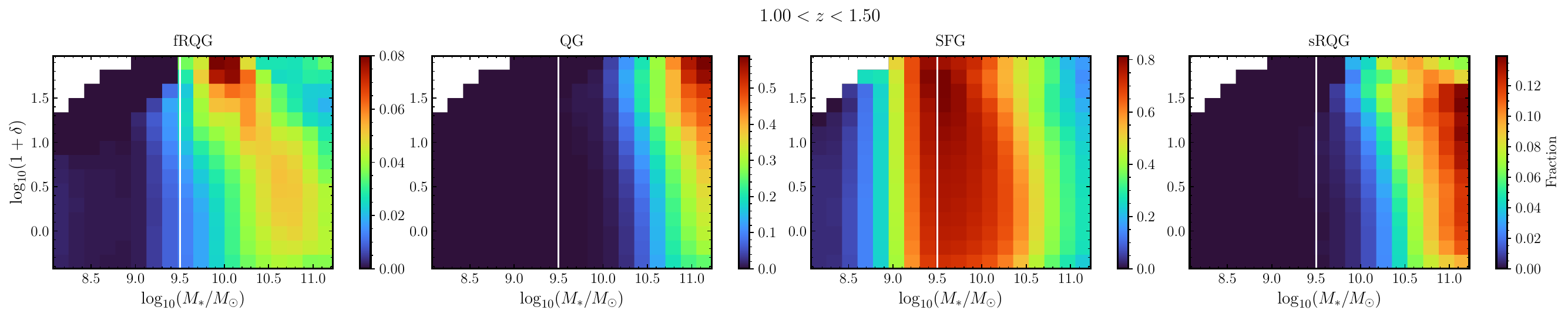}
\caption{Fractions of fRQG, QG, SFG, and sRQG on stellar mass-local density panels. Three rows show five redshift bins, $0.25<z<0.5$, $0.5<z<1$, and $1<z<1.5$. {The solid lines show the QG mass completeness limit in each redshift bin.}}
\label{fig:Fraction-mass-env}
\end{figure*}

The distributions of galaxies in various environments are displayed in terms of relative abundance {in Fig.\,\ref{fig:Fraction-mass-env} mass-local density panel for the three redshift intervals at $z<1.5$}. The fRQGs have a higher relative abundance in dense environments at $z>0.5$. This result for fRQGs is consistent with \cite{Q1-SP069}, finding that quiescent discy galaxies (which could be the later stage of our fRQGs) are more abundant in the low-mass regime of high-density environment. 
For comparison, the sRQGs have a higher relative abundance in dense environments at $z>1.25$, and gradually move to low-density environments at $z<0.75$.

\section{\label{sc:stackmethod}Stacking method} 
We carry out the stacking procedure using the \texttt{SpectraPyle} code introduced in \citet{EP-Quai}. Spectra are first divided 
in redshift bins ($z_{\rm min}<z<z_{\rm max}$), shifting each of them to the rest-frame and normalising at the common rest-frame wavelength range, that is, $\num{12100}\,\AA/(1+z_{\rm min})<\lambda_{\rm rest}<\num{18900}\,\AA/(1+z_{\rm max})$, after removing strong line feature ranges. In the case where the common range without strong lines is too narrow, we adopt a wider range for normalisation to minimise the effects of lines. 
For QGs, we stack spectra in five redshift bins in the range $0<z<1.5$.
For sRQGs, we stack spectra in three redshift bins in the range $0<z<1.5$.
For fRQGs, we stack in redshift bins $0<z<0.35$, $0.35<z<1.0$, and $1.0<z<1.5$, to have similar number of spectra in the first two bins while the spectra in each bin still share a common wavelength range.
All spectra are resampled in the rest frame with $7\,\AA$ per pixel (corresponding to an NISP pixel of $\sim 13.4\,\AA$ in the rest frame at $z\sim1$). We apply sigma clipping at $4\sigma$, using spectra with a minimum of two dithers over four observed, and remove bad pixels with the bad pixel mask (\texttt{flag}=0). For each spectrum in stacking, ten pixels are removed at both edges.

\begin{figure}[htbp!]
\centering
\includegraphics[angle=0,width=1.0\hsize]{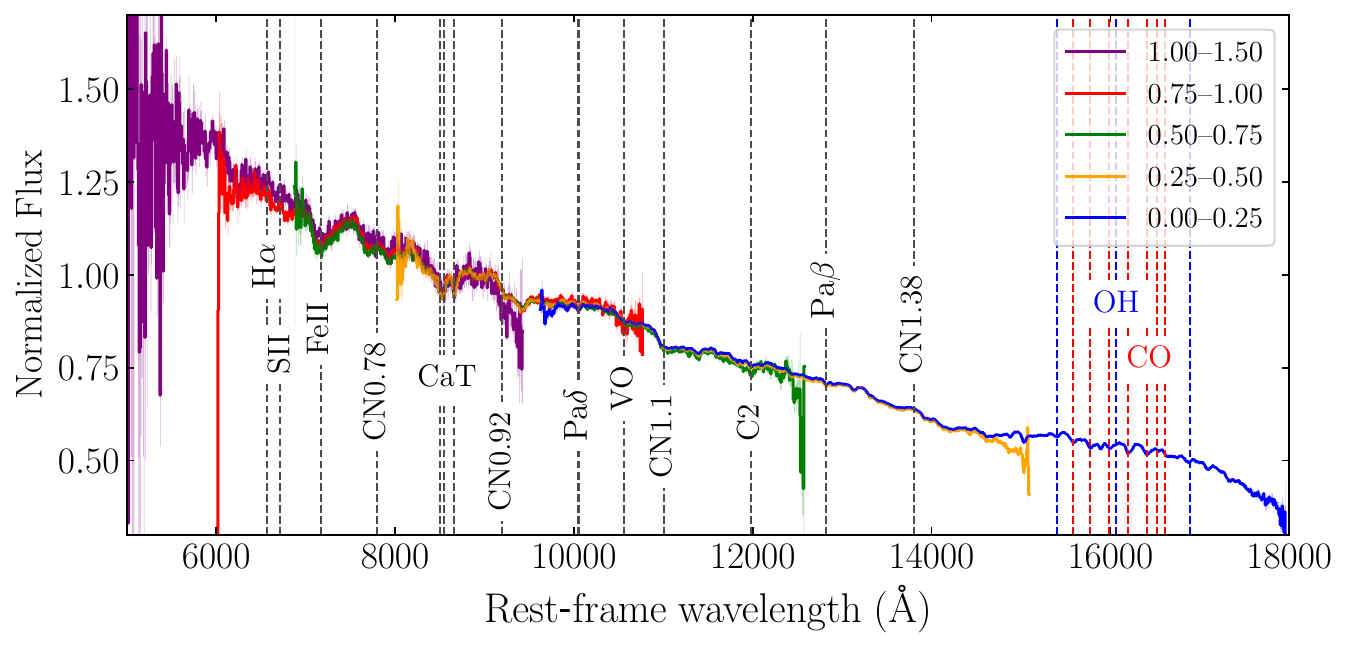}
\includegraphics[angle=0,width=1.0\hsize]{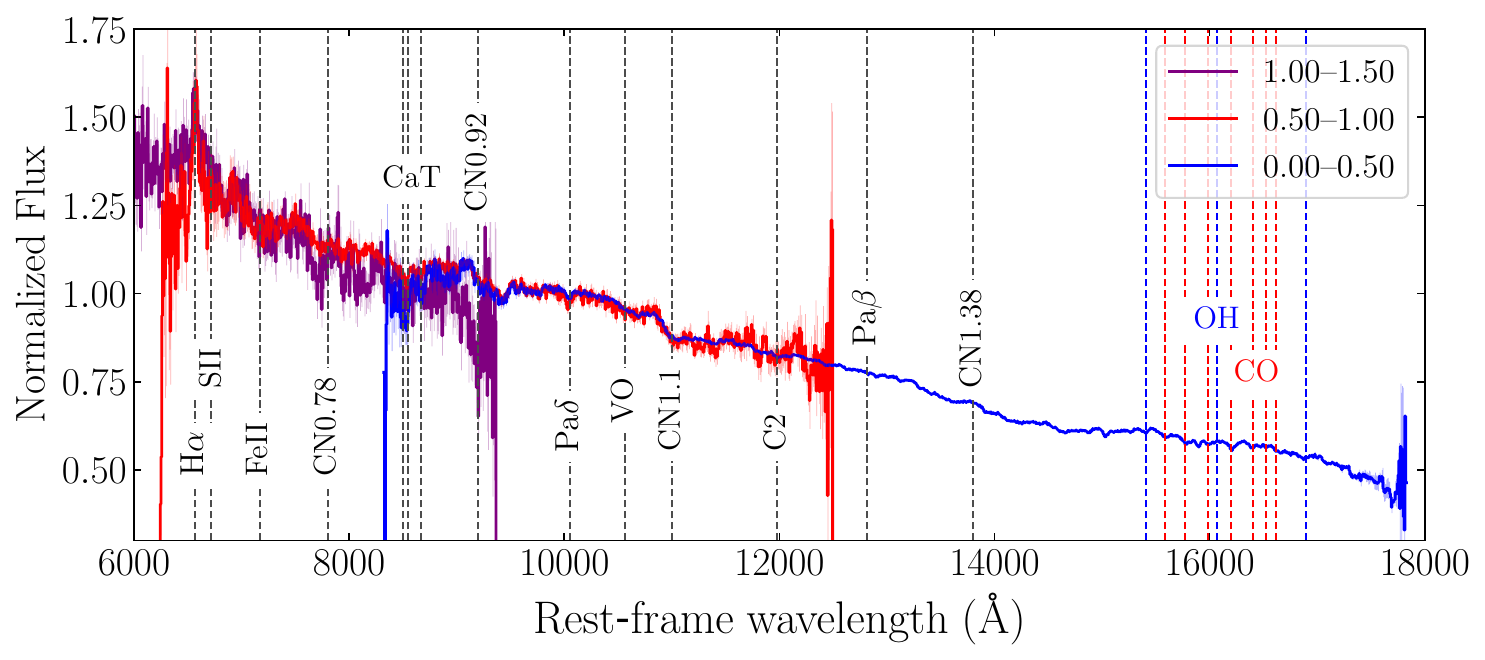}
\includegraphics[angle=0,width=1.0\hsize]{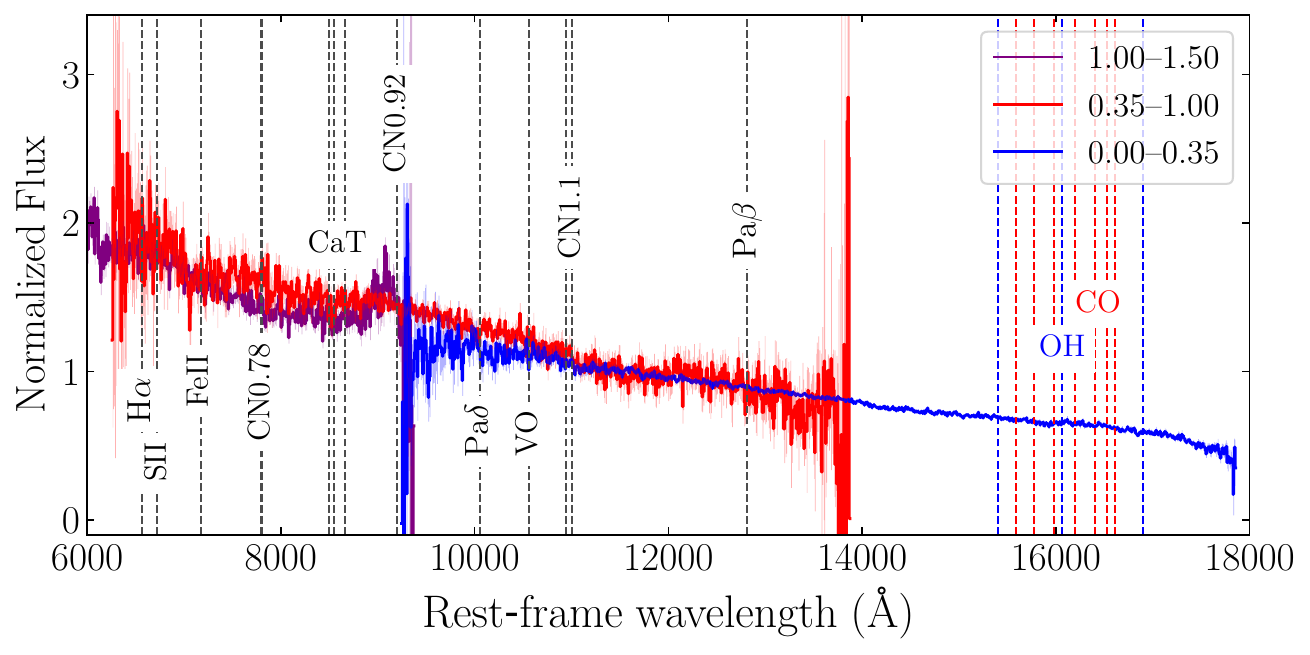}
\caption{Median stacked spectra for QG ({\it top panel}), sRQG ({\it mid panel}) and fRQG ({\it bottom panel}) sample in different redshift bins. The redshift bin information is in the legends. All red vertical lines mark CO lines, and blue lines mark OH lines.}
\label{fig:stack-in-z-bins}
\end{figure}

We first stack QGs and RQGs population, in different redshift bins.
We derive the dispersion of each stacked spectra to demonstrate the homogeneity of the populations.
We find that the mean dispersion is about 16\% for QGs and sRQGs, being minimal in the normalisation range and in the lowest redshift bin, mainly due to the spectral noise, and increases by only 2\% in the outskirts of the stacked spectra. For fRQGs, due to the lower number of spectra, the mean dispersion is of the order of 30\% and increases by 5\% in the outskirt of the templates.
Figure\,\ref{fig:stack-in-z-bins} shows the stacked spectra of each population across different redshift bins, renormalised so that the spectra share a common continuum level. We note that, based on the \Euclid wavelength coverage and the current sample with a reliable external $z_{\rm spec}$, they do not show any significant differences for all the three populations.

\begin{figure}[htbp!]
\centering
\includegraphics[angle=0,width=1.0\hsize]{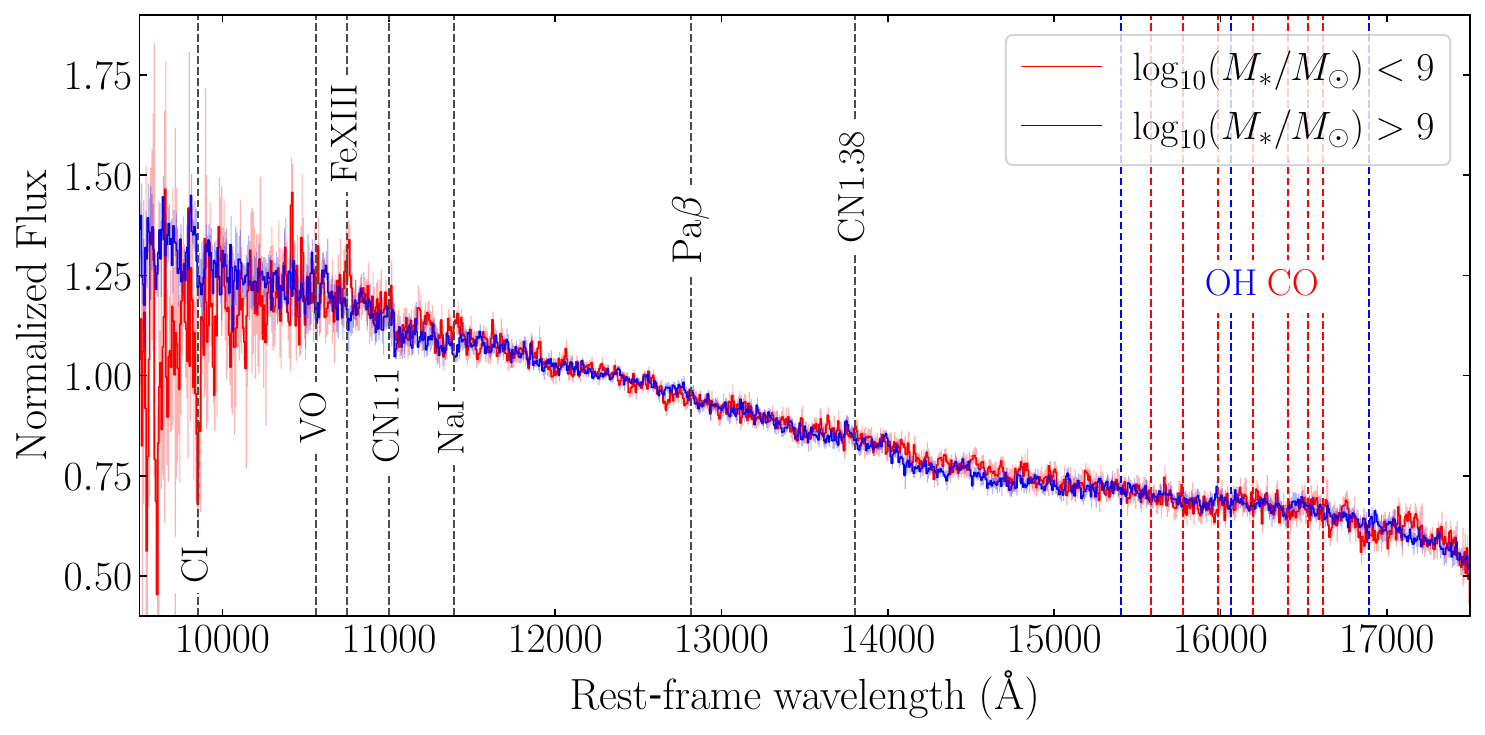}
\caption{Stacked spectra in two stellar mass bins for fRQG at $0<z<0.5$. All red vertical lines mark CO lines, and blue lines mark OH lines.}
\label{fig:fRQG_in_mass_bin}
\end{figure}

\begin{figure}[htbp!]
\centering
\includegraphics[angle=0,width=1.0\hsize]{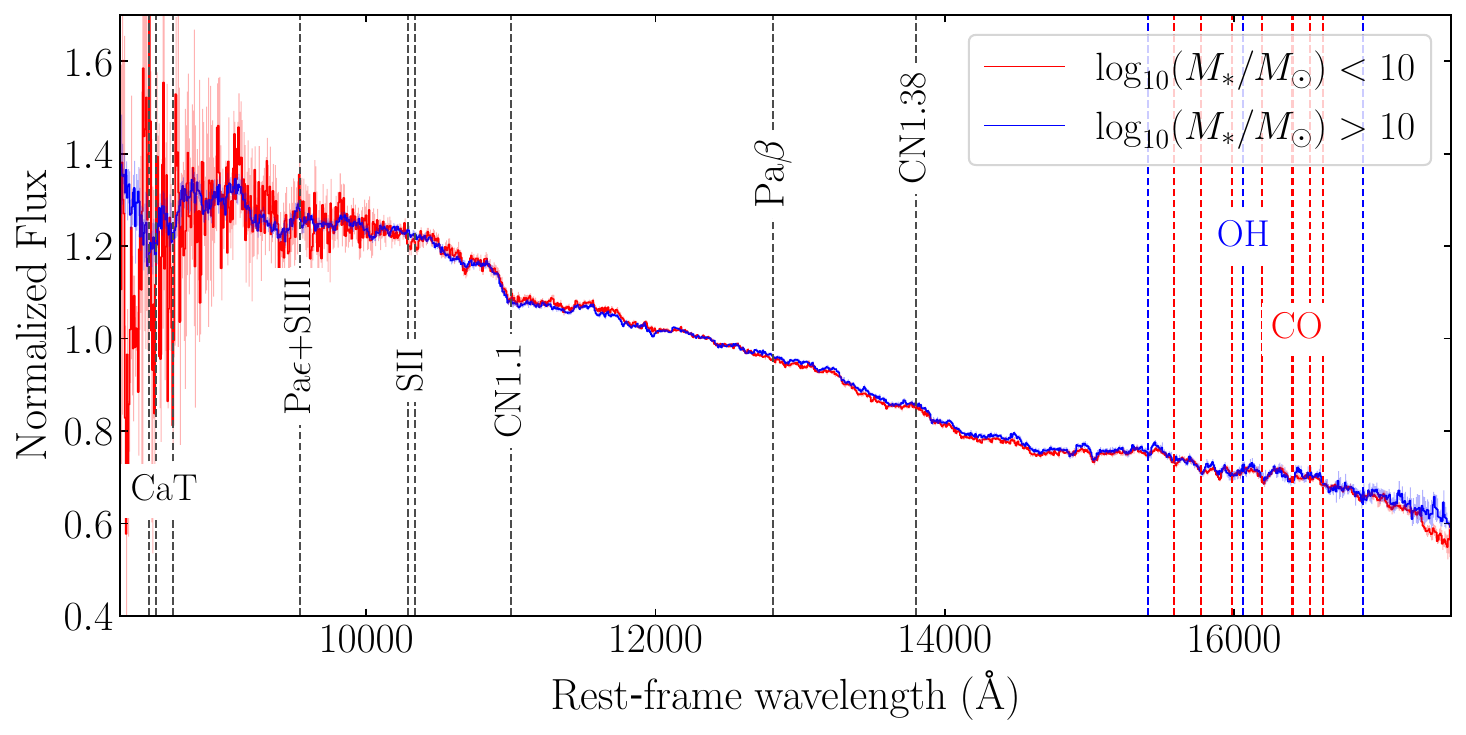}
\caption{This figure shows stacked spectra in two stellar mass bins for sRQG at $0<z<0.5$.}
\label{fig:sRQG_in_mass_bin}
\end{figure}

\begin{figure}[htbp!]
\centering
\includegraphics[angle=0,width=1.0\hsize]{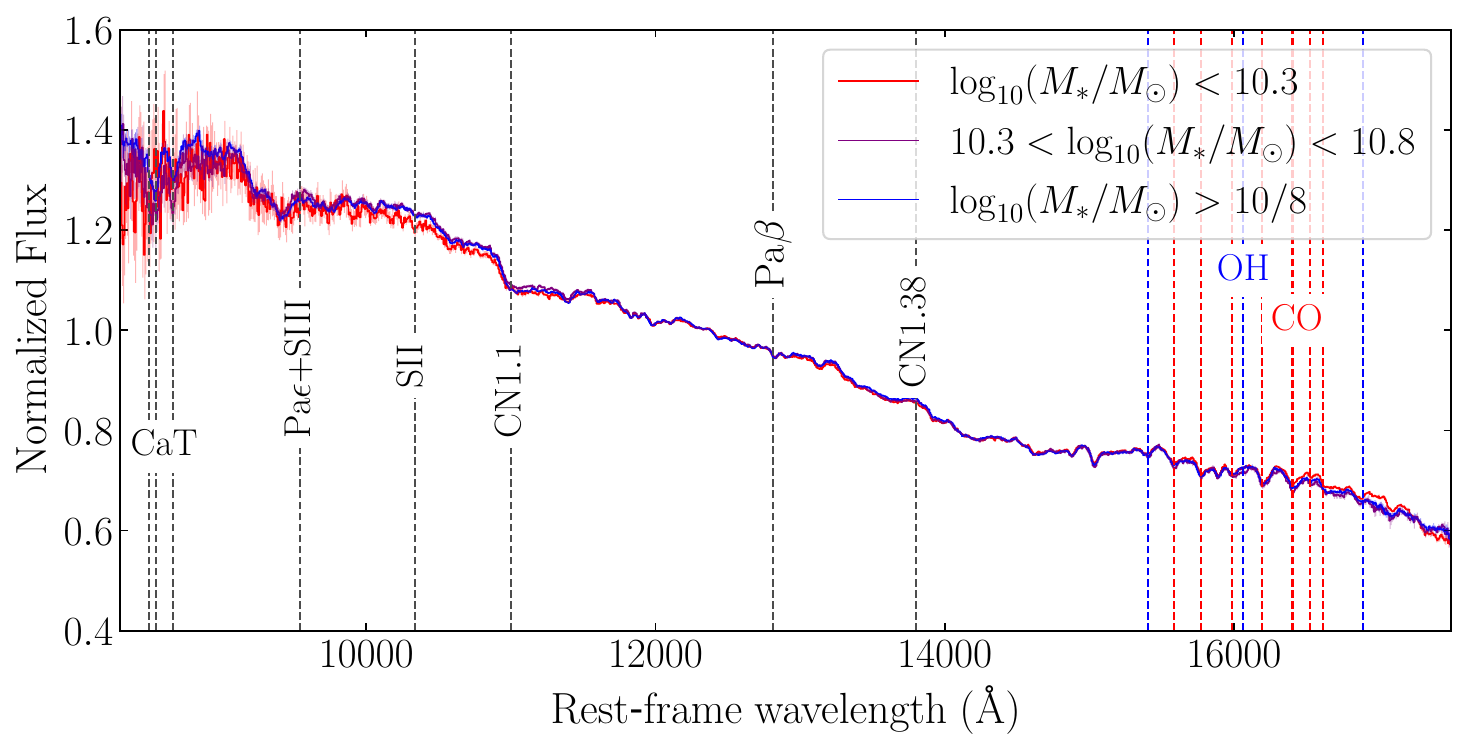}
\includegraphics[angle=0,width=1.0\hsize]{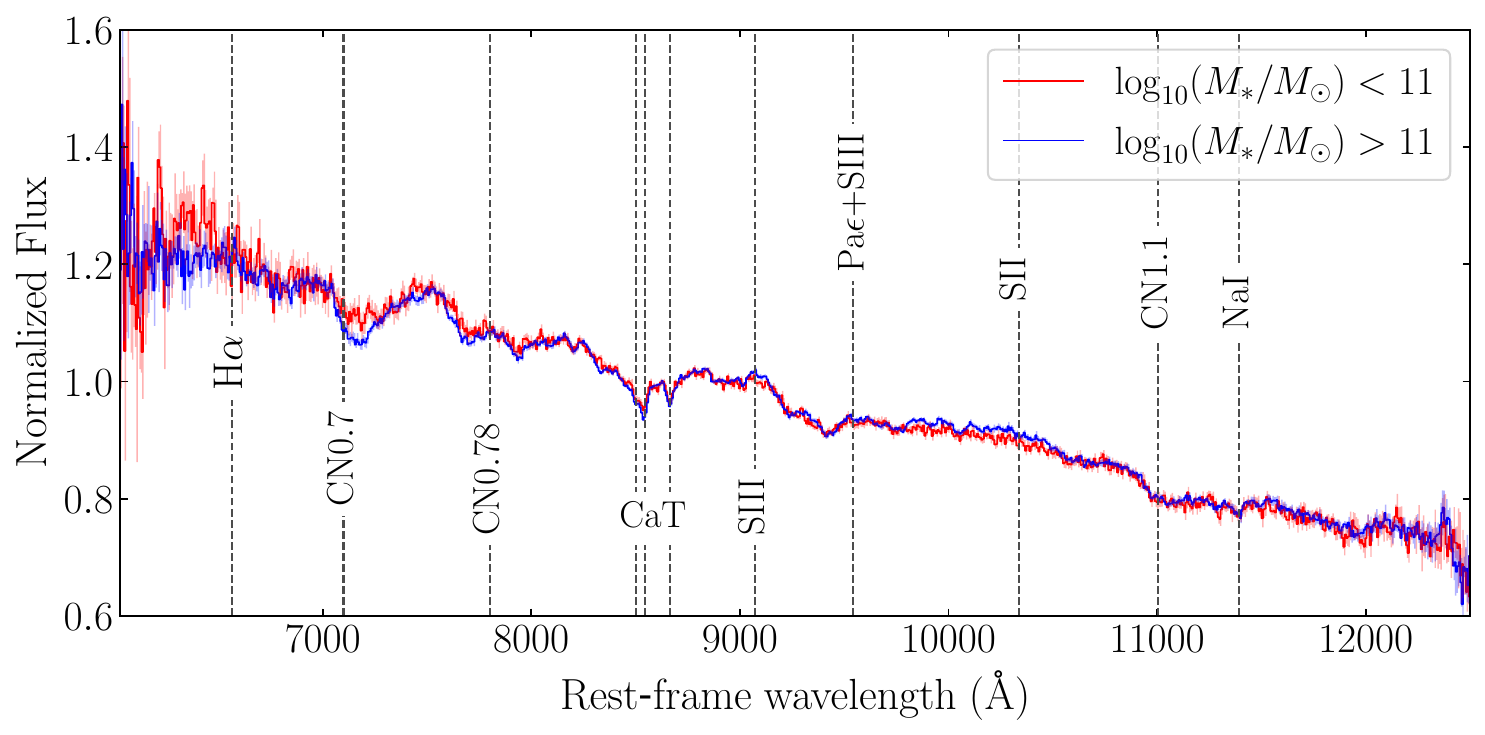}
\caption{Stacked QG spectra in different stellar mass bins. The top panel is at $0<z<0.5$, stacked in three bins, and the bottom panel is at $0.5<z<1$, stacked in two bins. All red vertical lines mark CO lines, and blue lines mark OH lines.}
\label{fig:QG_in_mass_bin}
\end{figure}

In addition, we stack spectra in each redshift bin further dividing into various stellar mass bins, and finding no significant difference between the spectra in each mass bin. 
Figure\,\ref{fig:fRQG_in_mass_bin} compares stacked fRQG spectra in two stellar mass bins at $0<z<0.5$. We note that the \ion{Fe}{xiii} line in the two bins shows a possible difference, being in emission in the low mass bin spectra, with EW(\ion{Fe}{xiii}) $= (6.1\pm1.2)\,\AA$, whereas it is not evident in the high mass spectra, with very low S/N $<2$. 
Figure\,\ref{fig:sRQG_in_mass_bin} shows the differences between the stacked sRQG spectra in two stellar mass bins at $0<z<0.5$. Figure\,\ref{fig:QG_in_mass_bin} shows the differences between stacked QG spectra in stellar mass bins at $0<z<0.5$ (top panel) and $0.5<z<1$ (bottom panel).
There are no major differences between the stacked spectra in each mass bin.

\section{\label{apdx:DESI_stack}DESI stacked spectra}
\begin{figure}[htbp!]
\centering
\includegraphics[angle=0,width=1.0\hsize]{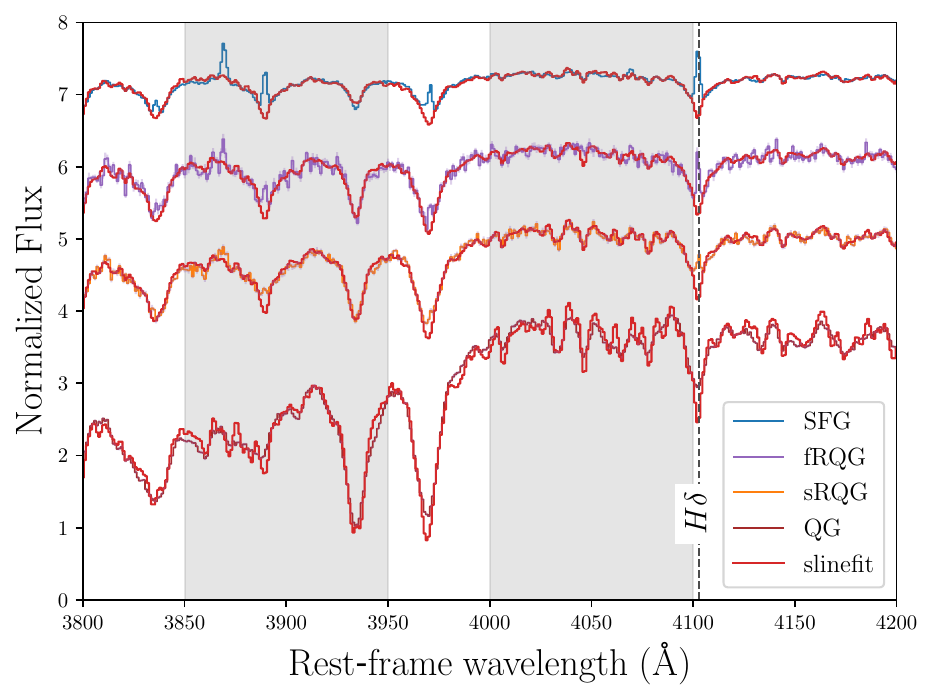}
\caption{Comparison between stacked DESI spectra of SFGs, QGs, fRQGs, and sRQGs around $4000\,\AA$ break and $\mathrm{H}\,\delta$ absorption. Shadow areas marks wavelength range used to calculate ${\rm Dn4000}$. Red lines are fitted with \texttt{slinefit} to remove impact of emission line.
}
\label{fig:desi_stack}
\end{figure}

We cross-match Q1 spectroscopic sample with DESI DR1 and find \num{6652} QGs, \num{16859} SFGs, 876 sRQGs, and 504 fRQGs. We stack these galaxies in three populations following the same method in Sect.\,\ref{sc:stackmethod}. The stack of DESI spectra are presented in Fig. \ref{fig:desi_stack}. After removing emission line component, fitting them using \texttt{slinefit}, we measure with \texttt{PyLick} the Balmer absorption features and $4000$\,\AA\ break.
The fRQG has EW$(\mathrm{H}\,\delta) = 5.7$\,\AA, and ${\rm Dn4000}=1.3$. The sRQG has EW$(\mathrm{H}\,\delta) = 3.5$\,\AA, and ${\rm Dn4000}=1.4$. The QG has EW$(\mathrm{H}\,\delta) = 1.1$\,\AA, and ${\rm Dn4000}=1.7$. These values are consistent with what expected for RQGs and QGs (see figure\,A.2 in \citealt{Mao_2022}). For comparison, SFGs has EW$(\mathrm{H}\,\delta) = 4.1$\,\AA, and ${\rm Dn4000}=1.2$. 

\end{appendix}

\label{LastPage}
\end{document}